\documentclass{emulateapj} 
\def\em{\it} 
 
\shorttitle{Galactic Positron Annihilation} 
 
\shortauthors{Higdon, Lingenfelter and Rothschild}

\begin{document}

\title{The Galactic Positron Annihilation Radiation \& the 
Propagation of Positrons \\ 
in the Interstellar Medium} 
 
\author{J. C. Higdon} 
\affil{W. M. Keck Science Center, Claremont Colleges, Claremont, 
CA 91711-5916 \\ 
and California Institute of Technology, Pasadena, CA 
91125\footnotemark[1]} 
 \email{jimh@lobach.JSD.claremont.edu} 
 
\and 
 
\author{R. E. Lingenfelter and R. E. Rothschild} 
\affil{Center for Astrophysics and Space Sciences, University of 
California San Diego, La Jolla, CA 92093} 
\email{rlingenfelter@ucsd.edu, rrothschild@ucsd.edu} 
 
\begin{abstract} 
 
The ratio of the luminosity of diffuse 511 keV positron 
annihilation radiation, measured by INTEGRAL in its four years, 
from a Galactic ``positron bulge" ($<$ 1.5 kpc) compared to that 
of the disk is $\sim$ 1.4. This ratio is roughly 4 times larger 
than that expected simply from the stellar bulge to disk ratio of 
$\sim$ 0.33 of the Galactic supernovae, which are thought to be 
the principal source of the annihilating positrons through the 
decay of radionuclei made by explosive nucleosynthesis in the 
supernovae. This large discrepancy has prompted a search for new 
sources. 
 
Here, however, we show that the measured 511 keV luminosity ratio 
can be fully understood in the context of a Galactic supernova 
origin when the differential propagation of these $\sim$ MeV 
positrons in the various phases of the interstellar medium is 
taken into consideration, since these relativistic positrons must 
first slow down to energies $\leq$ 10 eV before they can 
annihilate. Moreover, without propagation, none of the proposed 
positron sources, new or old, can explain the two basic properties 
on the Galactic annihilation radiation: the fraction of the 
annihilation that occurs through positronium formation and the 
ratio of the broad/narrow components of the 511 keV line. 
 
In particular, we show that in the neutral phases of the 
interstellar medium, which fill most of the disk ($>$ 3.5 kpc), 
the cascade of the magnetic turbulence, which scatters the 
positrons, is damped by ion-neutral friction, allowing positrons 
to stream along magnetic flux tubes. We find that nearly 1/2 of 
the positrons produced in the disk escape from it into the halo. 
On the other hand, we show that within the extended, or 
interstellar, bulge ($<$ 3.5 kpc), essentially all of the 
positrons are born in the hot plasmas which fill that volume. We 
find that the diffusion mean free path is long enough that only a 
negligible fraction annihilate there and $\sim$ 80\% of them 
escape down into the HII and HI envelopes of molecular clouds 
that lie within 1.5 kpc before they slow down and annihilate, 
while the remaining $\sim$ 20\% escape out into the halo and the 
disk beyond. This propagation accounts for the low observed 
annihilation radiation luminosity of the disk compared to the 
bulge. 
 
In addition, we show that the primary annihilation sites of the 
propagating positrons in both the bulge and the disk are in the 
warm ionized phases of the interstellar medium. Such annihilation 
can also account for those two basic properties of the emission, 
the fraction ($\sim 93\pm7$\%) of annihilation via positronium 
and the ratio ($\sim$ 0.5) of broad ($\sim$ 5.4 keV) to narrow 
($\sim$ 1.3 keV) components of the bulge 511 keV line emission. 
Moreover, we expect that the bulk of this broad line emission 
comes from the tilted disk region ($0.5 < R < 1.5$ kpc) with a 
very large broad/narrow flux ratio of $\sim$ 6, while much of the 
narrow line emission comes from the inner bulge ($R < 0.5$ kpc) 
with a negligible broad/narrow flux ratio. Separate spectral 
analyses of the 511 keV line emission from these two region 
should be able to test this prediction, and further probe the 
structure of the interstellar medium. 
 
Lastly, we show that the asymmetry in the inner disk annihilation 
line flux, which has been suggested as added evidence for new 
sources, can also be fully understood from positron propagation 
and the asymmetry in the inner spiral arms as viewed from our 
solar perspective without any additional sources. 
\end{abstract} 
 
 
\footnotetext[1]{Sabbatical 2008-2009} 
 
\newpage

\section{Introduction} 
 
The discovery (Johnson et al.\ 1972; Leventhal et al.\ 1979) of the 
diffuse Galactic positron annihilation radiation at 511 keV from the 
inner Galaxy has led to extensive studies of the possible origin of 
the annihilating positrons. These studies have shown (e.g.\ Ramaty 
\& Lingenfelter 1979; Kn\"odlseder et al.\ 2005) that, of all the 
various potential sources, the positrons from the decay of 
radioactive nuclei produced by explosive nucleosynthesis in 
supernovae are the largest and, hence, most likely source. 
 
Recent measurements, however, have raised new questions about the 
source. Analyses of the measurements by the gamma-ray spectrometer 
(SPI) on ESA's INTEGRAL spacecraft of the distribution of the 
diffuse Galactic positron annihilation radiation have shown that 
the luminosity of a Galactic positron bulge within 1.5 kpc of the 
Galactic Center is $\sim$ 1.4 times that of the Galactic disk 
(Kn\"odlseder et al. 2005; Weidenspointner et al. 2007, 2008a). 
Their best fit models are either concentric shell sources, or 
Gaussian spheroidal sources, in a bulge extending out to $\sim$ 
1.5 kpc combined with a thick disk with a half thickness of $z 
\sim$ 0.4 kpc, that extends from the Galactic Center to well 
beyond the Solar distance of 8 kpc. They also find a large 
additional contribution from a spherical halo extending beyond 
the bulge. 
 
In order to compare with these observations, we use these model 
fitting definitions of the positron bulge, thick disk, and halo, 
following Kn\"odlseder et al.\ (2005) and Weidenspointner et al. 
(2007, 2008a) in all of the discussion that follows, although 
both have been defined in various other ways in the literature. 
 
\subsection{ 511 keV Line Bulge/Disk Luminosity Ratio } 
 
The best-fit analysis of the SPI/INTEGRAL first year's data 
(Kn\"odlseder et al.\ 2005, Shells+D1) gave a 511 keV line 
luminosity of $L_B \sim (0.78\pm0.04)\times10^{43}$ photons 
s$^{-1}$ in a spherical bulge from 0 to 1.5 kpc and an old, thick 
disk luminosity of $L_D \sim (0.26\pm0.07)\times10^{43}$ photons 
s$^{-1}$, scaled to a Solar distance of $R_o \sim$ 8 kpc. This 
gave a bulge to disk 511 keV luminosity ratio of $L_B/L_D \sim 
3.0\pm0.8$. The subsequent best-fit analyses of the first two 
years' data, which included more coverage of the disk 
(Weidenspointner et al.\ 2007, for 4 Nested Shells and Old Disk), 
gave a bulge flux of $\sim (0.79\pm0.07)\times10^{-3}$ photons 
cm$^{-2}$ s$^{-1}$ in spherical shells out to 1.5 kpc, a thick 
disk flux of $\sim (1.41\pm0.35)\times10^{-3}$ photons cm$^{-2}$ 
s$^{-1}$, and a marginally detected spherical halo flux of $\sim 
(0.86\pm0.59)\times10^{-3}$ photons cm$^{-2}$ s$^{-1}$ from 1.5 
to 8 kpc. To infer the corresponding 511 keV line luminosities, 
we use the flux to luminosity ratios from their earlier modeling 
(Kn\"odlseder et al.\ 2005 Table 3), which are essentially just 
$4\pi R^2$ where $R \sim R_o$ for the bulge, about 0.57$R_o$ for 
the thick disk, and about 0.93$R_o$ for the halo. From the 
best-fit fluxes, we infer a 511 keV positron bulge luminosity 
$L_B \sim (0.60\pm0.05)\times10^{43}$ photons s$^{-1}$ in the 
spherical shells out to 1.5 kpc, and a thick disk luminosity of 
$L_D \sim (0.36\pm0.09)\times10^{43}$ photons s$^{-1}$ for a 
$L_B/L_D \sim 1.7\pm0.5$. 
 
Most recently from more than four years of observation, 
Weidenspointner et al.\ (2008a, model BD) fit the 511 keV fluxes 
and luminosities to a two component narrow (FWHM of 3$^o$, or 
0.42 kpc) and wide (FWHM of 11$^o$, or 1.5 kpc) spheroidal 
Gaussian distribution and thick disk from more extensive 
observations that increased the effective sampling distance to 
0.75$R_o$ for the disk. They found best-fit fluxes of $\sim 
(0.75\pm0.09)\times10^{-3}$ photons cm$^{-2}$ s$^{-1}$ from the 
bulge and $\sim (0.94\pm0.16)\times10^{-3}$ photons cm$^{-2}$ 
s$^{-1}$ from the disk. Scaled to $R_o \sim$ 8 kpc, these fluxes 
give a best-fit 511 keV luminosity $L_B \sim (0.57\pm0.08) 
\times10^{43}$ photons s$^{-1}$ in the spherical bulge $<$ 1.5 
kpc, and a thick disk luminosity of $L_D \sim (0.40\pm0.06) 
\times10^{43}$ photons s$^{-1}$. This gives a bulge to disk 511 
keV luminosity ratio of $L_B/L_D \sim 1.4\pm0.3$. They also found 
a best-fit combined halo and bulge luminosity of $\sim 
(1.55\pm0.1)\times10^{43}$ photons s$^{-1}$ for similar scaling, 
and subtracting the bulge luminosity, suggests a spherical halo 
luminosity of $L_H \sim 1\times10^{43}$ photons s$^{-1}$ beyond 
the bulge, for a total Galactic 511 keV luminosity $L_G \sim 
(2.0\pm0.2)\times10^{43}$ photons s$^{-1}$. 
 
Best-fit SPI/INTEGRAL spectral analyses by Weidenspointner et al.\ 
(2008b) also show an asymmetry in the positron annihilation line 
flux from two opposing inner disk components at 0$^o$ to 50$^o$ 
to either side of the Galactic bulge. These show an $80\pm40$\% 
excess in the 511 keV line flux at the negative longitudes 
compared to the positive. 
 
Assuming simply that the positron annihilation and production 
rates are in local equilibrium, these observations of the 
luminosities have been taken to imply a similar bulge/disk ratio 
and asymmetric spatial distribution for Galactic positron 
production. This suggested production ratio and distribution has 
challenged supernova source models, since it is much larger than 
that expected from the distribution and mean bulge/disk ratio of 
Galactic supernovae (e.g. Kn\"odlseder et al.\ 2005; 
Weidenspointner et al.\ 2008b). As we show, however, these spatial 
properties can be fully explained by positron propagation. 
 
\subsection{511 keV Line Width \& Positronium Fraction } 
 
We also consider the other two fundamental spectral properties of 
the Galactic positron annihilation: the ratio of the broad to 
narrow 511 keV line emission and the fraction ($f_{Ps}$) of the 
annihilation that occurs via positronium (Ps) formation. As shown 
by Jean et al. (2006) neither of these can be explained by any of 
the proposed positron sources, either old or new, without 
extensive positron propagation, since all of these sources are 
expected to produce most of the positrons in the hot tenuous 
phases of the interstellar medium, which cannot be their primary 
annihilation site. For this reason alone a detailed treatment is 
required of the production, propagation and annihilation in each 
phase and region of the Galaxy. 
 
Studies (e.g. Guessoum, et al.\ 1991; Guessoum, et al.\ 2005) 
show that positron annihilation in different phases of the 
interstellar medium occurs in differing ratios of direct 
annihilation on both free and bound electrons, to indirect 
annihilation via positronium formation, depending on the 
ionization fraction and temperature. Direct annihilation produces 
two photons at 511 keV, while positronium annihilation produces 
either two 511 keV photons or a three photon continuum depending 
on the spin state. Positronium is formed 25\% of the time in the 
singlet state, parapositronium, which annihilates with a mean 
life of $\sim 1.25\times10^{-10}$ s into two 511 keV photons, 
while 75\% of the time it forms in the triplet state, 
orthopositronium, which annihilates with a mean life of $\sim 
1.4\times10^{-7}$ s into a three photon continuum between 0 and 
511 keV (e.g.\ Guessoum et al.\ 1991). 
 
The SPI/INTEGRAL measurements of the ratio of the Galactic 
2$\gamma$ 511 line flux to that of the 3$\gamma$ continuum in the 
bulge and disk shows that the bulk of the positrons annihilate via 
positronium (Ps) with $f_{Ps} \sim 94\pm6$\% (Churazov et al.\ 
2005), $\sim 92\pm9$\% (Weidenspointner et al.\ 2006), and $\sim 
95\pm3$\% (Jean et al. 2006). The weighted mean positronium 
fraction of $\sim 94\pm4$\% also allows us to determine the total 
Galactic positron annihilation rates, $A$, in the bulge and disk, 
since $A = (e^+/\gamma_{511})L$. From the above we see that 
$(e^+/\gamma_{511}) = [2(f_{Ps}/4) + 2(1 - f_{Ps})]^{-1}$, which 
equals 1.69$\pm$0.17. Thus the inferred best-fit positron bulge 
and Galactic disk annihilation rates are $A_B \sim 
(0.96\pm0.17)\times10^{43}$ e$^+$ s$^{-1}$ and $A_D \sim 
(0.68\pm0.12)\times10^{43}$ e$^+$ s$^{-1}$ using the previously 
derived L$_{B}$ and L$_{D}$. In the hot tenuous plasma of the 
halo, however, the positronium fraction depends on the refractory 
grain abundance, ranging from only about 18\% with narrow (FWHM 
$\sim$ 2 keV) 511 keV line emission, if all the refractories are 
in grains, to 42\% with broad (FWHM $\sim$ 11 keV) if the grains 
all disintegrated (Jean et al. 2006). Thus we assume a possible 
range of $(e^+/\gamma_{511}) \sim 0.65\pm0.07$ there. From the 
best-fit halo flux, we thus infer a halo annihilation rate of 
$A_H \sim (0.65\pm0.11)\times 10^{43}$ e$^+$ s$^{-1}$, implying a 
total Galactic positron production rate of $\sim 
(2.3\pm0.2)\times10^{43}$ e$^+$ s$^{-1}$. 
 
The width of the 511 keV line also depends on the temperature of 
the medium and its state of ionization which effects the fraction 
of positronium formed by charge exchange with H, H$_2$, and He by 
superthermal, $\sim$10 eV, positrons as they slow down (Guessoum 
et al.\ 1991; Guessoum et al.\ 2005). The prompt annihilation of 
this fast moving positronium produces a characteristic broad 
(FWHM $\sim$ 6 keV) component of the 511 keV line, while the 
subsequent thermal annihilation, either directly on free and 
bound electrons or via positronium formed by radiative 
combination or charge exchange give much narrower ($<$ 2 keV) 
lines. Even the annihilation in the hot ($T \sim 10^6$K), tenuous 
ionized medium which would produce a broader ($\sim$10 keV) line 
from the hot free elections, can instead produce a narrow line if 
most of the positrons annihilate on refractory dust grains 
(Guessoum et al.\ 1991; Guessoum et al.\ 2005). The dust, as it 
does with interstellar molecule formation, provides a 
2-dimensional regime with much higher interaction rates for 
surface chemistry. 
 
The width of the Galactic bulge 511 keV line measured by 
SPI/INTEGRAL from the first years' data has been recently fitted 
(Churazov et al.\ 2005; Jean et al.\ 2006) by two components with 
about 67$\pm$10\% of the emission in a narrow line with a width 
of 1.3$\pm$0.4 keV and the remaining 33$\pm$10\% in a broad line 
with a width of 5.4$\pm$1.2 keV by positronium formation in 
flight, giving a broad/narrow 511 keV line flux ratio of $\sim$ 
0.5. The combined positronium fraction and relative fluxes of the 
broad and narrow 511 keV line emission strongly constrain the 
bulge annihilation processes, as we discuss in detail below. 
 
\subsection{Positron Production \& Propagation } 
 
The large difference between the ratio of the 511 keV 
annihilation line bulge/disk luminosity and that of the expected 
positron production by supernovae has led to suggestions there is 
some new, unrecognized source of positrons in the Galactic bulge. 
This seems unlikely, however, since various recent reviews (e.g.\ 
Dermer \& Murphy 2001; Kn\"odlseder et al.\ 2005; Guessoum, Jean 
\& Prantzos 2006) of the potential Galactic positron sources all 
conclude that supernovae are still the most plausible source and 
that other suggested sources, including cosmic-ray interactions, 
novae, and various exotic processes, all seem to be weaker and 
much less certain. 
 
However, as we have shown (e.g.\ Guessoum, Ramaty \& Lingenfelter 
1991), the 511 keV annihilation emission only illuminates the 
annihilation sites, not the sources, of the positrons, since 
these positrons must first slow down to energies $\leq$ 10 eV 
before they can annihilate. Thus, we examine in detail the 
propagation and annihilation of the $\beta^+$-decay positrons in 
the bulge, disk, and halo, and we show here that the expected 
positron propagation and annihilation are very different in the 
bulge and disk, and the cloud and inter-cloud environments. When 
these differences are taken into consideration, we find that the 
spatial distribution, the 511 keV line widths and the positronium 
fraction can all in fact be clearly understood in the context of 
a Galactic supernova origin. 
 
Although positron propagation has previously been discussed, the 
arguments range from the simple suggestion (Prantzos 2006) that 
most of the disk positrons might diffuse along dipolar field 
lines and annihilate in the bulge, to Monte Carlo transport 
simulations (Gillard et al.\ 2007) suggesting that they travel 
less than a few hundred parsecs in all but the hot phase. We 
therefore also carefully examine both the physics of MeV electron 
propagation and the observational evidence of such propagation 
from extensive measurements of Solar flare and Jovian electrons 
in the heliosphere. 
 
Here, we examine in detail the mechanisms of the relativistic 
positron propagation in the various phases of the interstellar 
medium. 
 
Overall, we assume that the Galactic stellar and interstellar 
distributions are defined by two superimposed systems: a stellar 
bulge-disk-and-halo system of stars and an interstellar 
bulge-disk-and-halo system of gas and plasma, as shown in Figure 
1. The stellar bulge and disk populations essentially determine 
the distribution of supernovae from the decay of whose 
radionuclei the positrons are born. Following Ferri\'ere et al. 
(2007), the interstellar bulge, disk and halo define the 
distribution of gas and plasma where the positrons die by 
annihilation. 
 
As we discuss in more detail later, the stellar bulge is confined 
within the interstellar bulge, which is blown out by a bulge wind 
to about 3.5 kpc in the Galactic plane and above the disk it 
feeds and merges into the halo. The stellar disk extends not only 
throughout the interstellar disk out to at least about 15 kpc, 
but also all the way into the interstellar bulge. The 
interstellar disk, however, extends in only to the outer edge of 
the interstellar bulge, and starts in the so-called ``molecular 
ring" at around 3.5 kpc, which is defined primarily by the two 
opposing innermost spiral arms. 
 
Threading the combined stellar and interstellar systems, is the 
Galactic magnetic field, along whose flux tubes the positrons 
travel from their birth to their death. We assume that the flux 
tubes are nearly vertical (e.g. Beck 2001), perhaps dipolar, in 
the inner part of the interstellar bulge and are blown out by the 
bulge wind (e.g. Bregman 1980; Blitz et al. 1993) roughly 
radially in the outer bulge, out to the molecular ring, where 
they begin to be tightly wound into the spiral arms of the 
interstellar disk (e.g. Beck 2001). There, in giant star formation 
regions along these spiral arms, hot, massive OB star 
associations are formed, whose collective supernovae generate 
superbubbles, that blow out plasma as well as flux tubes into the 
overlying halo (e.g. Parker 1979). 
 
Within this framework we model the various aspects of the 
production, propagation and annihilation of Galactic positrons. 
 
We show that roughly half of the Galactic positrons are born in 
the extended interstellar bulge (e.g. Ferri\'ere, Gillard \& Jean 
2008) in the inner $\sim$ 3.5 kpc of the Galaxy, and nearly all 
result from SNIa occurring in the hot, tenuous high-pressure 
plasma fills nearly all of that volume. We show that $\sim$ MeV 
positrons created in these SNIa supernovae are expected to escape 
from the remnants into the surrounding tenuous plasma. There they 
propagate along magnetic flux tubes by 1-dimensional diffusion, 
resonantly scattered in pitch angle by turbulently generated 
magnetohydrodynamic (MHD) waves that cascade down to the thermal 
electron gyroradius. In these plasmas the diffusion mean free 
path is long enough that only a negligible fraction annihilate 
there and $\sim$ 80\% of the positrons escape down into the dense 
HII and HI shells of neighboring molecular clouds within 1.5 kpc, 
where they quickly slow down and annihilate, while the remaining 
$\sim$ 20\% escape out into the halo and disk beyond. 
 
In the Galactic disk beyond 3.5 kpc, the filling factor of hot 
($\sim 10^6$ K) plasma is more modest ($\sim$ 20\%). Here the 
plasma is mostly in superbubbles, created by the correlated 
core-collapse supernovae of massive stars. Consequently, the bulk 
of positron production via $^{26}$Al decay, synthesized by these 
stars, occurs preferentially in the superbubbles. The larger 
bubbles blow out into the halo, sweeping the magnetic flux tubes 
up with them. Again the diffusion mean free paths of the 
positrons are such that the positrons escape along the 
flux tubes, either up into the halo, or down into the warm 
ionized envelopes, surrounding the base of the superbubble 
generated chimneys, where they slow down and annihilate. 
 
In the warm ($\sim 10^4$ K), but essentially neutral HI gas, 
which fills the largest fraction of the disk $>$ 3.5 kpc, we show 
that the cascade of magnetic turbulence, created at parsec 
scales, is damped by friction between the ions and neutral gas 
long before the cascading turbulence reaches the positron 
resonant gyroradii. Thus the positrons suffer much less 
scattering, and stream along the flux tubes with an isotropic 
pitch angle distribution. Therefore the positrons also mostly 
stream out into adjoining phases, either into the halo or into 
those warm ionized envelopes, separating the ubiquitous warm HI 
gas from the hot superbubbles, where they slow down and 
annihilate. 
 
Therefore we find that roughly 1/2 of the positrons, produced 
anywhere in the disk $>$3.5 kpc, slow down and annihilate there, 
primarily in the warm ionized gas, while the rest escape from the 
disk into the halo. This escape thus accounts for the low observed 
annihilation radiation luminosity of the disk and explains the 
observed ratio of annihilation rates in the bulge and disk. 
 
Moreover, the efficient diffusion of positrons out of the hot 
tenuous plasmas in both the bulge and disk, where nearly 70\% are 
born, and into the warm ionized envelopes and cold neutral 
interiors of molecular clouds, where they annihilate, also 
explains both the observed positronium fraction and broad to 
narrow 511 keV line flux ratio. 
 
The positrons escaping into the halo from the bulge and disk also 
lead to the annihilation of roughly 1/3 of all the positrons in 
the halo, even though there in no significant production there. 
 
\subsection{Outline } 
 
Although the processes of Galactic positron production, 
propagation and annihilation are rather straightforward, all 
the details still make the determination of the expected 511 
keV line emission both lengthly and complicated. 
 
We consider positron production from the decay of three separate 
long-lived radioisotopes, each synthesized by distinct classes of 
stars: the $^{56}$Ni and $^{44}$Ti, respectively, by SNIa and 
SNIp from thermonuclear explosions of $\sim$ Gyr-old accreting 
white dwarves in close binary systems, and the $^{26}$Al by young 
($<$ 40 Myr), massive ($>$ 8 M$_{\odot}$) stars, most likely 
through Wolf Rayet (WR) winds and core collapse SNII and SNIbc. 
The supernovae, SNIa and SNIp, from older stellar populations are 
known to occur in galactic bulges and disks, while WR stars, SNII 
and SNIbc, formed in extremely young stellar populations, occur 
in active, or recent, galactic star-formation sites, such as 
spiral arms, molecular-cloud complexes, and galactic nuclei. Thus 
in section 2 we briefly re-examine the production of 
$\beta^+$-decay positrons by Galactic supernovae and their 
spatial and temporal distribution in the light of recent 
observations and theory. 
 
In section 3 we then explore in detail the expected energy loss, 
scattering, and propagation processes of the relativistic 
positrons from these supernovae in the different phases of the 
interstellar medium. We consider collisionless scattering of 
these positrons by small-scale fluctuations generated by 
magnetohydrodynamic turbulence in the ionized interstellar 
phases, based on extensive studies and observations of 
relativistic electrons in the analgous turbulence in the 
interplanetary medium. Within this context we adopt a 
self-consistent transport model tied to field-aligned turbulence 
dependent on the properties of the ambient interstellar phases.

In sections 4, 5, and 6 we determine the expected spatial 
distribution of positron production, propagation, slowing down 
and annihilation in the various phases of the positron bulge, 
disk and halo. In each of these we review the fundamental 
properties of the interstellar medium, which constrain positron 
transport as well as leave signatures in the positron 
annihilation line emission features. Here we explore the very 
large differences between the interstellar properties, and the 
resulting positron propagation and annihilation, in the 
well-studied local interstellar medium compared to those in the 
inner $\sim$ 1.5 kpc of our Galaxy, where the diffuse 511 keV 
emission is most intense. 
 
Finally, in sections 7 \& 8 we give a summary and conclusions, 
comparing the expected 511 keV line emission with the SPI/INTEGRAL 
observations. We show that the observed spatial and spectral 
features of the Galactic positron annihilation radiation can all be 
fully understood in the context of a supernova radionuclide origin, 
when the differential propagation of these $\sim$ MeV positrons 
in the various phases of the interstellar medium is taken into 
consideration. 
 
\section{Galactic Nucleosynthetic Positron Sources} 
 
Here we briefly re-examine the production of positrons by 
$\beta^+$-decay radioactive nuclei produced by different types of 
Galactic supernovae. We also re-examine the rates of such 
supernovae in the Galaxy and the resulting spatial and temporal 
distribution of their production of positrons in both the Galactic 
bulge and disk in the light of recent observations and theory, in 
order to determine the expected positron bulge/disk ratio of the 
such positron production, $Q_B/Q_D$. 
 
\subsection{ Positrons from Supernovae} 
 
Positrons resulting from the decay chains of the longer-lived 
radioactive nuclei, $^{56}$Ni, $^{44}$Ti, $^{26}$Al, all produced 
by explosive nucleosynthesis in supernovae, have long been thought 
(Colgate 1970; Burger, Stephens \& Swanenburg 1970; Clayton 1973; 
Ramaty \& Lingenfelter 1979; Chan \& Lingenfelter 1993; Milne, The 
\& Leising 1999) to be the major source of Galactic positrons. The 
total Galactic positron production rates from these radionuclei 
depend on their supernova mass yields $M$, the supernova occurrence 
rates $\nu$ and the survival fraction, $f$, of their positrons in 
the supernova ejecta as it expands into the interstellar medium. 
 
\subsubsection{$^{56}$Ni } 
 
The decay chain of 
$^{56}$Ni$\rightarrow$$^{56}$Co$\rightarrow$$^{56}$Fe produces a 
positron only 19\% of the time, since $^{56}$Ni decays solely by 
electron capture with a mean life of 8.8 days into $^{56}$Co, 
which also decays primarily by electron capture, and just 19\% by 
$\beta^+$ emission into $^{56}$Fe with a mean life of 111.4 days. 
Thus the production rate of positrons from $^{56}$Co decay is 
$Q_{56} = 0.19f_{56} \dot{N}_{56}$, or $ 130\times10^{43}f_{56} 
\nu M_{56}$ e$^+$ s$^{-1}$, where $f_{56}$ is the time-integrated 
survival fraction of $^{56}$Co decay positrons in the expanding 
supernova ejecta, $\dot{N}_{56}$ is the production rate of 
$^{56}$Ni atoms per s, $M_{56}$ is the $^{56}$Ni yield in 
M$_{\sun}$ per SN, and $\nu$ is the Galactic supernova rate in SN 
per 100 yr. This Galactic rate alone would greatly exceed the 
observed rate of positron annihilation, if all of the positrons 
escaped from the supernova ejecta, but most do not. 
 
The nucleosynthetic yields of $^{56}$Ni, whose decay energy powers 
the visible light from the exploding ejecta, have been extensively 
calculated in all different types of supernovae. This is 
particularly true of the cosmologically important Type Ia 
thermonuclear supernovae in accreting white dwarves, which are 
best described by the deflagration model W7 of Nomoto, 
Thielemann, \& Yoko (1984) that gives a $^{56}$Ni yield of 0.58 
M$_{\sun}$. The peculiar subclass of Type Ia(bg), also called 
Type Ip, supernovae in accreting sub-Chandrasakar white dwarves 
are expected to produce 0.44 M$_{\sun}$ of $^{56}$Ni from the 
helium detonation model (Woosley, Taam \& Weaver 1986). 
 
Because of the relatively short decay mean life of $^{56}$Co, most 
of these positrons lose their energy and annihilate in the ejecta 
before it becomes either thin enough for them to escape or 
sufficiently rarefied for them to survive. Indeed, quantitative 
studies of the survival of positrons in the expanding supernova 
ejecta (Chan \& Lingenfelter 1993; Milne, The \& Leising 1999), 
for various supernova models, have shown that essentially all of 
the positrons produced by its decay in the most frequent, but 
massive, core collapse SNII supernovae slow down and annihilate 
unobserved in their much denser ejecta and do not 
escape into the interstellar medium. 
 
In particular, Chan \& Lingenfelter (1993) have calculated 
$f_{56}$ of positrons from $^{56}$Co decay expected for various 
supernova model distributions of the ejecta density, velocity and 
magnetic fields. They found that in the limit where the magnetic 
field is thoroughly tangled, the positron survival fraction for 
SNIa deflagration model, $f_{56}$, ranged from 0.1\% to 2.5\% for 
unmixed versus uniformly mixed ejecta. Alternatively in the limit 
that the magnetic field is fully combed-out (e.g.\ Colgate, 
Petschek \& Kriese 1980) the SNIa $f_{56}$ was much greater, 
ranging from 5\% to 13\% for unmixed and uniformly mixed ejecta. 
Following this same procedure Milne, The \& Leising (1999) have 
further shown that a mean survival fraction $f_{56} \sim 
3.5\pm2$\% for the deflagration model gave the best fits to SNIa 
light curves at late times ($>$1 yr), when the positrons from 
$^{56}$Co are the dominant source of energy. Using the 
deflagration model (Nomoto, Thielemann, \& Yoko 1984) $^{56}$Ni 
yield of 0.58 M$_{\sun}$, we calculate the expected positron 
production rate $Q_{56} = 75\times10^{43} f_{56} \nu_{Ia}$ e$^+$ 
s$^{-1}$. 
 
\subsubsection{$^{44}$Ti } 
 
Chan \& Lingenfelter (1993) have shown that the positrons from 
$^{44}$Ti$\rightarrow$$^{44}$Sc$\rightarrow$$^{44}$Ca decay, which 
produces a positron 95\% of the time from $\beta^+$ emission of 
$^{44}$Sc to $^{44}$Ca, is a significant source of the 
annihilating positrons. The Solar system abundance ratio of 
$^{44}$Ca to $^{56}$Fe of 1.23$\times$10$^{-3}$ (e.g.\ Lodders 
2003) requires a similar relative nucleosynthetic yield ratio of 
$^{44}$Ti to $^{56}$Ni, since these radionuclei are the primary 
sources of $^{44}$Ca and $^{56}$Fe (Woosley \& Pinto 1988). The 
much longer $^{44}$Ti decay mean life of 89 yr would allow 
essentially all ($>$ 97\%) of its positrons to survive in the 
expanding ejecta of even the most massive supernovae (Chan \& 
Lingenfelter 1993). Therefore the total Galactic positron 
production rate from $^{44}$Ti decay can be scaled by the Solar 
system abundance ratio of $^{44}$Ca to $^{56}$Fe, assuming that 
half of the Galactic $^{56}$Fe is produced by SNIa (e.g. Timmes, 
Woosley \& Weaver 1995) and $f_{44} \sim 1$, so that $Q_{44} = 
0.95f_{44} \dot{N}_{44},$ or $1.2\times10^{43} \nu_{Ia}$ e$^+$ 
s$^{-1}$. 
 
The most likely sources of these $^{44}$Ti decay positrons are 
the peculiar SNIp supernovae, typified by SN 1991bg (Filippenko 
et al.\ 1992; Turatto et al.\ 1996), so that we assume the same 
spatial distribution as SNIa. In particular, the primary source of 
Galactic $^{44}$Ca appears to be He detonations in accreting 
sub-Chandrasakar mass white dwarves, whose calculated (Woosley, 
Taam \& Weaver 1986; Woosley \& Weaver 1994) $^{44}$Ti yields of 
$M_{44} \sim 0.02 M_{56}$ are $\sim$ 16 times the Solar system 
$M_{44}/M_{56}$ ratio. And such models also give the best fits to 
the light curves of these peculiar supernovae (e.g. Milne, The \& 
Leising 1999; Blinnikov \& Sorokina 2004; The et al.\ 2006). The 
$M_{44}/M_{56}$ ratios calculated for other types of supernovae 
are much less than the Solar system ratio. Thus we assume the 
spatial distribution of the SNIp is the same as that of the SNIa 
and we also scale the $^{44}$Ti production to that of $^{56}$Ni 
in SNIa. 
 
\subsubsection{$^{26}$Al } 
 
The additional contribution of positrons, produced 82\% of the 
time from long-lived ($\tau_{26} \sim 1.04\times10^6$ yr) decays 
of $^{26}$Al to $^{26}$Mg, can be determined much more directly 
from the measured Galactic luminosity of the 1.809 MeV line, which 
accompanies that decay. This luminosity implies a steady-state 
Galactic mass $M_{26} \sim 2.8\pm$0.8 M$_\odot$ of $^{26}$Al 
(Diehl et al.\ 2006), or a positron production $Q_{26} \sim 0.82 
M_{26}/26m_p \tau_{26} \sim (0.34\pm0.10)\times10^{43}$ e$^+$ 
s$^{-1}$. Thus they can contribute $\sim$ 15$\pm$4\% of the total 
Galactic positron annihilation rate of $\sim (2.3\pm0.2) 
\times10^{43}$ e$^+$ s$^{-1}$ inferred from the best-fit bulge, 
disk and halo annihilation luminosity discussed above. 
 
Kn\"odlseder (1999) has shown that the $^{26}$Al 1.809 MeV line 
luminosity is strongly correlated with the distribution of massive 
stars in the disk, which confirms that it is produced in Wolf 
Rayet winds and/or in core collapse supernovae, SNII and SNIb/c. 
 
With these calculated yields of $^{56}$Ni, $^{44}$Ti and $^{26}$Al 
in various types of supernovae, we now estimate the total Galactic 
positron production rate from the decay of radionuclei synthesized 
in supernovae in the bulge and disk, assuming no significant 
production in the halo, 
 
\begin{eqnarray} 
Q_{B+D} = Q_{56} + Q_{44} + Q_{26}\nonumber \\ 
= (75 f_{56} \nu_{Ia} + 1.2 \nu_{Ia} + 0.3)\times10^{43}~~{\rm 
e}^+ {\rm s}^{-1} 
\end{eqnarray} 
\vspace*{0.1cm} 
 
\noindent 
scaled to the mean Galactic occurrence rate, $\nu_{Ia}$ of SNIa 
supernovae per 100 yrs.

\subsection{Mean Supernova Rates in Our Galaxy} 
 
Galactic supernova occurrence rates depend on the Hubble class of 
the galaxies and are commonly defined (e.g.\ van den Bergh \& 
Tammann 1991; Cappellaro, Evans \& Turatto 1999) in units of SNu, 
equal to 1 SN per 100 yr, times a factor of $h_{o}^2L^{BL}_{Gal}$, 
where $h_{o}$ is the Hubble constant in units of 75 km s$^{-1}$ 
Mpc$^{-1}$ and $L^{BL}_{Gal}$ is the blue luminosity of the 
galaxy in units of $(10^{10} L^{BL}_{\sun})^{-1}$. The Hubble 
classification of our Milky Way Galaxy is Sbc (Binney \& 
Merrifield 1998; Kennicutt 2001). From a study of the 
extra-galactic observations Cappellaro, Evans \& Turatto (1999) 
find that the SNIa rate is 0.21$\pm$0.08 SNu in Sbc galaxies. 
Thus from the Galactic blue luminosity, $L^{BL}_{Gal}$ of 
$(1.9\pm0.3)\times10^{10} L_{\sun}$ (Freeman 1985), and this 
extragalactic rate for Sbc galaxies, we estimate the {\em mean} 
supernova rate in our Galaxy for SNIa to be 0.40$\pm$0.16 per 100 
years. 
 
However, recent studies of the secular evolution of the Galactic 
bulge suggest that episodic ($\sim$Gyr) bursts of star formation 
could cause modest variations in the supernova bulge/disk ratio. 
From observations of oxygen-rich, cool giants, Sjouwerman et al. 
(1998) conclude that the most recent burst of star formation in 
the inner bulge occurred roughly 1 Gyr ago and produced $\sim$ 
10$^6$ SNII and SNIb/c supernovae. Such a burst would then be 
followed by a much more extended period of enhanced SNIa and 
SNIp, occurring with a range of delay times between 0.1 and 1 
Gyr, calculated from evolutionary models by Greggio (2005). Thus, 
although the SNII and SNIbc supernovae all occurred during the 
burst of star formation a Gyr ago, we would expect the SNIa and 
SNIp to still be occurring at the present time. Using the average 
ratio of SNIa to SNII for our Galaxy of 0.25 (e.g.\ van den Bergh 
\& Tammann 1991; Cappellaro, Evans \& Turatto 1999), we would 
expect 2.5$\times10^5$ SNIa from this star burst spread over 
$\sim$1 Gyr, or an added contribution of 0.025 SNIa per 100 
years. Thus we estimate the {\em present} supernova rate in our 
Galaxy for SNIa to be 0.43$\pm$0.16 per 100 years. 
 
Using this SNIa rate $\nu_{Ia}$ we then estimate from equation (1) 
the expected mean Galactic positron production from the decay of 
supernova produced $^{56}$Co, $^{44}$Ti and $^{26}$Al, 
 
\begin{eqnarray} 
Q_{B+D} = [(32\pm13)f_{56} + (0.5\pm0.2)\nonumber\\ 
 + (0.3\pm0.1)] \times10^{43}~~{\rm e}^+ {\rm s}^{-1}. 
\end{eqnarray} 
\vspace*{0.1mm} 
 
Thus, of the Galactic positron production of $\sim (2.3\pm0.2) 
\times 10^{43}$ e$^+$ s$^{-1}$, inferred from the observed bulge, 
disk and halo annihilation radiation as discussed above, $^{26}$Al 
decay accounts for $\sim 0.3\times10^{43}$ e$^+$ s$^{-1}$, and 
$^{44}$Ti decay produces $\sim 0.5\times10^{43}$ e$^+$ s$^{-1}$, 
nearly all from SNIp, leaving $\sim (1.5\pm0.3)\times10^{43}$ 
e$^+$ s$^{-1}$, which can come from $^{56}$Ni decay positrons, if 
they have a survival fraction in SNIa ejecta of $f_{56} \sim 
5\pm2$\%, including the uncertainty in the Galactic SNIa rate. A 
survival fraction of $\sim$ 5\%, was in fact just what was 
calculated (Chan \& Lingenfelter 1993 Fig.\ 3) for the standard 
deflagration model of SNIa (Nomoto et al. 1984 W7) with a combed 
out magnetic field in the ejecta. It is also quite consistent with 
the mean value of $\sim 3.5\pm2$\% inferred (Milne, The \& Leising 
1999) from SNIa light curves at late times, when the positrons 
become a major source of ejecta heating. 
 
\subsection{Spatial Distribution of Galactic Supernovae} 
 
Here we investigate both the \textsl{stellar} bulge/disk ratio 
($B^*/D^* \sim (\nu_B/\nu_D)_{Ia+Ip}$) of Galactic supernovae and 
their radial and transverse distributions within the stellar bulge 
and disk. Of primary concern are the SNIa and SNIp, which appear 
to be the source of $\sim$ 85\% of the annihilating positrons. 
 
We first determine the relative contributions of SNIa occurring 
in the Galactic stellar bulge and disk. It is well known that the 
SNIa occurrence rate is essentially the same across the Hubble 
sequence of galaxy types from ellipticals (E) through late-type 
spirals (Sd), when such rates are defined in SNu, the number of 
supernovae per year per unit of blue luminosity of the parent 
galaxy (e.g.\ Cappellaro, Evans \& Turatto 1999). Across this 
broad Hubble sequence, the fraction of galactic blue-band 
luminosities contributed by Galactic disks range from zero in 
ellipticals to 95\% in late-type Scd galaxies (Simien \& de 
Vaucouleurs 1986), yet the SNIa birth rate per unit blue 
luminosity of parent galaxies remains the same. 
 
Since the SNIa occurrence rates per blue luminosity is 
independent of the relative contributions of galactic bulges and 
disks, we scale the ratio of SNIa birthrates in the stellar bulge 
and disk of our Galaxy by the corresponding ratio of stellar 
bulge to disk blue luminosities, $L_B^{BL}/L_D^{BL}$. In their 
classic study Simien \& de Vaucouleurs (1986) investigated the 
systematics of bulge-to-disk ratios in the blue-band. The mean 
value of their tabulations of $L_B^{BL}/L_D^{BL}$ for our Galaxy 
is 0.25. Thus, we expect that the Galactic stellar luminosity 
bulge/disk ratio implies a similar time-averaged, mean stellar 
bulge/disk ratio $(\nu_B/\nu_D)_{Ia} \sim$ 0.25 for Galactic SNIa 
and SNIp. Adding the star burst contribution of 0.025 SNIa per 
100 years in the bulge, which is a 30\% enhancement over the mean 
bulge rate of 0.08 SNIa per 100 years, gives a current stellar 
Galactic bulge/disk ratio $(\nu_B/\nu_D)_{Ia+Ip} \sim 
0.25\times1.3 \sim$ 0.33 for SNIa and SNIp supernovae. 
 
To model the spatial distribution of SNIa in the Galaxy, we use 
the representations of Dehnen \& Binney (1998) who parameterized 
the large-scale stellar distributions within the Galaxy by 
comparing models to the Galactic rotational and other 
observational constraints. Thus the number density of SNIa in the 
Galactic disk, $n_{D*}(R,z)_{Ia}$, and the number density of SNIa 
in the Galactic bulge, $n_{B*}(R,z)_{Ia}$, expressed in 
cylindrically symmetric Galactocentric coordinates, are 
 
\begin{eqnarray} 
n_{D*} (R, z)_{Ia} = n_{o_D}e^{-R/R_d -|z|/z_d},\nonumber\\ 
n_{B*} (R,z)_{Ia} = \frac{n_{o_B}}{\eta ^{1.8}} 
e^{-\frac{\eta^2}{r_t^2}}, 
\end{eqnarray} 
\vspace*{0.1cm} 
 
\noindent where $\eta = \sqrt{R^2 + (z/0.6)^2}$, $R$ represents 
the planar distance from the Galactic center in kpc, $z$ 
represents the distance normal to the Galactic plane in kpc, and 
the constant $r_t$ = 1.9 kpc (Dehnen \& Binney 1998). Here the 
normalization constants, $n_{o_D}$ and $n_{o_B}$ are found from 
the total Galactic SNIa birthrate and the $(\nu_B/\nu_D)_{Ia}$ 
ratio. Dehnen and Binney ascertained that the more than 90\% of 
the stellar disk population could be represented by $z_d$ of 
0.180 kpc. They concluded that the most important model parameter 
was the ratio of the disk scale length, $R_d$, to $R_{\odot}$, the 
distance of the Sun from the Galactic center. They found that the 
observational constraints were best satisfied by any of four 
models with $R_{\odot}$ = 8 kpc and $R_d/R_{\odot}$ of 0.25, 0.3, 
0.35, \& 0.4. Here we will use their second model with $R_d$ of 
2.4 kpc. We also include a star burst contribution to the bulge 
located in the CMZ (Sjouwerman et al. 1998). 
 
\subsection{Spatial Distribution of Galactic Positron Production} 
 
In Figure 2 we show the relative spatial distributions, $F(<r)$, 
as functions of distance, $r$, from the Galactic center for SNIa 
residing in the stellar bulge and disk employing the relative 
number densities, equation (3) for $n_{B*}(R, z)_{Ia}$ and 
$n_{D*}(R, z)_{Ia}$, expressed a functions of $R$ \& $z$ in a 
Galactocentric cylindrical coordinate system, 
 
\begin{eqnarray} 
F_{B*}(<r) = 2\pi\int^{r}_{0} n_{B*}(R(r), z(r))_{Ia}r^2dr, \nonumber\\ 
F_{D*}(<r) = 2\pi\int^{r}_{0} n_{D*}(R(r), z(r))_{Ia}r^2dr, \nonumber\\ 
F_{OB*}(<r) = 2\pi\int^{r}_{0} n_{OB*}(R(r), z(r))_{II\&Ibc}r^2dr. 
\end{eqnarray} 
 
$F_{OB*}(<r)$ is the relative spatial distribution of massive 
stars, the expected Wolf-Rayet and SNII \& SNIbc progenitor 
sources of $^{26}$Al. Although the great majority of massive 
stars, which synthesize $^{26}$Al, are located in the outer ($>$ 
3.5 kpc) Galactic disk (McKee \& Williams 1997), some massive 
stars found in the inner ($<$ 0.2 kpc) stellar bulge. Since 
Galactic Lyman continuum radiation is also produced by a massive 
star population (e.g. McKee \& Williams 1997), we use the ratio 
of Lyman continuum radiation emitted by the inner stellar bulge 
to that from the outer disk to estimate the corresponding 
$^{26}$Al ratio. G\"usten (1989) estimated a luminosity of 
ionizing photons from massive stars in the inner bulge of $\sim 
1.5\pm0.5\times10^{52}$ photons s$^{-1}$. Comparing this to the 
estimated (McKee \& Williams 1997) total Galactic value of $\sim 
1.9\times10^{53}$ photons s$^{-1}$, we expect that roughly 10\% 
of the massive stars and their generated $^{26}$Al occur in the 
inner bulge, amounting to $\sim 0.03\times10^{43}$ e$^+$ s$^{-1}$. 
 
For the remaining disk portion we use the spatial distribution of 
OB associations, $n_{OB}(R,z)_{II\&Ibc}$, of (McKee \& Williams 
(1997, eq. 35) with $z_{d}$ = 0.15 kpc to represent that of the 
$^{26}$Al made by massive stars. These spatial distributions have 
been normalized at large $r$ to their expected relative 
contributions to the total Galactic positron production rate 
fractions: $F_{B*}$ = 0.21, $F_{D*}$ = 0.64, and $F_{OB}$ = 0.15, 
based on the SNIa and SNIp bulge/disk ratio of 0.33 in the 85\% 
of positrons produced by them. 
 
As we see from Figure 2, of the positrons produced from SNIa and 
SNIp in the disk, 13\% are born within 1.5 kpc. Although the best 
fit disk model used in the SPI/INTEGRAL analyses (Weidenspointner 
et al. 2007, 2008a) of the annihilation radiation also includes 
12\% occurring within that radius, in order to treat the 
propagation of the positrons before the annihilate, we obviously 
need to include them in the total production within 1.5 kpc. Once 
we have determined the distribution of the subsequent annihilation 
of all of the positrons, we then adjust the estimated bulge ($<$ 
1.5 kpc) and disk ($>$ 1.5 kpc) annihilation luminosities to 
compare with the SPI/INTEGRAL bulge and disk components, as we 
discuss in section 7. 
 
Therefore, the positron bulge production ($<$ 1.5 kpc), $Q_B \sim 
[F_{B*} + F_{D*}(<1.5) + F_{OB}(<1.5)]Q_{B+D} \sim [0.21 + 
0.64(0.13) + 0.1(0.15)]Q_{B+D} \sim 0.31 Q_{B+D}$, and the disk 
production, including the OB contribution, $Q_D \sim [F_{D*}(1 - 
F_{D*}(<1.5)) + 0.9F_{OB}]Q_{B+D} \sim [0.64(0.87) + 
0.9(0.15)]Q_{B+D} \sim 0.69 Q_{B+D}$. Thus the effective positron 
production ratio in the bulge and disk within and beyond 1.5 kpc 
is $(Q_B/Q_D) \sim 0.31/0.69 \sim$ 0.45. 
 
From the ratio of positron production rates in the bulge $Q_B$ 
and disk $Q_D$ we now determine the differential propagation of 
the positrons within and between those regions and the halo 
before they annihilate. In simplest terms we define the resulting 
positron bulge/disk annihilation ratio as, $A_B/A_D \sim 
(P_{B:B}Q_{B} + P_{D:B}Q_D)/(P_{B:D}Q_{B} + P_{D:D}Q_D),$ where 
$P_{B:B}$ and $P_{D:B},$ are respectively the fractions of the 
positrons, born in the bulge $B$ and disk $D$ beyond 1.5 kpc, 
which annihilate in the positron bulge $B < 1.5$ kpc, while 
$P_{B:D},$ and $P_{D:D}$ are the corresponding fractions of those 
positrons which annihilate in the disk $D$ beyond the positron 
bulge $> 1.5$ kpc. As we discuss below, however, we need to 
investigate the positron propagation and annihilation in each of 
the interstellar gas phases within each of the regions.

\section{Nature of Galactic Positron Propagation} 
 
We now look in detail at the nature of these positrons, and their 
propagation, slowing down and annihilation in the various phases 
of the interstellar medium in the bulge, disk and halo. In 
particular, we investigate the plasma processes that determine 
the rate of positron propagation, drawing upon the extensive 
studies and observations of relativistic electron propagation in 
the interplanetary medium. 
 
\subsection{Relativistic $\beta^+$-Decay Positrons from Supernovae} 
 
Assuming that the dominant source of Galactic positrons is the 
$\beta^{+}$ decay of $^{56}$Co, $^{44}$Sc and $^{26}$Al from 
supernovae, the detailed studies of their survival in and escape 
from the ejecta show that the great bulk of the surviving and 
escaping positrons are {\em relativistic}. 
 
Positrons emitted in the $^{56}$Co $\rightarrow$ $^{56}$Fe decay 
are distributed in kinetic energy with a maximum of 1.459 MeV and 
a mean of 0.630 MeV, and those from $^{44}$Sc $\rightarrow$ 
$^{44}$Ca decay have a very similar spectrum. The expected (Chan 
\& Lingenfelter 1993; Milne, The \& Leising 1999) mean energy of 
the escaping positrons from the shorter-lived (111.4 day mean 
life) decay of $^{56}$Co is close to 0.5 MeV, reduced from their 
initial energy by  ionization losses in the still dense ejecta, 
while the mean energy of those from the longer lived (89 yr 
$^{44}$Ti dominated meanlife) decay of $^{44}$Sc is essentially 
unchanged at about 0.6 MeV in the much less dense ejecta. The 
mean energy of positrons from the decay of the very long-lived 
(1.04$\times$10$^6$ yr) decay of $^{26}$Al is likewise about 0.5 
MeV. 
 
In order for these surviving relativistic positrons to annihilate 
in the interstellar medium they must first be {\em decelerated to 
energies, $\leq$ 10 eV.}  In order to slow down to thermal 
energies, these $\sim$ 0.5 MeV positrons, must run through $\sim$ 
0.12 g cm$^{-2}$ (Berger \& Seltzer 1964), or $nd_{sd} \sim 
6\times10^{22}$ H cm$^{-2}$, in unionized interstellar gas, and 
this range is reduced by a factor of $1/(1 + x_e)$ if the gas is 
ionized with an ionization fraction, $x_e$, relying solely on 
collisional ionization losses. In the process of traveling this 
slowing down distance, $d_{sd} \sim 20/n(1 + x_e)$ kpc, the 
average velocity of these positrons, $\bar{\beta}c$ is $\sim$ 0.7 
of the velocity of light, so their slowing down time, 
 
\begin{eqnarray} 
t_{sd} \sim d_{sd}/\bar{\beta}c \sim 9\times10^4 /n(1 + x_e)~~{\rm 
yr}, 
\end{eqnarray} 
\vspace*{0.1cm} 
 
\noindent where $n$ is the mean H density in cm$^{-3}$. 
 
Thus we see that since the slowing down distance of the 
radioactive decay positrons in the nominal phases of the 
interstellar medium is $>$ 0.5 kpc in HI clouds with typical $n 
\sim$ 40 cm$^{-3}$, $>$ 50 kpc in warm media with typical $n 
\sim$ 0.4 cm$^{-3}$, and $>$ 3 Mpc in hot phases with typical $n 
\sim 3\times10^{-3}$ cm$^{-3}$. Since all of these distances are 
far greater than the scales of their corresponding interstellar 
components, these fast positrons, streaming along the large-scale 
interstellar magnetic field, would all escape the Galactic disk 
before they even slow down, {\em unless there is sufficient 
plasma turbulence to efficiently isotropize their trajectories.} 
We thus investigate in detail under what conditions the 
interstellar turbulence can isotropize the trajectories of these 
relativistic positrons.

\subsection{Nature of the Turbulence which Scatters Interstellar Positrons} 
 
In ionized interstellar phases we expect that the interstellar 
MeV positrons are scattered by resonant interactions with ambient 
turbulent magnetic fluctuations. Consequently, the efficiency of 
these scatterings in turn depends on the fundamental properties of 
the MHD turbulence at very small spatial scales 
of the electron gyroradius, $r_{e}$. However, the basic properties 
of MHD turbulence is poorly understood. In a recent review of turbulence 
and magnetic fields in astrophysical plasmas, Schekochihin \& 
Cowley (2007) state ``despite over fifty years of research and 
many major advances a satisfactory theory of MHD turbulence 
remains elusive. Indeed, even the simplest (most idealized) cases 
are not fully understood." 
 
An important advancement was the realization that in many 
astrophysical sites MHD turbulence is anisotropic. Such 
anisotropy is a well-observed phenomenon in solar-wind plasmas 
(e.g.\ Belcher \& Davis 1971; Matthaeus, Golstein, \& Roberts 
1990; Osman \& Horbury 2007) as well as in increasingly more 
accurate, numerical simulations (Shebalin, Matthaeus, \& 
Montgomery 1983; Maron \& Goldreich 2001; Cho, Lazarian, \& 
Vishniac 2002; Mason, Cattaneo, \& Boldyrev 2006; Perez \& 
Boldyrev 2008). In a pioneering study Goldreich \& Sridhar (1995) 
developed a sophisticated theory of anisotropic incompressible 
MHD turbulence. They presented the concept of ``critical balance" 
which predicted in inertial regime of 
turbulent MHD flows, first, a filamentary shape for turbulent 
fluctuations, via a relationship between the wave numbers 
parallel ($\|$) and transverse ($\bot$) to the direction of the 
local mean magnetic field, k$_{\|} \propto$ k$_{\bot}^{2/3}$ and, 
second, the modeled energy spectra of transverse 
turbulent components possesses Kolmogorov (1941) scaling, 
$k_{\bot}^{-5/3}$, observed in low-speed, terrestrial flows. 
 
At present, a puzzling dichotomy exists between these model 
predictions and a plethora of solar-wind observations and 
numerical simulations. Power spectra of small scale 
electromagnetic fluctuations in solar-wind plasmas have been 
found to have the expected $-$5/3 spectral index (e.g. Horbury et 
al.\ 1996; Leamon et al.\ 1998; Bale et al.\ 2005), but 
corresponding spectra in many precise numerical simulations have 
spectral indices much closer to $-$3/2 (Mason \& Goldreich 2001; 
Beresnyak \& Lazarian 2006; Mininni \& Pouquet 2007; Perez \& 
Boldyrev 2008). Yet the critical balance prediction about the 
wavenumber relationship of turbulent eddies seems to be 
consistent with many numerical simulations (Cho \& Vishniac 2000; 
Maron \& Goldreich 2001; Cho, Lazarian, Vishniac 2002; Beresnyak 
\& Lazarian 2006). Unfortunately, analyses of single 
spacecraft measurements are unable to 
determine the full modal three-dimensional wavevector 
spectra of turbulent fluctuations (e.g. Horbury, Forman, \& 
Oughton 2005) and thus cannot confirm or refute the 
critical balance prediction about filamentary-shaped fluctuations 
(Beresnyak \& Lazarian 2008).

Although much progress has been made in the understanding and 
modeling of anisotropic MHD turbulence, such models, which assume 
that the frequency of the turbulent fluctuations is low compared to 
the proton gyrofrequency $\Omega_{p}$, still have difficulty in representing 
high-frequency collisionless turbulence evolving on frequencies, $\omega 
\approx \Omega_{e}$, which seem to be most efficient at resonantly scattering 
$\sim$ MeV positrons, where $\Omega_{e}$ represents electron 
gyrofrequency. The application of such models of anisotropic turbulence 
seems to imply that collisionless scattering of MeV electrons 
would be modest and MeV electron transport should be dominated 
by streaming at $\sim\bar{\beta}c$. Yet a great variety of independent 
investigations of MeV electron transport in the interplanetary medium 
irrefutably demonstrate that such particles propagate $\emph{diffusively}$ 
in the collisionless solar wind plasma, although the nature of the 
plasma fluctuations which actually scatter such electrons remains yet 
unknown. Thus, we employ a simple phenomenological model 
of collisionless MeV electron scarttering by turbulent 
interstellar fluctuations tied heavily to the well determined 
properties of electron scattering in turbulent solar-wind plasmas. 
 
The propagation of MeV electrons has been well studied in the 
interplanetary medium employing measurements of electrons 
accelerated by solar particle events as well as by the Jovian 
magnetosphere. The propagation of Jovian MeV electrons is, for 
the most part, dominated by diffusion (Ferreira et al.\ 2001a,b, 
2003). Fits to intensity and anisotropy time profiles of 
solar-flare electrons also illustrate the dominance of diffusive 
transport at these electron energies, and, consequently, lead to 
the straightforward determination of scattering mean free paths 
via fits to intensity and anisotropy time profiles (Kallenrode 
1993). 
 
Bieber, Wanner, \& Matthaeus (1996) 
related such phenomenologically-derived scattering mean free 
paths from solar flare particle measurements to the determination 
of $\emph{simultaneous}$ power spectra of magnetic fluctuations 
at MHD spatial scales. Since it is impossible to 
determine the full three-dimensional wavevector 
spectra of turbulent fluctuation from single spacecraft 
measurements (e.g.\ Fredricks \& Coroniti 1976), 
Bieber, Wanner, \& Matthaeus employed a simple two-component 
representation of the anisotropic interplanetary MHD turbulence. 
They found that their magnetic-field 
data consisted of two anisotropic populations, fluctuations with 
large correlation lengths transverse to the direction of the mean 
magnetic field (slab turbulence) and fluctuations with large 
correlation lengths parallel to the direction of the mean 
magnetic field (quasi-two-dimensional turbulence).  In such slab 
models the wave vectors of the fluctuations are aligned parallel 
to the direction of the mean magnetic field, and a simple 
interpretation is that in the inertial range slab turbulence 
consists of Alfv\'en wave propagating along the mean magnetic field. 
Later investigations (Oughton, Dmitruk, \& Matthaeus 2004, 2006; 
Oughton \& Matthaeus 2005) have demonstrated that this simple 
two-component model of anisotropic turbulence seems to model 
the basic properties of solar wind MHD turbulence. 
 
Bieber, Wanner, \& Matthaeus found that interplanetary turbulence 
is dominated\footnote{Recently, Horbury, Forman \& Oughton (2005) 
have investigated in more detail the properties of anisotropic solar 
wind turbulence, implementing a wavelet-based method to track the 
direction of the local mean magnetic field; such an approach reduces 
greatly the noise in the magnetic field measurements. 
Employing the two-component model, they ascertained that 
their results agreed ``remarkably well'' with 
the results of Bieber, Wanner, \& Matthaeus.} 
by quasi-two-dimensional turbulence, and the mean 
ratio of the power in slab turbulence to the power in 
two-dimensional turbulence is $\sim$0.15. Further, observing that 
the modeled two-dimensional fluctuations do not contribute to 
particle scatterings, they calculated scattering mean free paths 
expected by resonant interactions solely with turbulent slab 
fluctuations via quasi-linear theory (Jokipii 1969). They found 
that such scattering mean free paths agreed well with the values 
determined independently (of the analyses of the turbulent 
magnetic variations) from data fits to intensity and anisotropy 
time profiles. 
 
Based on these discussions, we model {\em interstellar} MHD 
turbulence as composite slab/two-dimensional turbulence. We 
assume here that the ratio of the power of interstellar slab 
turbulence along the mean magnetic field to the power of 
2-dimensional turbulence normal to the direction of the mean field 
is 0.15 following Bieber et al.\ (1996). Similarly, we assume 
that the spectrum of interstellar slab turbulence steepens from a 
spectral index of $-$5/3 to $-$3 at a wavenumber $k_d > 1/r_p$, 
where $r_p$ is the proton gyroradius, since Leamon et al.\ (1998) 
have found that at higher wave numbers the spectral index of 
solar-wind variations varied from $-$2 to $-$4.4 with an average 
index of $-$3. 
 
To quantify such scattering mean free paths for interstellar MeV 
positrons we use an analytic approximation (Teufel \& 
Schlickeiser 2002) calculated for slab MHD turbulence employing 
quasi-linear theory. Following Teufel \& Schlickeiser we assume 
that energy is injected into the interstellar medium at an outer 
spatial scale, $l_o$. For $kl_o \geq$ 1 and $k \leq$ k$_{d}$, the 
modeled turbulence possesses a spectral index, $s$. From the 
discussion above, we assume $s$ = 5/3 and at $k_{d}$ = 1/r$_{p}$, 
we assume the spectral shape steepens to $s$ = 3. Teufel \& 
Schlickeiser (2002, their eq. 58), find that for such slab 
turbulence the electron scattering mean free path along the 
magnetic field is 
 
\begin{eqnarray} 
\lambda_{\parallel}  = \frac{9} 
{2}\left( {{{B_o } \mathord{\left/ 
 {\vphantom {{B_o } {\delta B_{\parallel} }}} \right. 
 \kern-\nulldelimiterspace} {\delta B_{\parallel} }}} \right)^2 \frac{{J^2 }} 
{{k_{\min } a}}K(a,h,s, I, J), 
\end{eqnarray} 
\[ 
k_{\min }  = \frac{{2\pi }}{{l_o }},   k_d  = \frac{{1}} {{r_p }}, \,J = \frac{{\bar{\beta} ck_{min} }} 
{{\Omega _e }},  I = \frac{{\bar{\beta} ck_d }}{{\Omega_e }}, a = \frac{\bar{\beta} c}{{V_a }}, 
\] 
 
\[  f_{1}(s, h) = \frac{2}{h - 2}  + \frac{2}{2 - s}, 
\]

$\begin{array}{l} 
 \,\,\,\,\,\,\,\,\,\,\,\,\, \\ 
 \,\,\,\,\,\,\,\,\,\,\,\,\,\,\,\,\,\,\,\,\,\,\,\,\,\,\,\,\,\,\,\,\,\,\,\,\,\,\,\,\,\,\,\,\,\,\,\,\,\,\,\,\,\,\,\,\,\,\,\,\,\,\,\,\,\,\,\,\,\,\,\,J \ll I \ll 1 \ll a \\ 
  \\ 
 K = \frac{{a^2 }}{{f_1 J^s I^{2 - s} }}\left\{ \begin{array}{l} 
 \,\,\,\,\,\,\,\,\,_2 F_1 (1,\frac{1}{{h - 1}},\frac{h}{{h - 1}}, - \frac{{\pi a}}{{f_1 }}Q^{h - 2} )\, \\ 
 -\frac{1}{3}\,_2 F_1 (1,\frac{3}{{h - 1}},\frac{{h + 2}}{{h - 1}}, - \frac{{\pi a}}{{f_1 }}I^{h - 2} ) \\ 
 \end{array} \right\} \\ 
  \\ 
 \,\,\,\,\,\,\,\,\,\,\,\,\,\,\,\,\,\,\,\,\,\,\,\,\,\,\,\,\,\,\,\,\,\,\,\,\,\,\,\,\,\,\,\,\,\,\,\,\,\,\,\,\,\,\,\,\,\,\,\,\,\,\,\,\,\,\,\,\,\,\,J \ll 1 \ll I \ll a \\ 
  \\ 
 K = \frac{1}{\pi }\left[ {\frac{1}{{2 - s}} - \frac{1}{{4 - s}}} \right]{\rm{ + }}\frac{{a^2 }}{{f_1 J^s I^{3 - s} }}\,_2 F_1 (1,\frac{1}{{h - 1}},\frac{p}{{h - 1}}, - \frac{{\pi a}}{{f_1 I}}) \\ 
 \end{array}$ 
\vspace*{0.4cm}

\noindent where $\delta B_{\parallel}^2/8\pi$ is the energy density of 
the turbulent magnetic fluctuation in the direction of the mean 
magnetic field, $c$ is the speed of light, $\Omega_e$ is the 
gyro-frequency of the positron, the mean magnetic field is $B_o$, 
$V_{a}$ is the Alfv\'{e}n speed, $B_o/\sqrt{4\pi\rho}$, $\rho$ is 
the ion mass density, $r_e$ is the electron gyroradius; the 
average positron speed is $\bar{\beta}c$, and $K$ is a 
dimensionless quantity involving a Gauss hypergeometric function 
$_2F_1$ (Table 3 of Teufel \& Schlickeiser 2002). 
 
Thus, by such one-dimensional (1-D) diffusion the positrons in 
their slowing down time $t_{sd}$ would be distributed along a 
flux tube a mean length $l_{sd} \sim (2t_{sd}\lambda_{\|} 
\bar{\beta} c/3)^{1/2}$ in either direction from their point of 
origin.

\subsection{Ion-Neutral Damping} 
 
The bulk of the Galactic interstellar mass resides in primarily 
neutral phases (HI \& H$_{2}$), concentrated in a cloud population 
(e.g. Tielens 2005). Moreover, a major fraction of the disk ($>$ 
3.5 kpc) is filled with warm neutral HI (Kulkarni \& Heiles 1987). 
Consequently, the nature of MHD turbulence in predominantly 
neutral interstellar phases needs to be addressed. 
 
In partially neutral plasmas, where magnetic forces act directly 
on charges, and indirectly on neutral atoms via ion-neutral 
collisions, the turbulent NHD fluctuations are dissipated into 
heat when ion and neutral velocities differ significantly (Higdon 
1984; Goldreich \& Sridhar 1995; Lithwick \& Goldreich 2001). 
Turbulent MHD cascades are quenched at scales of roughly the 
collision mean free path of protons with neutral atoms, if the 
proton collision rate with neutral atoms exceeds the eddy cascade 
rate. The mean free path for proton-HI collisions is $l_{pH}= 
5\times10^{13}/n$ cm at 8000 K, or if hydrogen is fully ionized, 
and He is the dominant neutral species, the mean free path for 
proton-He collisions is $l_{pHe} \sim 1.5\times10^{15}/n$ cm 
where $n$ is the H number density, including both ions and 
neutrals (Lithwick \& Goldreich 2001). 
 
Consequently, they showed that MHD turbulence can cascade to 
small spatial scales, much less than these damping collision mean 
free paths, only if the HI fraction, $n_{HI}/n$, is less than a 
critical value, 
 
\begin{eqnarray} 
n_{HI}/n < f_{crit} \approx 5(l_{pH}/l_o)^{1/3}. 
\end{eqnarray} 
\vspace*{0.1cm} 
 
Therefore in predominantly neutral interstellar phases turbulent 
MHD cascades are halted by ion-neutral collisions at 
$\sim l_{pH}$, spatial scales far greater than the gyroradii of 
MeV positrons. Hence we expect collisionless scattering of 
positrons by such large-scale turbulent MHD fluctuations to be 
very inefficient. Similarly simple elastic collisions with ambient 
electrons, whose mean pitch angle scattering is 90$^o$ and mean 
energy loss is 50\%, have a mean free path much longer than 
slowing down distance due to ionization losses (e.g. Berger \& 
Seltzer 1964), so they too offer no significant scattering. 
 
In the absence of such scatterings, $\sim$MeV positrons might be 
expected to generate plasma waves by a resonant streaming 
instability, similar to the creation (Kulsrud 2005) of ion 
Alfv\'en waves by streaming relativistic cosmic ray nuclei. But 
these do not appear to be effective either. These $\sim$ MeV 
positrons generate resonant whistler waves (Schlickeiser 2002) at 
wavelengths significantly less than the scale of the thermal 
proton gyroradius, $r_p = u_p /\Omega_p$, at frequencies 
approaching those of thermal electrons, $\Omega_e,$ where $u_p$ 
is the thermal proton speed. The phase speed of whistler waves in 
the direction of the mean field is the \textsl{electron} Alfv\'en 
speed, $V_{ae} = (m_p/m_e)^{1/2}V_a.$ If the positron streaming 
velocity, $V_{s} > V_{ae}$, and if the whistler waves weren't 
damped so they grew to saturation, then the resulting magnetic 
field fluctuations in these waves would interact with the 
streaming positrons via quasi-linear wave-particle interactions 
that change their pitch angles, reducing $V_s$ to a level just 
less than $V_{ae}$. However, in predominantly neutral 
interstellar phases such whistler waves are also subject to 
severe ion-neutral damping (Kennel 2008 private communication). 
 
Thus, we expect that positron streaming velocities along the 
magnetic flux tubes are comparable to their particle velocities, 
$\bar{\beta} c$, similar to the very high drift speeds found 
(Felice \& Kulsrud 2001) for cosmic-ray nuclei in warm HI 
regions. Nonetheless, in turbulent media the flux tubes 
themselves may essentially random walk, so that the propagation 
is described by so-called ``compound diffusion" (Lingenfelter, 
Ramaty \& Fisk 1971). 
 
\section{Positron Propagation \& Annihilation in the Positron Bulge } 
 
Using these models of relativistic electron transport, we now 
investigate the propagation, slowing down, and annihilation of 
positrons produced in each of the different phases of the 
interstellar medium in the positron bulge, Galactic disk and 
halo. The schematic model of the bulge, disk and halo is shown in 
Figure 1. Because the positrons, produced by SNIa and SNIp, are 
widely distributed in the bulge and disk (equation 3), we assume 
that the fraction produced in the various phases of the 
interstellar medium are proportional to their appropriate filling 
factors. Although the supernovae themselves disturb the local 
medium, the $\sim$ MeV positrons from the decay of radionuclei 
escape from the remnant into the undisturbed surroundings. 
 
We estimate the relative probabilities, P, of positron 
annihilation within each phase and positron escape into 
neighboring phases by the following procedure. From the 
properties of the medium in each phase, we determine the 
propagation mode, diffusion or streaming, and calculate the 
diffusion mean free path $\lambda_{\|}$ (equation 6) in the MHD 
scattering mode, or the mean velocity, $\bar{\beta}c$, in the 
unscattered streaming mode. 
 
Although we have also made Monte Carlo simulations, we estimate 
the propagation of positrons during their slowing down in each 
phase of the interstellar medium, using a simple approximation. 
Since even the average properties and structure of the magnetic 
field especially are very poorly known in the various phases, no 
more sophisticated treatment seems justified. For the 1-D 
diffusion in a uniform medium with a mean free path 
$\lambda_{\|}$ along a magnetic flux tube, $N$ positrons produced 
at a point, $l$ = 0 and $t$ = 0, will be distributed in both 
directions along the flux tube with a density, $n(l) = (N/l_{sd}) 
e^{-(l/l_{sd})^2}$, where $l_{sd} = (2\lambda_{\|}\beta c 
t_{sd}/3)^{1/2}$, by the time $t_{sd}$ that they slow down and 
annihilate. This corresponds to a mean positron density $\bar{n} = 
N/2l_{sd}$ over the total mean length $2l_{sd}$. Similarly for 
positrons streaming at their velocity $\bar{\beta} c$ with an 
isotropic pitch angle distribution they will also have a mean 
density $N/2l_{sd}$ over a total mean length $2l_{sd}$, where 
$l_{sd} \sim \bar{\beta} c t_{sd}/2$, when they slow down and 
annihilate. 
 
Thus, if the positrons are produced uniformly along some mean 
length $_xl_B$ of the magnetic flux tubes threading through some 
phase $x$, from the above we expect that the probability, 
$P_{x:x}$, that such positrons will slow down and annihilate in 
that phase before they escape is crudely,

\begin{eqnarray} 
P_{x:x} ~ & \sim & ~_xl_B/2_xl_{sd}~~~{\rm for}~~_xl_B < _xl_{sd}\nonumber\\ 
{\rm and}\nonumber\\ 
P_{x:x} ~ & \sim & ~_xl_B/(_xl_B + _xl_{sd})~~~{\rm for}~~_xl_B > 
_xl_{sd}, 
\end{eqnarray} 
\vspace*{6pt} 
 
\noindent where $_xl_{sd}$ is the slowing down length in that 
phase. The remaining positrons have a probability, $1 - P_{x:x}$, 
of escaping from phase $x$ into a neighboring phase, and we 
estimate the relative fractions of the escaping positrons that go 
into each of the adjacent phases from simple geometric arguments, 
allowing large uncertainties of $\pm$ 50\%. 
 
The probability that positrons which diffuse into a phase $x$ 
will slow down and annihilate there is only slightly different 
from those born uniformly within it. These positrons are 
effectively born at their first scatter within the phase, roughly 
within a scattering mean free path of the boundary. So if $_xl_B 
< _xl_{sd}$, the probability of their slowing down and 
annihilating in the phase, $P_{x:x}$ is the same as for those 
born uniformly within, $\sim _xl_B/2_xl_{sd}$. But if $_xl_B > 
_xl_{sd}$, then $P_{x:x}$ remains at $\sim 1/2$, since after the 
first scatter near the boundary half would be expected to be 
spread over $_xl_{sd}$ in both directions along the flux tube by 
the time they slow down and annihilate. Thus only about half of 
them will slow down and annihilate in phase $x$, while the other 
half will escape, or be effectively reflected. As we show, 
however, in the dense labyrinth of clouds in the CMZ and tilted 
disk the half of the positrons that diffuse into clouds from the 
VH and HM and are reflected back out rather than annihilating 
within, quickly diffuse into another cloud where half are again 
reflected, and by the time they have encountered a half dozen 
other clouds only a small fraction, $\sim (1/2)^7,$ or $<$ 1\% 
remain. As we also show, however, in the diffuse disk beyond 1.5 
kpc that is not the case. 
 
From these simple approximations, we estimate the different 
propagation fractions $P$ that relate the positron production 
rates $Q$ and annihilation rates $A$, as defined in equation (4) 
for the expected bulge/disk annihilation ratio, expanding that 
equation to explicitly define these fractions in the different 
phases of the interstellar medium within the bulge, disk and halo. 
 
We consider here the Galactic positron bulge as it is defined by 
the spherical component ($<$ 1.5 kpc) of the 511 keV luminosity 
in the SPI/INTEGRAL analyses (Kn\"odlseder et al. 2005; 
Weidenspointner et al. 2007, 2008a). Also based on their model 
fitting we discuss it in two parts: the inner bulge region (R $<$ 
0.5 kpc) and the surrounding outer, or tilted-disk, bulge region 
(0.5 kpc $<$ R $<$ 1.5 kpc). 
 
In order to compare with the SPI/INTEGRAL measurements, we need 
to separate the positron production and annihilation in the bulge 
and disk into five regions, shown in Figure 1, defined primarily 
by their galacto-centic radius, although they also have very 
different mixes of interstellar phases, as recently reviewed by 
Ferri\`ere, Gillard \& Jean (2007). These are 1) the very hot, 
dense inner or nuclear bulge, $Bi$, ($<$ 0.5 kpc); 2) the hot, 
middle bulge and tilted disk, $Bm,$ (0.5 to 1.5 kpc); 3) the hot, 
tenuous outer bulge, $Bo$ between $\sim$ 1.5 kpc and the Galactic 
molecular ring at around $\sim$ 3.5 kpc, dominated by the bulge 
wind (e.g. Bregman 1980; Blitz et al. 1993); 4) the predominantly 
warm neutral outer disk beyond 3.5 kpc, $Do$, and 5) the hot, 
tenuous halo, $H$, beyond 3.5 kpc and above the disk. The 
properties of each of these regions, which we discuss below, are 
summarized in Tables 1 and 2. 
 
Based on the distributions of the surface density of the three 
positron source components, given in equation (3): the bulge and 
disk distributions of SNIa and SNIp occurrences, and the disk 
distribution of massive (OB) stars (McKee \& Williams 1997), we 
estimated the fraction of positron production in each of these 
regions. The cumulative production as a function of Galactic 
radius is shown in Figure 2 for each component. We find that 45\%, 
38\% and 17\% of the positrons from the stellar bulge $B^*$ 
component of SNIa and SNIp are produced in the regions $<$ 0.5 
kpc, 0.5 to 1.5 kpc, and 1.5 to 3.5 kpc, respectively, plus 10\% 
of the massive star $^{26}$Al component within 0.5 kpc. Of the 
positrons from the stellar disk $D*$ component of SNIa and SNIp, 
2\%, 11\% and 30\% are produced in those same regions, and the 
remaining 57\% are produced beyond 3.5 kpc, together with 90\% of 
the massive star $^{26}$Al component. 
 
Thus the total positron production in the inner bulge $Bi$ within 
0.5 kpc, the middle bulge $Bm$ between 0.5 and 1.5 kpc and the 
outer bulge $Bo$ between 1.5 and 3.5 kpc, including both stellar 
bulge and disk components, and the production in the outer disk 
$Do$ beyond 3.5 kpc, are 
 
\begin{eqnarray} 
Q_{Bi} = [(4.1\pm1.6) f_{56} + (0.10\pm0.05)]\times10^{43}~{\rm 
e}^+ {\rm s}^{-1}\nonumber\\ 
Q_{Bm} = [(5.7\pm2.3) f_{56} + (0.09\pm0.04)]\times10^{43}~~{\rm 
e}^+ {\rm s}^{-1},\nonumber\\ 
Q_{Bo} = [(8.6\pm3.5) f_{56} + (0.13\pm0.05)]\times10^{43}~~{\rm 
e}^+ {\rm s}^{-1},\nonumber\\ 
Q_{Do} = [(13.7\pm5.6) f_{56} + (0.48\pm0.24)]\times10^{43}~~{\rm 
e}^+ {\rm s}^{-1}.~~~ 
\end{eqnarray} 
\vspace*{0.1cm} 
 
\noindent And using the best-fit positron survival fraction, 
$f_{56} \sim 5\pm2$\% from $\S$ 2.2 above, the total positron 
production rates are 
 
\begin{eqnarray} 
Q_{Bi} = (0.31\pm0.07)]\times10^{43}~{\rm 
e}^+ {\rm s}^{-1}\nonumber\\ 
Q_{Bm} = (0.37\pm0.09)]\times10^{43}~~{\rm 
e}^+ {\rm s}^{-1},\nonumber\\ 
Q_{Bo} = (0.56\pm0.14)]\times10^{43}~~{\rm e}^+ {\rm s}^{-1},\nonumber\\ 
Q_{Do} = (1.16\pm0.20)]\times10^{43}~~{\rm e}^+ {\rm s}^{-1}.~~~ 
\end{eqnarray} 
\vspace*{0.1cm} 
 
Within each of these regions, $y$, we determine the expected 
positron production rates in each separate interstellar gas and 
plasma phase, $x$, by the relative filling factors of those 
phases, $f_{yx}$, times the production rate in that region, $Q_y$, 
such that $Q_{yx} \sim f_{yx} Q_y.$ We then estimate the final 
positron annihilation rates, $A_{y'x'}$, in each phase $x'$ 
within each region $y'$, as the sums over $x, y, x'$ and $y'$ of 
products of the production rates, $Q_{yx}$, times the propagation 
fractions, $P_{yx:y'x'},$ where each is the fraction of the 
positrons born in $yx$ that propagate to and annihilate in 
$y'x'$, as discussed above and given for $x:x$ in equation (8). 
 
In the next three sections, we estimate these propagation 
fractions from modelling the propagation, slowing down and 
annihilation of positrons in each of these regions and phases 
starting with the Galactic bulge and moving outward in the disk 
and ultimately into the halo, which is an important region of 
propagation and annihilation, despite its lack of local 
production.

\subsection{Inner Bulge (R $<$ 0.5 kpc) }

We model the complex interstellar phenomena of the inner region 
of the Galaxy (R $<$ 0.5 kpc), assuming that a very hot ($\sim$ 
10$^8$ K), high-pressure plasma permeates the region, pressure 
equilibrium exists among the interstellar phases (Spergel \& 
Blitz 1992; Carral et al.\ 1994), and the properties of the HII, 
HI, and H$_{2}$ regions are related to each other via the 
scenario of photodissociation regions (PDR) (Tielens \& 
Hollenbach 1985). 
 
O-star radiation is the primary photo-ionization source in the 
interstellar medium (McKee \& Williams 1997). Outside of galactic 
centers, ultraviolet radiation flux at an arbitrary location in 
galactic nuclear regions is generated by nearby OB associations, 
or by single O stars distributed randomly throughout the regions 
(Wolfire, Tielens, \& Hollenbach 1990). However, in the immediate 
vicinity of a galactic center ionizing radiation is dominated by 
the radiation contribution of active galactic nuclei. In the 
inner Galactic bulge the bulk of the ionizing radiation seems to 
be generated by randomly distributed O-stars which photo-ionize 
the outer layers of nearby interstellar clouds (Wolfire, Tielens, 
\& Hollenbach 1990; Carral et al. 1991). However, in the outer 
($>$ 3.5 kpc) Galaxy the great majority of O stars are clumped in 
OB associations, which in turn create dense, compact HII regions, 
which surround such OB associations (McKee \& Williams 1997). In 
the outer galaxy, only, small fraction ($<$ 15\%) of ionizing 
radiation generated in these OB associations escapes to maintain 
the diffuse HII. 
 
Thus the photo-ionized, as well as the neutral, phases in the 
centers of galaxies have been modeled as ensembles of PDRs 
(Wolfire, Tielens, \& Hollenbach 1990; Carral et al.\ 1991). Each 
cloud is viewed as having a spherical molecular core of radius, 
$r_{H_2}$, which contains the bulk of the cloud mass at a 
molecular hydrogen density $n_{H_{2}}$. Surrounding the molecular 
cores are cold atomic HI envelopes, $\Delta r_{HI}$ thick, which 
in turn generate infrared continua and fine-structure line 
emissions (Wolfire, Tielens, \& Hollenbach 1990). Finally, 
envelopes of photoionized HII plasma form outer shells, $\Delta 
r_{HII}$ thick, surrounding these HI shells (Carral et al.\ 
1994), ionized by Lyman continuum emission from randomly 
distributed hot, massive stars within the central 0.5 kpc. 
Following Caral et al., we assume that these HII envelopes 
constitute the primary photoionized gas component and a schematic 
model of the clouds is shown in Figure 3. Although the the hot, 
massive O stars that generate the HII ionizing emission are 
thought to occur only in the CMZ, we expect that clouds in the 
inner portion of the surrounding tilted disk are also likely to 
be irradiated with sufficient flux to maintain HII envelopes. The 
extent of such irradiation is not known, however, and here we 
arbitrarily assume that it extends only to 0.5 kpc. 
 
We also assume that the magnetic flux tubes nominally pass 
through the cold clouds in a parallel array, but with large 
turbulent perturbations superimposed, as suggested by magnetic 
field in giant molecular clouds. Li et al. (2006), using the 
SPARO 450 $\mu$m polarimetry observations, have found that within 
some such clouds in the disk, the direction of the magnetic field 
is roughly correlated with that of the local Galactic field. 
However, extensive Zeeman measurements of molecular clouds also 
indicate strong field perturbations from supersonic motions 
driven by MHD turbulence with energies comparable to that of the 
magnetic fields within the clouds (see reviews by Crutcher 1999; 
Falceta-Goncalves, Lazarian \& Kowal 2008). 
 
Thus, we assume a nominal field to calculate the base mean length 
of flux tubes through the clouds, and then from an estimate of 
the scale length of turbulent motions, we estimate the increase of 
that mean length resulting from the effective random walk or 
meandering of the flux tubes through the turbulent perturbations. 
The nominal mean flux tube length through cold cloud cores is 
$l'_B \sim 4r/3$, where $r$ is the core radius, while the nominal 
mean length through an overlying shell of thickness $\Delta r$, is 
$l'_B \sim (4/3)[(r + \Delta r)^3 - r^3]/(r + \Delta r)^2$. 
Conservatively taking the outer scale of turbulent motion $l_o$ 
as the step size, which minimizes the random walk, we expect a 
mean meandering flux tube length, $l_B \sim (l'_B/l_o) l'_B$. 
 
In the inner bulge the molecular clouds concentrate into two 
nested disks, the Central Molecular Zone (CMZ) and the 
surrounding Tilted Disk, which extends well beyond 0.5 kpc out to 
edge of the middle bulge at 1.5 kpc. The CMZ is a highly 
asymmetric region, which is an ellipsoidal disk, $\sim$ 40 pc 
thick vertically, with lateral axes of $\sim$ 500 pc by $\sim$ 
250 pc and an interstellar H$_2$ mass of 
1.9$\times10^7~M_{\odot}$(Ferri\`ere, Gillard, \& Jean 2007). In 
this region we estimate a mean molecular cloud core mass, $M_{H_2 
core} \sim 5\times10^4$ M$_{\odot}$ and a mean core radius, 
$r_{H_2} \sim$ 5 pc, from integrations of the mass spectrum 
(2$\times10^3$ M$_{\odot} \leq $ M$_{H_{2}core}$ $\leq$ 
2$\times10^6$ M$_{\odot}$) and the size spectrum (3.3 pc 
(resolution limit) $\leq r_{H_2} \leq 10^2$ pc) of the CMZ 
molecular cloud population determined by Miyazaki \& Tsuboi 
(2000). This implies a total of $\sim$ 400 such cloud cores with 
a H$_{2}$ density $n_{H_2} \sim 1900$ cm$^{-3}$. It also gives a 
total volume filling factor of these H$_2$ clouds in the CMZ is 
$f_{H_2} \sim$ 0.08, but in the full inner bulge $<$ 0.5 kpc 
volume, the filling factor is only $f_{H_2} \sim 4\times 10^{-4}$. 
 
The much larger Tilted Disk, which is described in more detail in 
the next subsection (4.2), surrounds the CMZ, extending out to 
$\sim$ 1.5 kpc with an estimated interstellar H$_2$ mass of 
3.4$\times10^7$ M$_{\odot}$ (Ferri\`ere, Gillard, \& Jean 2007). 
About 20\% of the Tilted Disk lies within the bulge source region 
$<$ 0.5 kpc. This inner portion of the Titled Disk contains $\sim 
7\times10^6$ M$_{\odot}$ of interstellar H$_2$. We assume that 
this H$_2$ gas is also concentrated in similar molecular cloud 
cores, which would number $\sim$ 140, if they have the same mean 
density and mass as those in the CMZ, and they would have an even 
smaller filling factor in the bulge source volume $<$ 0.5 kpc. 
 
Within and surrounding the CMZ, containing the molecular clouds, 
is a very hot tenuous plasma (VH). Observations of 6.7 keV line 
emission from the K$\alpha$ He-like transition of ionized Fe 
indicate that this VH plasma has a temperature of $\sim 10^8$ K, 
and fills most of this central region (Koyama 1989; Yamauchi et 
al.\ 1990; Spergel \& Blitz 1992; Koyama et al.\ 1996; Muno et 
al.\ 2004). Thus we assume that a hot ($\sim 9\times10^7$ K), 
plasma ($n \sim 0.04$ cm$^{-3}$) fills the inner $\sim$ 500 pc of 
the bulge (e.g.\ Yamauchi et al.\ 1990). The thermal pressure, 
$p_{th}/k = n_{VH}T_{VH}/\mu \sim 10^7$ K cm$^{-3}$, where $\mu$ 
represents the mean particle mass for fully ionized H, and $k$ is 
Boltzmann's constant. This value agrees well with previous 
estimates of the CMZ thermal pressure of $\sim 5\times10^6$ K 
cm$^{-3}$ by Spergel \& Blitz (1992). 
 
Analyzing the diffuse nonthermal radio emission of the Galactic 
center region, Spergel \& Blitz (1992) found that the magnetic 
field pressure is in approximate equilibrium with the thermal 
gas pressure. Consequently, we assume that both the intercloud 
and cloud field strengths are $\sim 100\mu$G, the value expected 
from such pressure arguments, 5$\times10^{-10}$ erg cm$^{-3}$. 
 
Between the VH plasma and the H$_{2}$ core of each cloud are the 
two surrounding envelopes, the upper warm, ionized HII gas and 
the lower, cold neutral HI gas. The HII density is estimated from 
pressure balance between HII gas and the VH (Carral et al.\ 
1994). Although the thermal pressure of the overlying VH is 
approximately matched by the total pressure of the cold medium 
(H$_2$), the latter pressure is dominated by turbulence (Tielens 
\& Hollenbach 1985; Spergel \& Blitz 1992; Oka et al.\ 1998). 
Similarly, the large line widths observed in the extended HII 
gases in the Galactic center region (e.g.\ Rodriguez-Fernandez \& 
Martin-Pintado 2005) imply that turbulence also dominates the HII 
pressure. We assume that the magnetic field strengths are the 
same in the VH plasma and the HII gas. Thus we find that a 
turbulent HII pressure, $p_{turb} = 
\frac{1}{2}\rho_{HII}\overline{(u^2_{HII})}$ (Tielens \& 
Hollenbach 1985), constituting $\sim$ 85\% of the total HII 
pressure equal to the total VH pressure, possesses a root mean 
square turbulent velocity, $\overline{(u^2_{HII})}^{1/2}$, of 
$\geq$ 25 km s$^{-1}$, for a mass density, $\rho_{HII}$, 
corresponding to $n_{HII}$ $ \sim$ 100 cm$^{-3}$. Note that the 
HII thermal pressure, $2n_{HII}T_{HII}$, $\sim 10^6$ K cm$^{-3}$, 
is small compared to that of the VH, since $T_{HII}$ is $ \sim$ 
5000 K (Mezger \& Pauls 1979). 
 
Based on Wolfire et al.\ (2003), we expect that at the high thermal 
pressures found in the Galactic nuclear region, warm HI is unstable 
and only cold HI is thermally stable. From analyses of fine structure 
emissions from PDRs of CMZ molecular clouds located far from thermal 
radio sources Rodr\'iguez-Fern\'andez et al.\ (2004) found typical HI 
densities, $n_{HI}$, of $\sim 1000$ cm$^{-3}$ with $T_{HI} \sim$ 
150 K. For R $<$ 0.5 kpc, Ferri\`ere, Gillard, \& Jean estimated 
an interstellar HI mass of 
1.7$\times10^6$ M$_{\odot}$, with a space averaged density, 
$<n_{HI}> \sim 25$ cm$^{-3}$ in the CMZ and a filling factor 
there of $f_{HI} \sim$ 0.025. At such a mean density, the typical 
thickness of cold atomic shells is expected to be $\Delta r_{HI} 
\sim$ 0.5 pc. The typical thickness of HII shells, surrounding 
such HI shells with a filling factor $f_{HII} \sim$ 0.045, is 
expected to be $\Delta r_{HII} \sim$ 0.7 pc. Thus, the outer 
radius of a typical, double-shelled CMZ cloud, $r_{c} \sim $ 6.2 
pc. The total cloud filling factor in the CMZ, $f_{c} = f_{H_2} + 
f_{HII} + f_{HI}$ is $\sim$ 0.15 and it further implies a mean 
distance between cloud centers $d_{cc} \sim 2r_c f_c^{-1/3}$ of 
only $\sim$ 23 pc and that between cloud surfaces, $d_{cc} - 
2r_c$ of just $\sim$ 11 pc. 
 
As we will show, the annihilation of the positrons takes place 
almost entirely in the cloud shells. However, only a negligible 
fraction ($<$ 1\%) of them are produced there, since the filling 
factors of the cloud phases are all negligible compared to the 
inner bulge volume. Thus we assume that all of the inner bulge 
positrons are born in the pervasive, very hot medium, $Q_{BiVH} 
\sim Q_{Bi} \sim (0.31\pm0.07)\times10^{43}$ e$^+$ s$^{-1}$, 
$Q_{BiHII} \sim Q_{BiHI} \sim Q_{BiH_2} \sim 0.$ 
 
The subsequent positron annihilation in these phases of the bulge 
is then defined by the propagation fractions, which we estimate 
below using this general model of the CMZ. Because, as we show, 
the positron slowing down and annihilation in the HII phase of 
CMZ is so efficient, the probability of positrons, formed in the 
VH, penetrating through the HII into the underlying HI and H$_2$ 
is negligible, so we only show here the terms for diffusion 
between the other adjacent phases with significant filling 
factors. We also find that the probability of positrons 
successfully escaping out of the inner disk is quite small. 
 
Thus the significant terms in each phase are, 
 
\begin{eqnarray} 
A_{BiVH} \sim P_{VH:VH}(Q_{BiVH} + Q_{Bm}P_{Bm:BiVH} \nonumber \\ 
+ Q_{Bo}P_{Bo:BiVH}),\nonumber\\ 
A_{BiHII} \sim P_{HII:HII}(Q_{BiHII} + Q_{BiVH}P_{VH:HII}\nonumber \\ 
+ Q_{Bm}P_{Bm:BiHII} + Q_{Bo}P_{Bo:BiHII}),\nonumber\\ 
A_{BiHI} \sim A_{BiH_2} \sim 0, 
\end{eqnarray} 
\vspace*{0.1cm} 
 
\noindent where, like those discussed above, the propagation 
fractions $P$, subscripted VH:VH is the fraction of the positrons 
produced in the VH phase that slow down and annihilate in that 
phase, VH:HII is the fraction that escaped from it into the 
adjacent HII, and other fractions are similar. The propagation 
fractions and the resultant annihilation rates are listed in 
Table 2. 
 
\subsubsection{Very Hot Medium (VH)} 
 
The bulk of the positrons from the inner bulge SNIa and SNIp are 
expected to be born in the VH, which includes both the high 
density and temperature plasma in the CMZ and all of the lower 
density and temperature hot plasma that fills nearly $\sim$ 100\% 
of the inner bulge $<$ 0.5 kpc. To determine the propagation 
fractions within the VH, and between it and the HII regions, we 
estimate the median distance from the positron production site to 
the HII envelopes around clouds in the CMZ and inner Tilted Disk. 
We consider separately the two distributed source components: the 
compact 57\% of the inner bulge SNIa and SNIp, resulting from the 
star burst and inner disk population, and the diffuse 43\%, 
resulting from the supernovae in the general bulge population 
that are expected to occur throughout the inner bulge. 
 
For the star burst and disk component with positron production 
uniformly distributed sources in the CMZ with a mean distance 
between HII shells would be $d \sim$ 10 pc, suggests that in a 
fairly regular magnetic field the mean flux tube length $l_B$ 
would also be $\sim$ 10 pc in the CMZ. On the other hand, for the 
bulge component with the production roughly uniform throughout 
the spherical volume, the positrons would be distributed along a 
mean distance $d \sim (4/3)\pi r^3/2\pi r^2 \sim (2/3)r \sim$ 300 
pc both above and below the CMZ with similar mean flux tube 
lengths, $l_B$, along the roughly vertical magnetic field in that 
region (e.g. Beck 2001). 
 
In order to determine what fraction of the positrons, born in the 
VH, can slow down and annihilate there and what fraction escape 
into the adjacent HII envelopes, we also need to estimate the 
diffusion mean free path of these positrons in the VH plasma. 
 
It is difficult to quantify the properties of hypothetical 
turbulent flows in the VH, since the VH origin and age are 
unknown. However, a plausible constraint on such turbulence is 
that the dissipation rate of the turbulent energy must be less 
than the cooling rate of the VH plasma. The rate, at which 
turbulent energy is dissipated is $\sim 
\rho_{VH}(\overline{u^2_{VH}})^{3/2}/l_o$, in units of erg 
cm$^{-3}$ s$^{-1}$, where $\rho_{VH}$ is the mass density of the 
VH, $(\overline{u^2_{VH}})^{1/2}$ is the root mean square 
turbulent velocity, and $l_o$ is the outer scale of turbulence 
(Townsend 1976). 
 
From the X-ray observations Muno et al.\ (2004) estimated that 
the VH cooling time, $t_{cool} \sim 10^8$ yr. Since the VH 
thermal energy density, $\epsilon_{th} \approx \frac{3}{2}p_{th}$, 
the cooling constraint becomes $(\overline{u^2_{VH}})^{1/2} \sim 
(l_o\epsilon_{th} / (t_{cool}\rho_{VH}))^{1/3}$. Taking the 
maximum estimate of $l_o \sim$ 10 pc, the mean separation between 
cloud surfaces, this relation limits 
$(\overline{u^2_{VH}})^{1/2}$ to less than 50 km s$^{-1}$, which 
is much less than the VH Alfv\'en speed, $V_a \sim$ 920 km 
s$^{-1}$. This cooling relation places a severe constraint on VH 
turbulence, if the VH is long lived on the time scale of 
$t_{cool}$. However, if the VH is a transient phenomenon, this 
constraint could be weakened and the turbulence would be stronger. 
 
Thus, if the mechanism generating the VH turbulence operated over 
a duration, $t_{m}$, less than $t_{cool}$, the dissipation rate 
of the turbulent energy can be greater, $\sim E_{th}/t_{m}$. Thus 
$(\overline{u^2_{VH}})^{1/2} \sim (l_o\epsilon_{th} / 
(t_m\rho_{VH}))^{1/3}$. If, very conservatively, $t_{m} \sim$ 25 
Myr, then $(\overline{u^2_{VH}})^{1/2}$ is less than 90 km 
s$^{-1}$. Since in strong turbulent MHD flows, where energy is 
equipartitioned between magnetic and velocity fields $<\delta 
B_{VH}^2> \sim 4\pi \rho_{VH}<u_{VH}^2>$ s (e.g.  Iroshnikov 1964; 
Kraichnan 1965; Biskamp 2003), constraining $<u_{VH}^2>$ limits 
$<\delta B_{VH}^2>$, and then $<\delta B_{\|}^2>^{1/2}$, since 
$<\delta B_{\|}^2>^{1/2} \sim 0.15<\delta B_{VH}^2>^{1/2}$. Thus 
$(B_o/\delta B_{\parallel})^2 \sim 800$, and, consequently a 
rather large $\lambda_{\|} \sim$ 25 pc in the CMZ from equation 
(6). 
 
In the VH the positron slowing down time is $t_{sd} \sim 
9\times10^4/n(1 + x_e) \sim 1.1\times10^6$ yr from equation (5) 
at $n \sim 0.04$ cm$^{-3}$. In the slowing down time, $t_{sd}$, 
the positrons diffusing along a flux tube in the VH will be 
distributed over a mean length $l_{sd} \sim 
(2\lambda_{\|}\bar{\beta} ct_{sd}/3)^{1/2} \sim$ 2000 pc, since 
the average positron velocity during deceleration is $\sim 0.7c$. 
 
Roughly 61\% of the inner bulge positrons  are expected to occur 
in the CMZ, predominantly from the SNIa and SNIp in the star burst 
and the inner disk, along with 10\% of the massive OB stars 
producing $^{26}$Al. The remaining 39\%, all from SNIa and SNIp, 
are expected to occur throughout the nuclear bulge $<$ 0.5 kpc. 
This mean diffusion length is much larger than the mean flux tube 
length, $l_B$, for positrons escaping into the HII. For those 
produced by the star burst and disk component of SNIa and SNIp in 
the CMZ and inner disk, where the mean distance between the HII 
envelopes of neighboring clouds, $l_B$ is only $\sim$ 10 pc, we 
would expect from equation (8) that $P_{VH:VH} \sim l_B/2l_{sd} 
<$ 0.01, so only a negligible fraction would slow down and 
annihilate in the VH, while nearly all would diffuse into the HII. 
 
The much more extended positron production by the bulge supernovae 
fills the inner bulge more uniformly within 0.5 kpc. Within the 
full volume the mean density of the VH is closer to $\sim$ 0.007 
cm$^{-3}$ and the temperature is much lower, $\sim 6\times10^6$ K 
(Almy et al. 2000). Assuming a mean $B_o \sim 17 \mu$G in 
equilibrium with the thermal pressure, and taking the same $t_m$ 
of 25 Myr, and a longer $l_o \sim$ 50 pc, we expect a mean free 
path $\lambda_{\|} \sim$ 30 pc from equation (6). Thus, with a 
slowing down time from equation (5) of $t_{sd} \sim 6\times10^6$ 
yr, the slowing down length $l_{sd} \sim$ 5000 pc, while the 
median flux tube length of $l_B \sim$ 300 pc between the CMZ and 
the middle bulge, so $P_{VH:VH} \sim$ 0.03. 
 
Thus, weighting these two probabilities with their relative 
(0.61:0.39 including $^{26}$Al) production fractions, we would 
expect the overall $P_{VH:VH} \sim$ 0.02. Monte Carlo simulations 
give the same result. They also suggest that less than 2\% of the 
positrons escape beyond the bulge ($>$ 1.5 kpc), since they tend 
to be reflected, because the scattering mean free path there is 
much shorter and the slowing down time is much longer, so there 
is little likelihood of their slowing down and annihilating there 
before they diffuse back out and eventually diffuse into the HII 
below. Although, as we discussed above, about half of the 
positrons entering the turbulent HII envelopes are also 
reflected, because of the closeness of clouds, positrons escaping 
from one cloud envelope quickly enter another, so after just half 
a dozen cloud encounters nearly all have slowed down and 
annihilated in the HII. Therefore, we conclude that the effective 
propagation factors are $P_{VH:VH} \sim 0.02$ and $P_{VH:HII} 
\sim 0.96,$ while $P_{VH:Be} \sim 0.02.$ 
 
{\em Thus, only 2\% of the positrons born in the VH and diffusing 
along meandering magnetic flux tubes, would be expected to slow 
down and annihilate there before they moved into the HII shells 
surrounding the molecular clouds.} 
 
\subsubsection{Photoionized Medium (HII)} 
 
From the source-weighted filling factors we expect that only 
$\sim$ 0.03 of the SNIa and SNIp positrons are born in the warm 
ionized HII phase, but over an order of magnitude more positrons 
are expected to diffuse into it from the surrounding VH. 
 
The highly turbulent nature of the HII medium follows from 
observations of the CMZ that reveal extreme radio-wave scattering 
by small-scale, turbulent density fluctuations (Lazzio \& Cordes 
1998; Goldreich \& Sridhar 2006). Although the origin of such 
intense small-scale turbulence is still uncertain, such 
small-scale fluctuations are expected to occur in the ionized 
interfaces between the dense, molecular cloud cores and the VH 
(Lazio \& Cordes 1998; Goldreich \& Sridhar 2006). 
 
As discussed previously, the HII shells around molecular cloud 
cores have a density $n_{HII} \sim 100$ cm$^{-3}$ and a 
thickness, $\Delta r_{HII} \sim$ 0.7 pc. The electron fraction 
high, $x_e \sim$ 1, in view of the expected presence of the 
small-scale, ionized density fluctuations. Thus, a positron 
slowing down time, $t_{sd}$, is $\sim$ 450 yr from equation (5). 
We model the spatial scale of the HII turbulence, $l_{o} \sim 
0.25\Delta r_{HII} \sim$ 0.2 pc, assuming that the properties of 
such turbulence can be approximated by those generated by shear 
flows in channels $\Delta r_{HII}$ wide (Pope 2000). Further, we 
find that $(\overline{u^2_{HII}})^{1/2} \sim$ 25 km s$^{-1}$ from 
our assumption that turbulent pressure is about 6 times that of 
the thermal pressure. From equation (6) we expect that 
$\lambda_{\|}$ is very small, $\sim 1.7\times10^{-3}$ pc. In the 
slowing down time, $t_{sd}$, these positrons diffusing along a 
flux tube in the HII plasma will be distributed over a mean 
length $l_{sd} \sim (2\lambda_{\|}\bar{\beta} ct_{sd}/3)^{1/2} 
\sim$ 0.3 pc. 
 
This positron slowing down distance is less than the expected 
nominal mean length of a flux tube through the shells, $l'_B \sim 
(4/3)[(r + \Delta r)^3 - r^3]/(r + \Delta r)^2 \sim$ 1.2 pc, and 
much less than that expected for the meandering flux tubes in the 
turbulent ionized gas, which effectively random walk on the scale 
of the turbulence, $l_o$. For a turbulent scale equal to a quarter 
of the HII shell thickness or $\sim$ 0.2 pc, as discussed above, 
the mean meandering flux tube length would be $l_B \sim (l'_B/l_o) 
l'_B) \sim$ 7.2 pc, or more than 20 times the slowing down length. 
 
From the filling factor we expect only a negligible fraction of 
the positrons to be born in the HII, while, as we saw, 
essentially all ($\sim$ 96\%) of the positrons born in the inner 
bulge, are expected to diffuse into the HII envelopes from the VH. 
There in each encounter, as we discussed above, roughly 50\% are 
expected to quickly slow down and annihilate within the outer 
$l_{sd}$, which is much less than the mean flux tube length $l_B$ 
through the HII. Since these positrons are effectively born close 
to the boundary, the other half are expected to escape back into 
the VH. But there, because of the close proximity of the clouds, 
the escaping positrons quickly diffuse into another HII envelope, 
where the process is repeated, and soon virtually all are slowed 
down and annihilated in the HII. Thus the net propagation 
fractions are effectively $P_{HII:HII} \sim$ 1 and $P_{HII:VH} 
\sim P_{HII:HI} \sim$ 0. 
 
{\em Thus we expect essentially all of the positrons that are 
either born in the inner bulge or diffuse into it from beyond, 
slow down and annihilate in the HII envelopes, making them the 
annihilation trap of the region.} 
 
\subsection{Middle Bulge (0.5 $<$ R $<$ 1.5 kpc)} 
 
As Ferri\`ere, Gillard, \& Jean (2007) have pointed out, 
``reliable observational information" in this region of the 
Galaxy is very sketchy and the gas distributions are very 
uncertain. The middle bulge from 0.5 to 1.5 kpc is dominated by a 
tilted elliptical interstellar gas disk, observed primarily in CO 
and HI lines, with a hole in the middle of sufficient size to 
envelop the CMZ (Ferri\`ere, Gillard, \& Jean 2007). They suggest 
that the most plausible disk model is that of Liszt \& Burton 
(1980) with a semi-major axis of 1600 pc and a semi-minor axis of 
600 pc, and a central hole are 800 pc and 260 pc, all tilted by 
29$^o$ relative to the Galactic plane. Thus, as can be seen in 
our Figure 1, this disk appears in a more face-on projection on 
the plane of the sky and extends $\sim$ 800 pc above and below 
that plane. This is twice the nominal 400 pc scale height of the 
overall Galactic thick disk modeled in the SPI/INTEGRAL analyses 
of the 511 keV line emission (Weidenpointner et al.\ 2007, 2008). 
Thus, we might expect that much of emission from positron 
annihilation in the Tilted Disk would not have been counted as 
part of the modeled thick planar disk fitted from the 
SPI/INTEGRAL data and would instead have been counted as part of 
their modeled extended spherical bulge from which it is 
essentially indistinguishable. 
 
Here we model the interstellar phenomena of this outer portion of 
the tilted interstellar gas disk, again assuming that a hot 
tenuous plasma permeates the region, pressure equilibrium exists 
among all the interstellar phases, magnetic pressure is in 
equilibrium with gas thermal pressure, and the properties of the 
HI, and H$_2$ regions are related to each other via the PDR 
scenario. 
 
This disk contains an estimated $\sim 3.4\times10^7$ M$_{\odot}$ 
of H$_{2}$ and the corresponding disk-averaged gas density is 
$<n_{H_2}> \sim$ 4.8 cm$^{-3}$ (Ferri\`ere, Gillard, \& Jean 
2007). In the 80\% of the disk beyond 0.5 kpc, we assume that the 
H$_2$ also resides in compact molecular clouds which are roughly 
similar to those in the CMZ with a nominal molecular core radius 
of 5 pc, but because of the lower external pressure, these clouds 
have a lower mean H$_2$ density $n_{H_2} \sim$ 1000 cm$^{-3}$. 
Thus, the typical core mass, $M_{H_2core}$ is $\sim 
2.6\times10^4$ M$_{\odot}$, and the number of such cloud cores, 
$N_{H_2core}$, is $ \sim 1300$. For this H$_2$ component we 
expect a cloud H$_2$ filling factor in the outer part of the 
tilted disk, $f_{H_2} \sim 4.8/1000 \sim$ 0.005. 
 
PDR shells surrounding molecular cloud cores must be sufficiently 
thick ($\sim$1.5 A$_{v}$) to shield the underlying H$_{2}$ gas 
from dissociation via FUV absorption (Wolfire, Hollenbach \& 
Tielens 1993). A visual extinction, A$_{v} \sim$ 1.5 corresponds 
to an HI column density $N(HI) \sim 3\times10^{21}$ cm$^{-2}$ 
(Tielens 2005). Thus to shield the tilted disk molecular cloud 
cores from dissociation, $n_{HI}\Delta_{HI}$ of the overlying HI 
shells, $\sim 3\times10^{21}$ cm$^{-2}$. If we assume that the 
nominal value of the HI density $n_{HI} \sim 1000$ cm$^{-3}$, 
half the density of the H$_{2}$ cores, then $\Delta r_{HI} \sim$ 
1 pc. Thus, for the HI interstellar component we expect a filling 
factor, $f_{HI} \sim [(1 + \Delta r_{HI}/r_{H_2})^3 - 
1]\times(4.8/1000) \sim$ 0.004 in the disk. 
 
These clouds, like those in the CMZ, are likely to be highly 
turbulent with the turbulent pressure balanced by the thermal 
pressure of the external hot medium (HM). Also as with the CMZ, 
we rely on analyses of X-ray observations to supply us with the 
expected HM pressure. In their analysis of the \textsl{ROSAT} 0.75 
keV all-sky survey, Almy et al.\ (2000) found that this region of 
the Galactic bulge is filled with a hot ($\leq 10^7$ K), tenuous 
($\leq 10^{-2}$ cm$^{-3}$) plasma. However, this thermal HM 
pressure is significantly greater than the thermal pressure at 
which \textsl{both} cold HI and warm HI can coexist, based on 
Wolfire et al.\ (2003). Thus, we expect that the cold HI is the 
only stable HI phase in the tilted disk region 
 
Unlike the CMZ and inner tilted disk region, the outer tilted disk 
beyond around 0.5 kpc does not appear to have any significant HII 
component. Although a population of molecular clouds exists in 
the Tilted Disk, no giant HII regions, standard tracers of massive 
star formation, have been resolved in this region (Smith, 
Biermann, \& Mezger 1978). Moreover, no stars younger than 200 
Myr have been found here by van Loon et al.\ (2003) in the mid-IR 
survey data from DENIS and ISOGAL. So we conclude that the stellar 
Lyman continuum flux and its resulting HII component are 
negligible in this region. 
 
Therefore the filling factors in the tilted disk are $f_{H_2} 
\sim$ 0.5\%, $f_{HI} \sim$ 0.4\%, and $f_{HM} \sim$ 99\%. The net 
cloud filling factor ($f_c = f_{HI} + f_{H_2}$) implies a mean 
distance between neutral cloud centers $d_{cc} \sim 2r_c 
/f_c^{1/3}$ of $\sim$ 60 pc, and a mean distance from a random 
point in the disk intercloud medium to a cloud surface of 
$(d_{cc}/2) - (r_c + \Delta r_{HI}) \sim$ 25 pc. 
 
We do not expect the SNIa and SNIp to be concentrated in this 
highly tilted disk which extends up to $\sim$ 800 pc above the 
Galactic plane. For as we saw from equation (3) and Figure 2, 
55\% of these supernovae are in the stellar bulge population, 
distributed throughout the spherical shell from 0.5 to 1.5 kpc, 
and the other 45\% are in the disk population, concentrated in 
the Galactic plane with a scale height of $\sim$ 0.18 kpc. Thus, 
the median distance from positron sources in the HM to the 
labyrinth of HI cloud envelopes in the disk is $\sim$ 0.5 kpc for 
both populations. With the roughly vertical magnetic field in 
that region (e.g. Beck 2001), the positrons are born along flux 
tubes above and below the disk with a mean length $l_B \sim$ 1 
kpc, twice the median value. We also expected that essentially 
all of the positrons are produced in the HM with the filling 
factor in the entire middle bulge, $f_{HM} \sim$ 1, and the other 
filling factors in the full middle bulge volume are $f_{H_2} \sim 
5\times10^{-5}$ and $f_{HI} \sim 4\times10^{-5}$. Thus $Q_{HM} 
\sim Q_{Bm} \sim (0.37\pm0.09)\times10^{43}$ e$^+$ s$^{-1},$ and 
$Q_{H_2} \sim Q_{HI} \sim$ 0. 
 
Since the HM filling factor is essentially unity, the subsequent 
positron propagation and annihilation in each of the phases of 
the outer titled disk is determined predominantly by the 
propagation fractions, which we estimate below, for the diffusion 
of positrons from the HM into the HI shells and on into H$_2$ 
cloud cores, where they slow down and annihilate. 
 
\begin{eqnarray} 
A_{HM} \sim Q_{HM}P_{HM:HM},\nonumber\\ 
A_{HI} \sim Q_{HM}P_{HM:HI}P_{HI:HI},\nonumber\\ 
A_{H_2} \sim Q_{HM}P_{HM:HI}P_{HI:H_2}P_{H_2:H_2}, 
\end{eqnarray} 
\vspace*{0.1cm} 
 
\noindent where, like those discussed above, the propagation 
fractions $P$, subscripted HM:HM, HI:HI and H$_2$:H$_2$, are the 
fractions of the positron that slow down and annihilate in that 
phase, independent of whether they are produced in that phase or 
escaped into it from an adjacent phase, as given by the other $P$ 
combinations of escape from one phase to another adjacent phase. 
These fractions and annihilation rates are also listed in Table 2. 
 
\subsubsection{Hot Medium (HM) } 
 
Since the hot medium fills effectively all of the middle bulge, 
we expect that all of the positrons produced between 0.5 and 1.5 
kpc are born in the HM. Based on the analysis of Almy et al.\ 
(2000) at a Galactic planar radius of $\sim$ 1 kpc, which is the 
midpoint of this region, hot gas temperature $T_{HM} \sim 
5\times10^6$ K, the density $n_{HM} \sim 5\times10^{-3}$ 
cm$^{-3}$. Thus the positron slowing down time $t_{sd} \sim 
9\times10^6$ yr from equation (5). 
 
As with the VH, we estimate the diffusion mean free path of the 
positrons in the HM as follows. We assume, as did Spergel \& 
Blitz (1992) for the CMZ, that the magnetic field pressure is in 
equilibrium with the gas pressure. Thus we expect $B_o \sim 
\sqrt{8\pi p_{th}} \sim$ 12 $\mu$G. As with the CMZ a plausible 
constraint on the nature of such turbulent flows is that the rate 
at which turbulent energy is dissipated is less than the rate at 
which the hot plasma cools. Almy et al.\ (2000) estimate the 
cooling time of this hot plasma, $t_c \sim 3\times10^8$ yr, but 
we again consider a conservative value of $\sim 5\times10^7$ yr. 
Approximating an outer scale, $l_o \sim$ 50 pc for a hypothetical 
turbulent HM flow, this cooling time limits 
($\overline{u^2})^{1/2}$ to be less than 45 km s$^{-1}$ for a 
thermal energy density, $\epsilon_{th} \approx 1.5$p$_{th} 
\approx 10^{-11}$ erg cm$^{-3}$. Since the estimated Alfv\'en 
speed, $V_a \sim$ 310 km s$^{-1}$, this limit on 
($\overline{u^2})^{1/2}$ corresponds to $\lambda_{\|} \sim$ 45 pc 
by equation (6). In the slowing down time, $t_{sd}$ of 
$9\times10^6$ yr, HM positrons can diffuse a mean distance 
$l_{sd} \sim (2\lambda_{\|}\bar{\beta} ct_{sd}/3)^{1/2} \sim$ 7 
kpc. 
 
This distance is much larger than the mean flux tube length in 
the HM, $l_B \sim$ 1 kpc, along which the positrons are born above 
the labyrinth of HI covered clouds in the tilted disk. Thus as we 
saw with the VH, we expect from equation (8) that the fraction of 
positrons born in the HM and slowed down and annihilated there, 
$P_{HM:HM} \sim l_B/2l_{sd} \sim$ 0.07. As Monte Carlo 
simulations also suggested with the VH, only a few percent of the 
positrons are likely to escape into the tenuous region above 1.5 
kpc, while the bulk ($\sim$ 91\%) diffuse into the cloudy 
labyrinths of both the tilted disk and the CMZ below. There due 
to the closeness of nearby clouds within these disks, essentially 
all of the positrons from the HM that enter these two labyrinths 
rapidly diffuse on into the turbulent outer envelopes of clouds 
where they slow down and annihilate. 
 
We further estimate that the relative fractions of positrons born 
in the middle bulge that enter into the labyrinth of clouds in the 
outer tilted disk and annihilate in their HI envelopes, compared 
to those that escape down into the inner bulge and enter into the 
labyrinth of clouds in the CMZ and inner tilted disk and 
annihilate in their HII envelopes, are proportional to the 
relative surface areas of the two disks, or about 75\% to 25\%, 
respectively. Thus, of the 91\% of the positrons from the HM that 
enter these regions, we expect $P_{HM:HII} \sim 0.22$, diffusing 
into the inner bulge, and the $\sim$ 0.69\% remaining in the 
middle bulge are all expected to diffuse into the HI shells of 
clouds in the titled disk, and as we discuss below, essentially 
all slow down and annihilate in those shells, giving an effective 
$P_{HM:HI} \sim$ 0.69 and $P_{HM:H_2} \sim$ 0. 
 
\subsubsection{Cold Medium, (HI) } 
 
As with the molecular clouds in the CMZ, we assume that only a 
fraction of the pressure is thermal in the tilted disk molecular 
cloud cores, so the mean temperature $T_{H_2} \sim$ 20 K and 
density $n_{H_2} \sim$ 1000 H$_2$ cm$^{-3}$, although we assume 
they also have nominal radii of $\sim$ 5 pc. But because of the 
negligible filling factors of these cloud cores, no significant 
positron production should occur within them, and we show that 
essentially all of the positrons born elsewhere are expected to 
slow down and annihilate in the overlying HI envelopes before 
they can get into the H$_2$ cores. 
 
In the overlying HI envelopes between the cold H$_2$ cloud cores 
and the hot HM plasma, we assume a mean density of $n_{HI} \sim 
1000$ H cm$^{-3}$ at a temperature $T_{HI} \sim$ 100 K, giving it 
thickness $\Delta r_{HI} \sim$ 1 pc, as discussed above. 
Therefore the positron slowing down time in the HI, $t_{sd}$ is 
$\sim$ 90 yr from equation (5), and since the MHD cascade is also 
damped by ion-neutral friction, the isotropic streaming positrons 
are distributed along a flux tube over a mean length $l_{sd} \sim 
\bar{\beta}ct_{sd}/2 \sim $ 9 pc in both directions in their 
slowing down time. 
 
This is longer than the assumed nominal mean length of $l'_B \sim 
(2/3)[(r + \Delta r)^3 - r^3]/(r + \Delta r)^2 \sim$ 1.7 pc of 
unperturbed parallel flux tubes passing through a HI shell. But 
it is less than the mean meandering flux tube length $l_B \sim 
(l'_B/l_o) l'_B) \sim$ 12 pc, expected for an turbulent scale 
$l_o \sim 0.25\Delta r_{HI} \sim$ 0.25 pc, if we assume the same 
shear flow scaling (Pope 2000), as discussed above. 
 
Only a negligible fraction ($\sim$ 0.4\%) of middle bulge 
positrons are expected to be born in the HI envelopes, but as we 
saw, $\sim$ 69\% of all these positrons are expected to diffuse 
into the HI from the HM. As with the positrons that diffuse into 
the HII envelopes of CMZ clouds, these positrons are all 
effectively born within a scattering mean free path of the 
surface. Thus, since $l_{sd} < l_B$, we expect also that $\sim$ 
0.5 slow down an annihilate in the HI, before they can penetrate 
into the H$_2$ cores below, and the remainder escape back into 
the HM above. There, again because of the close proximity of the 
clouds, the escaping positrons quickly diffuse into another HI 
envelope, where the process is repeated, and soon virtually all 
are slowed down and annihilated in the HI. Thus the net 
propagation fractions are effectively $P_{HI:HI} \sim$ 1 and 
$P_{HI:VH} \sim$ 0. 
 
Although the density of the HI envelopes are quite uncertain, the 
expectation that they are thick enough to slow down and 
annihilate the positrons is rather robust. That results from the 
fact that the thickness of the HI and the slowing down length are 
both inversely proportional to the density, because the column 
depth of the HI, which shields the H$_2$ from dissociation, is 
fixed and so is the required column depth for the slowing down. 
Therefore, $l'_B$, $l_{sd}$, $l_o$, $l_B$ and $l_B/l_{sd}$ all 
nominally scale proportionally. 
 
{\em Thus, we expect most ($\sim$ 70\%) of the positrons that are 
either born in the moddle bulge or diffuse into it from the outer 
bulge, slow down and annihilate in the HI envelopes of the clouds 
in the tilted disk, making them the annihilation trap of the 
region, as the HII envelopes are in the inner bulge.} 
 
As we discuss in the summary ($\S$7), nearly all of the positrons 
that slow down in the HI envelopes in the middle bulge annihilate 
via fast positronium formation in charge exchange, producing broad 
511 keV line emission, while those that slow down in the HII 
envelopes in the inner bulge annihilate predominantly via thermal 
radiative positronium formation, producing narrow 511 keV line 
emission.

\subsection{Outer Bulge (1.5 $<$ R $<$ 3.5 kpc)} 
 
Based on the stellar disk distribution of SNIa and SNIp, as 
discussed above, we see from Figure 2 that $\sim$ 30\% of all the 
stellar disk positrons and $\sim$ 22\% of all the stellar bulge 
positrons produced by them should be made in the outer bulge 
between 1.5 kpc and 3.5 kpc, and, as we now show, they become a 
major contributor to the modeled positron bulge component of the 
observed annihilation radiation within 1.5 kpc. 
 
This region has been known for some time (Faber \& Gallagher 
1976; G\"usten 1989; Dehnen \& Binney 1998) to contain little HI, 
as traced by 21 cm radio line emission (Lockman 1984), and little 
H$_{2}$, as traced by 2.6 mm line of CO (Sanders, Solomon \& 
Scoville 1984; Robinson et al.\ 1988). The lack of neutral gas 
results from a ``high pressure galactic wind" (e.g. Blitz et al. 
2007, Bregman 1980), driven primarily by stellar bulge supernovae 
within $\sim$ 0.5 kpc, that blows out to at least 3 kpc, before 
it can recombine to form new neutral gas. 
 
In this region, therefore, we consider only the hot plasma, with 
a mean density $n \sim 2\times10^{-3}$ cm$^{-3}$ averaged over the 
scale height of $\sim$ 2 kpc, a conservative mean magnetic field 
strength, $B_o \sim$ 4 $\mu$G, an outer scale for MHD turbulence, 
$l_{o} \sim$ 75 pc, based on simulations of supernova-driven MHD 
turbulence (de Avillez and Breitschwerdt 2004), and, finally, we 
assume that the root mean square strength of the turbulent 
magnetic field $\delta B/B_o \sim$ 0.5. From equation (6), we 
then find that the scattering mean free path parallel to the mean 
magnetic field, $\lambda_{\|} \sim$ 14 pc. Assuming that the bulge 
wind blows the magnetic field out radially, the mean length of a 
flux tube through the region is $l_B \sim 2$ kpc. 
 
In this ionized plasma the slowing down time, $t_{sd} \sim$ 23 
Myr from equation (5). In that time the positrons diffuse both 
directions along a flux tube a mean length $l_{sd} \sim 
(2\lambda_{\|}\bar{\beta}ct_{sd}/3)^{1/2} \sim $ 6 kpc. Comparing 
this length with that of the flux tube, we expect from equation 
(8) that the probability that positrons, born uniformly in the Bo, 
slow down and annihilate there is $\sim l_B/2l_{sd} \sim$ 0.17. 
But since the contribution from this region was also included in 
the model halo of Weidenspointner et al. (2008a), we include Bo 
there instead and take 1.5 kpc as the inner boundary of the halo. 
 
The remainder escape both into the bulge below and the disk and 
halo above. As we see from the distribution of the supernovae in 
Figure 2, the 70\% of the Bo supernovae in the disk are rather 
uniformly distributed between the two boundaries and we assume 
that equal fractions escape up and down. However, since the 
remaining 30\% from the bulge component are concentrated along an 
$l_B <$ 1 kpc close the inner boundary at 1.5 kpc, we expect that 
they escape primarily into the bulge and have a smaller chance of 
annihilating in the Bo. Thus, we estimate that the net fraction 
of Bo positrons annihilating there is $P_{Bo:Bo} \sim$ 0.14 and 
that escaping down into the middle bulge $P_{Bo:Bm} \sim$ 0.57. 
 
As with the positrons produced in the middle bulge, we further 
expect that the $\sim$ 57\% of the outer bulge positrons going 
down into the Bm from the Bo will also meet the same fates in the 
same proportions, mostly diffusing into the labyrinths of the 
tilted disk and the CMZ, annihilating in the HI and HII envelopes 
of the molecular clouds. Thus, of all the positrons produced in 
the Be we expect $P_{Bo:HI} \sim 0.40$ and $P_{Bo:HM} \sim 0.04$ 
in the middle bulge and $P_{Bo:HII} \sim 0.13$ in the inner bulge. 
 
Those positrons going beyond 3.5 kpc are also expected to 
annihilate and not return, and within large uncertainties we 
assume that they have roughly equal likelihood of going into the 
halo or the disk. Thus $P_{Bo:Do} \sim$ 0.14, while since the Be 
anniilation was also included in the halo, the net $P_{Bo:H} 
\sim$ 0.29.

\subsection{Positron Bulge ($<$ 1.5 kpc) Annihilation} 
 
We now estimate the total positron annihilation rate within the 
1.5 kpc bulge, using the contributions of the various phases and 
the propagation fractions within and between them, summarized in 
Table 2. 
 
Using these fractions and combining equations (9) through (13), 
the total annihilation rate in the positron bulge $<$ 1.5 kpc and 
its components are thus expected to be, 
 
\begin{eqnarray} 
A_{HII} \sim (0.46\pm0.08)\times10^{43}~~{\rm e}^+ {\rm 
s}^{-1},\nonumber \\ 
A_{HI} \sim (0.48\pm0.17)\times10^{43}~~{\rm e}^+ {\rm 
s}^{-1},\nonumber \\ 
A_{VH+HM} \sim 0.06\times10^{43}~~{\rm e}^+ {\rm 
s}^{-1},\nonumber \\ 
A_{Bi} \sim (0.47\pm0.08)\times10^{43}~~{\rm e}^+ {\rm 
s}^{-1},\nonumber \\ 
A_{Bm} \sim (0.53\pm0.13)\times10^{43}~~{\rm e}^+ {\rm 
s}^{-1},\nonumber \\ 
A_B \sim (1.00\pm0.15)\times10^{43}~~{\rm e}^+ {\rm s}^{-1}.~~~ 
\end{eqnarray} 
\vspace*{0.1cm} 
 
As we discuss in detail in the summary ($\S$ 7), this is quite 
consistent with the best-fit bulge component annihilation rate 
from the SPI/INTEGRAL analyses (Weidenspointner et al. 2008a). 
Although we assumed large uncertainties of $\pm$50\% in the 
estimated propagation fractions between interstellar gas phases, 
which produce comparably large uncertainties in the individual 
annihilation rates in each, their sums are more tightly 
constrained. 
 
We also see (Table 2) that although the hot phase of the 
interstellar medium fills most of the bulge, because of 
propagation it is not the site of most of the annihilation. We 
find (Table 2 and equation 13) from the expected propagation and 
slowing down, that $\sim 46\pm$8\% of the bulge positrons 
annihilate in the HII shells surrounding the molecular clouds in 
the highly ionizing radiation environment of the inner bulge $<$ 
0.5 kpc, via thermal positronium formation resulting in the 
narrow ($\sim$ 1.2 keV), 511 keV line emission (Guessoum, Jean \& 
Gillard 2005) from the bulge. We also see that $\sim 48\pm$12\% 
of the annihilation is expected to occur in the middle bulge, 0.5 
to 1.5 kpc, in the neutral HI shells surrounding the molecular 
clouds which lack significant HII shells, via in-flight 
positronium formation resulting in the broad ($\sim$ 5.8 keV), 
511 keV line emission from the bulge. As we discuss in detail 
below (\S 7), these two annihilation sites can account for the 
observed broad/narrow 511 keV line ratio of $\sim$ 0.5 (Churazov 
et al. 2005; Jean et al. 2006) in the positron bulge. We further 
point out that since these two components are expected to be 
essentially formed in separate, resolvable regions, spectral 
analyses of the two regions can test this prediction and further 
probe the structure of the bulge interstellar medium. 
 
Only a fraction $P_{Bo:Bo} \sim$ 14\% of the positrons are 
expected to slow down in the outer bulge, producing an 
annihilation rate $A_{Bo} \sim 0.09\times 10^{43}$ e$^+$ s$^{-1}$ 
in an essentially spherical shell contribution between 1.5 and 
3.5 kpc. This is consistent with the effective limit of $< 
(0.2\pm0.2)\times 10^{43}$ e$^+$ s$^{-1}$ from that particular 
region, inferred from the best-fit modeling of the SPI/INTEGRAL 
data (Kn\"odlseder et al. 2005). But since the contribution from 
that region has subsequently been included as part of the model 
halo in the analyses of Weidenspointner et al. (2008a), as 
discussed above, we include $A_{Bo}$ in the halo annihilation 
instead. Even though the annihilation fraction in the outer bulge 
is small, its contribution to the observed 511 keV line flux is 
still significant, because of the high $\gamma_{511}/e^+$ yield of 
$\sim 1.7$ expected (Jean et al. 2006) from positron annihilation 
in dusty hot plasma, as discussed above ($\S$ 1.2). 
 
In the next section we follow the remaining $\sim (0.16\pm0.10) 
\times 10^{43}$ e$^+$ s$^{-1}$ of the outer bulge positrons that 
escape beyond 3.5 kpc into both the halo and the inner spiral 
arms that define the Galactic molecular ring of the interstellar 
disk. As we also show below ($\S$ 5.4), the inner disk 
annihilation dominated by the these arms, which are viewed 
asymmetrically from our solar perspective, can fully account for 
the disk flux asymmetries (Weidenspointner et al. 2008b).

\section{Positron Propagation \& Annihilation in the Galactic Disk ($>$ 3.5 kpc)} 
 
Having estimated the expected production, propagation and 
annihilation of positrons in the positron bulge ($<$ 1.5 kpc) and 
in the outer bulge from 1.5 to 3.5 kpc, we now turn to the 
various components of the positron production, propagation and 
annihilation in the outer disk ($Do)$ beyond 3.5 kpc. 
 
As we have seen, the total expected positron production rate in 
this region consists of those born there, $Q_{Do}$, and those that 
diffuse into it from the outer bulge between 1.5 and 3.5 kpc, 
$Q_{Bo}P_{Bo:Do}$, we estimate 
 
\begin{eqnarray} 
Q_{Do} \sim (1.16\pm0.20)\times10^{43}~~{\rm e}^+ {\rm s}^{-1},\nonumber\\ 
Q_{Bo}P_{Bo:Do} \sim (0.08\pm0.04)\times10^{43}~~{\rm e}^+ {\rm 
s}^{-1}. 
\end{eqnarray} 
\vspace*{0.1cm} 
 
We again focus on the major phases of the interstellar medium, 
which fill essentially all of the outer disk volume in which 
essentially all of the supernovae in this region occur, namely: 
the warm predominantly neutral HI medium, (HI), the hot tenuous 
medium (HT), and the photoionized medium produced by O stars: the 
dense dense HII regions (HII) and the diffuse warm ionized medium 
(WI). The dense ionized phases together with the molecular 
clouds, OB star associations and superbubbles are all concentrated 
along the spiral arms (e.g Drimmel \& Spergel 2001; Stark \& Lee 
2006), which dominate the structure of the outer disk. Their 
properties, discussed below, are summarized in Table 3. 
 
We assume nominal average values of filling factors of these 
phases beyond 3.5 kpc of $f_{HI} \sim$ 0.50, $f_{HT} \sim$ 0.20, 
$f_{HII} \sim$ 0.15 and $f_{WI} \sim$ 0.15, based roughly on 
recent models (Ferri\`ere 1998, 2001) of a thick disk of $\sim$ 
400 pc half thickness. There are large uncertainties in these 
numbers, but since they provide only an estimate of where the 
positrons are born, the uncertainties do not have a significant 
impact on where they eventually slow down and annihilate, for as 
we have seen elsewhere, that is dominated by their propagation. 
Although we do not consider independently the other interstellar 
phases of the disk, namely the molecular H$_2$ and HI clouds, 
because of their very small filling factors ($<$ 1\% Ferri\`ere 
1998), we do consider their effects in the diffuse neutral medium 
(HI). 
 
As we saw from the integration of equation (3) and Figure 2, 57\% 
of the SNIa and SNIp in the disk occur $>$ 3.5 kpc, producing 
$\sim (0.86\pm0.30)\times10^{43}$ e$^+$ s$^{-1}$. We assume that 
they occur randomly in the various phases of the disk 
proportional to their filling fractions and that, as discussed 
above, $\sim$ 90\% of the $\sim (0.3\pm0.1)\times10^{43}$ e$^+$ 
s$^{-1}$, produced by $^{26}$Al decay from massive stars in the 
disk, are born in the HT phase of hot superbubbles. Thus we 
expect that the positron production rates in each of the major 
phases of the disk beyond 3.5 kpc are 
 
\begin{eqnarray} 
Q_{HT} \sim (0.45\pm0.12)]\times10^{43}~~{\rm e}^+ {\rm s}^{-1},\nonumber\\ 
Q_{HI} \sim (0.45\pm0.15)]\times10^{43}~~{\rm e}^+ {\rm s}^{-1},\nonumber\\ 
Q_{HII} \sim Q_{WI} \sim (0.13\pm0.04)]\times10^{43}~~{\rm e}^+ 
{\rm s}^{-1}.~~~ 
\end{eqnarray} 
\vspace*{0.1cm} 
 
From the positron production in these phases, we estimate the 
propagation fractions, listed in Table 4, and the resulting 
positron annihilation in the disk beyond 3.5 kpc as follows, 
 
\begin{eqnarray} 
A_{HT} \sim P_{HT:HT}(Q_{HT} + Q_{HII}P_{HII:HT} + Q_{WI}P_{WI:HT}),\nonumber \\ 
A_{HI} \sim P_{HI:HI}(Q_{HI} + Q_{HII}P_{HII:HI} + Q_{WI}P_{WI:HI}),\nonumber \\ 
A_{HII} \sim P_{HII:HII}(Q_{HII} + Q_{HI}P_{HI:HII} + Q_{WI}P_{WI:HII}\nonumber \\ 
 + Q_{HT}P_{HT:HII} + Q_{Bo}P_{Bo:HII},)\nonumber\\ 
A_{WI} \sim P_{WI:WI}(Q_{WI} + Q_{HI}P_{HI:WI} + Q_{HII}P_{HII:WI}\nonumber \\ 
+ Q_{HT}P_{HT:WI}),~~~ 
\end{eqnarray} 
\vspace*{0.1cm} 
 
\noindent where, like those discussed above, the propagation 
fractions $P$, subscripted HT:HT, HI:HI, HII:HII, and WI:WI, are 
the fractions of the positrons that are either produced in that 
phase or escape into it from an adjacent phase, that then slow 
down in and annihilate in that phase, while the other $P$s are all 
combinations of escape from one phase to another adjacent phase. 
That excludes only direct escape from the HT into the HI, or vise 
versa, because they are separated from one another by the HII and 
WI, and any return from the halo, which we assume is negligible. 
 
Because of the large vertical scale height of the hot tenuous 
plasma in the large blown out superbubbles, annihilation there is 
also effectively part of the halo, not a resolvable component of 
the $\pm$400 pc disk, so we add $A_{HT}$ to $A_H$. 
 
\subsection{Hot Tenuous Medium (HT) } 
 
Ferri\`ere (2001) found that the mean HT filling factor in the 
outer disk was a modest, $f_{HT} \sim$ 0.2. She also found that 
the HT was dominated by the contribution of superbubbles (e.g. 
Tomisaka \& Ikeuchi 1986; Heiles 1987, 1990; Mac Low \& McCray 
1988) created by core-collapse supernovae, which in turn are 
correlated in space and time. We further found (Higdon \& 
Lingenfelter 2005) that the majority ($\sim$ 85\%) of 
core-collapse supernovae (SNII \& SNIb/c) occur in superbubbles 
and an additional fraction ($\sim$ 20\%) of the SNIa and SNIp 
occur there randomly with a probability equal to the superbubble 
filling factor. 
 
Positron production in the superbubbles therefore comes both from 
the decay of $^{26}$Al from the massive star Wolf Rayet winds and 
SNII, which create them, and from the decay of $^{56}$Co and 
$^{44}$Ti from the random SNIa and SNIp. Thus, as discussed above, 
the production in the HT is expected to be $Q_{HT} \sim 
(0.45\pm0.12)\times10^{43}$ e$^+$ s$^{-1}$, with $Q_{26} \sim 
(0.27\pm0.09)\times10^{43}$ e$^+$ s$^{-1}$ from clustered massive 
stars and $Q_{Ia+Ip} \sim (0.18\pm0.04)\times10^{43}$ e$^+$ 
s$^{-1}$ from the random SNIa and SNIp. 
 
The typical properties of the hot medium (e.g. Ferri\`ere 1998; 
Yan \& Lazarian 2004) are: a nominal temperature $T_{HM} \sim 
1\times10^6$ K, a scale height averaged density, $n_{HM} \sim 
2\times10^{-3}$ cm$^{-3}$, and a mean magnetic field, $B_o \sim$ 4 
$\mu$G. As discussed in \S 4.1.1, the rate at which turbulence is 
dissipated must be less than the radiative rate at which the 
plasma cools. However, the cooling time of HM in outer Galaxy is 
very uncertain. Therefore we used $t_{cool} \sim 3\times10^7$ yr, 
the log average cooling time for a hot medium, whose evolution is 
dominated by time-dependent cloud evaporation following the 
prescription of McKee \& Ostriker (1977), and a cooling time from 
Tielens (2005) for the above values of $T_{HM}$ and $n_{HM}$. 
Using an outer scale, $l_o \sim$ 75 pc (de~Avillez and 
Breitschwerdt 2004), the above cooling time limits 
($\overline{u^2})^{1/2}$ to be less than 40 km s$^{-1}$. With 
$V_a \sim$ 80 km s$^{-1}$ and the root-mean-square $\delta B/B_o 
\leq$ 0.5, equation (6) implies $\lambda_{\|} \geq$ 10 pc. 
 
In this very tenuous plasma the positron slowing down time is 
$t_{sd} \sim 23$ Myr from equation (5) and the slowing down 
length $l_{sd} \sim (2\lambda_{\|} \bar{\beta}c t_{sd}/3)^{1/2} 
>$ 6 kpc. Assuming a typical flux tube length of $l_B <$ 1 kpc in 
the superbubbles blowing out into the halo, the effective 
propagation fraction, $P_{HT:HT} \sim l_B/2l_{sd} < 0.08$. Since, 
as noted above, these scales are much larger than the disk 
supernova distributions, there is no resolvable disk component of 
annihilation in the HT. Thus the $A_{HT} \sim P_{HT:HT}Q_{HT} < 
0.04\times10^{43}$ e$^+$ s$^{-1}$ is included in the halo $A_H$. 
 
As with the bulge wind dominated hot medium between 1.5 and 3.5 
kpc, we again assume that the escaping positrons go into both the 
overlying halo and the photoionized, WI and HII phases in the 
disk. For although the upper ends of these flux tubes are blown 
out into the overlying halo, so that the positrons could all 
escape, their roots feed into the surrounding HII and WI 
envelopes that separate the HT from the HI. Thus within large 
uncertainties, we also assume the nominal case where escape up 
into the halo or down into the surrounding photoionized gas is 
equally probable ($\sim$ 46\% each), and that of those entering 
the photoionized phases the WI and HII are also equally likely 
($\sim$ 23\% each). As we show below, however, we expect that 
only 66\% of the positrons that enter either the WI or the HII, 
slow down and annihilate before there, while the remainder escape 
into the halo, where they annihilate. Thus, we expect that the 
effective net propagation factors are $P_{HT:WI} \sim 
0.15\pm0.07$, $P_{HT:HII} \sim 0.15\pm0.07,$ and $P_{HT:H} \sim 
0.70\pm0.35,$ including $P_{HT:HT}$. 
 
\subsection{Warm Neutral Medium (HI) } 
 
Studies by Wolfire et al. (2003) find that over most of the 
Galactic disk $>$ 3.5 kpc neutral HI gas resides in two stable, 
co-existing phases: a cold ($\sim$ 100 K) neutral medium and a 
warm ($\leq 10^4$ K) neutral medium, but the filling factor of 
cold HI is negligible (Ferri\`ere 1998). Further, Wolfire et al. 
found relative electron densities in the predominantly neutral 
HI, $n_{HI}/n_{i} \sim 1/x_{e} \sim$ 50 which greatly exceeds 
$f_{crit}$ of equation (7), so turbulent Alfv\'enic cascades are 
damped by ion-neutral friction (e.g.\ Kulsrud \& Pearce 1969). 
Thus we expect the positrons to stream isotropically along 
magnetic flux tubes at $\bar{\beta}c/2$. 
 
The mean HI gas density seen by the positrons escaping into the 
interstellar medium from SNIa is 0.13 cm$^{-3}$, using the HI gas 
density distribution from Ferri\`ere (1998) applicable at $>$ 3.5 
kpc and for our modeled SNIa and SNIp spatial distribution, 
equation (3). Thus the slowing down time, $t_{sd} \sim 
7\times10^5$ yr from equation (5), and in that time they are 
spread both directions along a flux tube over a slowing down 
length $l_{sd} \sim \bar{\beta}c/2 \sim$ 90 kpc. 
 
We estimate the flux tube lengths in the HI in the Galactic disk 
from the mean separation between superbubbles, since are the 
primary agents for disrupting the interstellar magnetic field, 
which otherwise tends to lie in the plane of the disk. From an 
analysis of the Galactic distribution of OB associations by McKee 
\& Williams (1997), we estimate that the mean distance between 
superbubbles in the Galactic disk, $s \sim 0.42 e^{-R/H_R}$ kpc 
as a function of galactocentric distance with $H_R \sim$ 3.5 kpc. 
Thus, the flux tube length $l_B \sim s$ may range from only 1.4 
kpc in the inner part of the warm disk at $R$ of 3.5 kpc to about 
4 kpc at the solar radius, giving a mean $l_B$ of about 3 kpc. 
Since these distances are far less that the slowing down length, 
we would expect that only a very small fraction ($\sim 
l_B/2l_{sd} \sim 3/180 \sim$ 2\%) of the positrons, born in the 
HI, slow down and annihilate there. 
 
{\em Thus, we expect that nearly all ($\sim 0.98$) of the SNIa 
positrons, streaming along the magnetic flux tubes, escape from 
the HI into the overlying halo or the neighboring warm ionized 
HII and WI, going into the latter phases with equal likelihood.} 
As we noted above, we expect that about 66\% of the positrons 
entering either the WI or the HI, slow down and annihilate there, 
while the remainder escape into the halo and annihilate. Therefore 
with large uncertainties, we take for the net factors, $P_{HI:HI} 
\sim 0.02$, and $P_{HI:WI} \sim 0.16\pm0.08,$ $P_{HI:HII} \sim 
0.16\pm0.08$, and $P_{HI:H} \sim 0.66\pm0.33$. 
 
\subsection{Photoionized Medium in the Galactic Disk} 
 
OB associations ionize the interstellar medium, and create 
localized, high-density (HII) regions as well as an extended, 
diffuse warm ionized medium (WIM) (e.g. McKee \& Williams 1997). 
Ionizing radiation from such clustered association stars, which 
are embedded initially in their parent molecular cloud, heat as 
well as ionize the surrounding dense molecular gas. Ultimately 
such ionized gas expands, streams away from the molecular cloud, 
evacuating large cavities, finally destroying the molecular cloud 
(Tenorio-Tagle 1979; Whitworth 1979). Thus OB associations create 
dense ($\geq$ 1000 cm$^{-3}$), compact ($\leq$ 1 pc) cores of 
photoionized gas, observable as luminous thermal radio sources 
and visible nebulae, but $\sim$ 2/3 of the ionizing radiation 
emitted by massive association stars escapes these dense cores 
and create extended ($\geq$ 100 pc), more tenuous ($\leq$ 10 
cm$^{-3}$) envelopes of photoionized gas (Anantharamaiah 1985; 
Heiles, Reach, \& Koo 1996; McKee \& Williams 1997). Following 
McKee \& Williams we model Galactic-disk HII regions as two 
components: the dense central cores surrounding the parent OB 
associations and the extended HII envelopes.  As noted by McKee 
\& Williams the tenuous HII envelopes completely dominate the 
HII filling factor, and also about half the volume of an HII 
region is filled with a hot ($geq 10^6$ K) tenuous, superbubble 
plasma. 
 
Analyses of pulsar dispersion measures and diffuse, optical line 
emissions require the existence of a widespread (volume filling 
factor $\sim$ 20\%), warm ($\sim$ 8000 K), ionized medium (WI) of 
a modest density  ($\sim$ 0.1 cm$^{-3}$) distributed throughout 
the Galactic disk and lower halo (Reynolds 1984, 1991; Tielens 
2005).  The extended WIM has a large scale height, $\sim$ 1 kpc, 
transverse the Galactic plane (Reynolds 1991; Taylor \& Cordes 
1993), compared to $\sim$ 0.05 kpc, the scale height of O stars 
(Mihalas \& Binney 1981). However, a dilute O-star radiation 
field seems to explain many of WI line features (Mathis 1986; 
2000).  Thus it seems likely that a modest fraction, $\sim$ 0.15, 
of Lyman-continuum radiation emitted by the planar OB 
associations escapes from cloud-debris HII regions to ionize the 
high-latitude WI (e.g. Bregman \& Harrington 1986; Mathis 1986; 
Dove \& Shull 1994; Dove, Shull, \& Ferrera 2000; Mathis 2000; 
Tielens 2005). Additionally, Slavin, McKee, \& Hollenbach (2000) 
found that the cooling of shock-heated gas, created by old 
supernova remnants located at high Galactic latitudes, creates a 
significant fraction of WI at these large kpc heights above the 
Galactic plane. 
 
 \subsubsection{Warm Ionized Medium (WI) } 
 
Based on thermal pressure balance and the assumption that the 
WI is highly ionized, the mean electron density in the WI, $n_e 
\sim$ 0.26 cm$^{-3}$ (Kulkarni \& Heiles 1987). The neutral 
fraction of the WI is not well constrained observationally 
by optical-line diagnostics, but neutral fractions $\sim$ 0.1 seem 
likely (Tielens 2005). However, analyses of pulsar radio signals 
provide a stronger constraint on the WI ionization fraction. 
Radio-wave scatterings increase roughly linearly with dispersion 
measures $\sim$ 80 pc cm$^{-3}$, a primary WIM tracer (Cordes 
et al. 1991). Thus Cordes et al. ascertained that the thermal 
electrons, which for create the pulsar dispersion measures, also 
generate corresponding radio scatterings. Since such radio 
scatterings are created by electron density fluctuations on 
spatial scales, $\sim 10^9 - 10^{11}$ cm (Cordes et al. 1991), 
significantly smaller than the mean free path for proton-HI 
collisions, $\sim 5\times10^{13}$ cm in unit-density media, means 
that efficient turbulent radio scattering exists in WI and, 
thus, HI fraction must be less than the critical value, 
equation (7). Consequently, we assume that in WI $x_e \sim$ 0.9. 
 
A lower limit to the typical thickness of such WI regions is the 
mean free path of 13.6 eV photons, 
$(n_{HI}\bar{\alpha}_{HI})^{-1}$, where $\bar{\alpha}_{HI} \sim 
3\times10^{-19}$ cm$^{2}$ (Tielens 2005) and the neutral HI 
density, $n_{HI} \sim (1 - x_e)n_e/x_e$. Taking a nominal value 
of $x_e \sim$ 0.9 and $n \sim n_e/x_e \sim$ 0.3 cm$^{-3}$ gives a 
typical WI thickness, $\Delta l_{WI} \sim $ 40 pc, and a positron 
slowing down time $t_{sd} \sim 1.6\times10^5$ yr from equation 
(5). 
 
Recent simulations (Esquivel et al.\ 2006) of the turbulent 
mixing generated by shear flow instabilities in the WI layers 
between the HT and HI suggest an outer turbulent scale $l_o \sim 
10$ pc and $(\overline{u^2})^{1/2} \sim$ 20 km s$^{-1}$. Assuming 
$B_o \sim 4 \mu$G, ($\delta B^2_{\bot})^{1/2}/B_o \sim 1$, and 
$T_{WI} \sim $ 8000 K, we find from equation (6) that 
$\lambda_{\|} \sim$ 0.2 pc. Thus in the slowing down time of 
1.6$\times10^5$ yr we would expect the positrons to be 
distributed both directions along a flux tube over a distance 
$l_{sd} \sim (2\lambda_{\|} \bar{\beta} c t_{sd}/3)^{1/2} \sim $ 
65 pc in both directions along the flux tubes. 
 
Again assuming that the mean length $l'_B$ of nominally parallel 
flux tubes through these thick turbulent phases is equal to the 
typical thickness of the region, $\Delta l_{WI} \sim 40$ pc, we 
expect that with the turbulent scale of $\sim$ 10 pc, the mean 
meandering tube length $l_B \sim l'^{2}_B/l_o \sim 40^2/10 \sim 
160$ pc. The probability that those positrons initially born in 
the WI, slow down and annihilate there before they escape into 
the neighboring HI and HT is $\sim l_B/(l_B + l_{sd}) \sim 
160/225 \sim$ 71\%. We expect that the escaping positrons enter 
into the HI and HT with equal probability and end up annihilating 
in the WI, HII or halo in the same relative proportions as those 
born in the HI and HT. Thus the effective propagation fractions 
for those born in the WI are $P_{WI:WI} \sim 0.76\pm0.38$, 
$P_{WI:HII} \sim 0.05\pm0.02$, and $P_{WI:H} \sim 0.19\pm0.09$. 
 
On the other hand, the positrons, born in the HI and HT, that 
diffuse into the WI from are effectively reborn in the WI at 
their first scatter just inside the boundary, and since $l_{sd} < 
l_B$, we expect that $\sim$ 0.5 of them slow down an annihilate 
in the WI, while the other half escape equally into the adjacent 
HI and HT. We also expect some of the escaping positrons to 
subsequently enter other WI phases and annihilate there, or in 
the HII or halo in the same relative proportions as those born in 
the HI and HT. Thus we expect that for the positrons that diffuse 
into the WI, their effective propagation fractions are $P_{WI:WI} 
\sim 0.58\pm0.29$, $P_{WI:HII} \sim 0.08\pm0.04$, and $P_{WI:H} 
\sim 0.34\pm0.17$. 
 
\subsubsection{HII Regions} 
 
Although the direct positron production by supernovae in the HII 
regions is modest because of their small filling factor, a large 
fraction of positrons made in the neighboring HT and HI escape 
into it. Thus, it becomes the major site of positron annihilation 
in the disk, because of its high density ($\sim$ 10 times that in 
the HI), which reduces their slowing down time, so that most of 
the positrons slow down and annihilate before they can escape. 
 
McKee \& Williams (1997) derived the Galactic volume filling 
factor of the HII regions, as a function of galacto-centric 
radius, $R$, assuming that these HII regions reproduced the 
observed Galactic plane dispersion measures. Further, they 
assumed the electron density is constant across the Galactic 
disk, and all the $R$-dependence resides in the filling factor, 
$f_{HII}(R)$. They ascertained that $f_{HII}(R) = 
f_{HII}(0)e^{-R/H_R}$, where $H_R$ = 3.5 kpc, and in such HII 
regions $n_e$ = 2.6 cm$^{-3}$, and $f_{HII}(0) =$ 0.13. In these 
HII regions, the positrons slow down in $t_{sd} \sim 
1.7\times10^4$ yr from equation (5). 
 
The typical thickness an HII envelope can be determined from the 
average dispersion measure contributed by the walls of a single 
large superbubble HII envelope, $\Delta DM \sim$ 70 cm$^{-3}$ pc 
(McKee \& Williams 1997). For a mean density $n_e \sim$ 2.6 
cm$^{-3}$, this corresponds to a typical wall thickness, $\Delta 
l_{HII} \sim 0.5\Delta DM/n_e \sim$ 13 pc. So, again assuming 
nominally parallel flux tubes, the mean length through one of 
these highly turbulent walls $l'_B$ is also $\sim$ 13 pc. 
Although the nature of turbulent processes in the HII is poorly 
known, Haverkorn et al.\ (2004) have determined from the analysis 
of rotation measures that the outer scale of turbulent flows in 
disk HII regions, $l_o$, is relatively small, $\sim$ 2 pc. From 
this, we then estimate the mean meandering tube length $l_B \sim 
l'^{2}_B/l_o \sim 13^2/2 \sim$ 85 pc. 
 
To compare that with the expected slowing down length $l_{sd}$, 
we assume $B_o \sim 4 \mu$G and $T_{HII} \sim$ 8000 K, and 
estimate the diffusion mean free path in the HII using the 
cooling constraint, as we have done in other ionized phases. 
Following Tielens (2005), we find that the cooling time is short, 
$t_{cool} \sim 1.1\times10^4$ yr. So taking $l_o \sim$ 2 pc and 
the constraint that the dissipation rate of the turbulence must 
be less than the cooling rate of the plasma, we find that the 
root-mean-square turbulent velocity, $(\overline{u^2})^{1/2} <$ 
23 km s$^{-1}$, which is greater than the Alfv\'en speed in this 
plasma, $V_a \sim$ 5 km s$^{-1}$, so ($\delta B^2)^{1/2} \sim 
B_o$, and from equation (6) we thus find that $\lambda_{\|} \sim$ 
0.1 pc. 
 
In the slowing down time of $\sim 1.8\times10^4$ yr, we would 
thus expect the positrons to be distributed over a length $l_{sd} 
\sim (2\lambda_{\|} \bar{\beta} c t_{sd}/3)^{1/2} \sim $ 15 pc in 
both directions along a flux tube. Thus we expect from equation 
(8) that $\sim l_B/(l_B + l_{sd}) \sim 85/(85+15) \sim$ 85\% of 
the positrons born in one of these HII envelopes, would slow down 
and annihilate there before they escape into the neighboring HI 
and HT. We again assume that half of the escaping positrons go 
into each of these two phases, and that from there some diffuse 
into other HII and WI regions or into the halo and annihilate in 
the same proportion as those initially born in the HI and HT. 
Thus the effective propagation fractions for those born in the 
HII are $P_{HII:HII} \sim 0.87\pm0.43$, $P_{HII:WI} \sim 
0.03\pm0.01$, and $P_{HII:H} \sim 0.10\pm0.09$. 
 
Again, however, most of the positrons in the HII were born in the 
HI and HT, and were reborn in the HII within a scattering mean 
free path of the boundary. As we have discussed above, with 
$l_{sd} < l_B$, we expect that $\sim$ 0.5 of the positrons slow 
down an annihilate in the HII, and the other half escape equally 
into the adjacent HI and HT. From there, we again expect them to 
eventually annihilate in the HII, WI or halo in the same relative 
proportions as those born in the HI and HT. Thus, as with the WI 
discussed above, we expect that for the positrons that diffuse 
into the HII, their effective propagation fractions are 
$P_{HII:HII} \sim 0.58\pm0.29$, $P_{HII:WI} \sim 0.08\pm0.04$, 
and $P_{HII:H} \sim 0.34\pm0.17$.

\subsection{Disk Annihilation Flux Asymmetry} 
 
The asymmetry in the 511 keV line flux (Weidenspointner et al.\ 
2008b) from the inner disk provides further evidence of positron 
propagation and annihilation in the HII and WI. We show that this 
flux asymmetry can be fully accounted for by the apparent 
asymmetry of positron annihilation in the inner spiral arms, as 
viewed from our Solar perspective, without any asymmetry in the 
positron production. 
 
Modeling the inner Galactic disk in two components, 
Weidenspointner et al.\ (2008b) find a best-fit SPI/INTEGRAL 511 
keV line flux of $(4.3\pm0.5)\times10^{-4}$ photons cm$^{-2}$ 
s$^{-1}$ from the component between $-50^o < l < 0^o$, compared 
to a flux of $(2.4\pm0.5)\times10^{-4}$ photons cm$^{-2}$ 
s$^{-1}$ from that between $+50^o > l > 0^o$. This amounts to an 
excess of $80\pm40$\% in the 511 keV flux from negative 
longitudes compared to that from positive longitudes, and it 
amounts to an asymmetry of about 20$\pm$8\% in the total observed 
disk flux (Weidenspointner et al.\ 2008a). 
 
The asymmetry in the Galactic spiral arms, as viewed from our 
solar perspective, can be seen in Figure 4, showing the current 
mapping of the four spiral arms determined from the study by 
Vall\'ee (2005a,b). As summarized in Table 4, we expect that the 
positron annihilation in the disk occurs almost entirely in the 
warm ionized HII and WI phases, which are concentrated along the 
spiral arms (e.g. Drimmel \& Spergel 2001). Thus, we assume that 
equal fractions of the positrons produced at any particular radius 
propagate to and annihilate in the ionized gas in their nearest 
spiral. 
 
We estimate the positron annihilation rate per unit length along 
each of the four arms, using the radial distributions of the 
positron production surface density in the disk as a function of 
Galacto-centric radius, $R$, given in equation (3), multiplied by 
$\pi RdR/2$ and by the disk annihilation/production fraction (Table 
4). We further weight the annihilation distribution along the flux 
tubes by a unit normalized surface density distribution of 
interstellar gas, since positrons propagating along the flux 
tubes are more likely to encounter the dense ionized gas, where 
they annihilate, in denser regions. We also include the positrons 
escaping into the arms from the outer bulge. In accordance with 
observations (e.g. Drimmel \& Spergel 2001) that the Sagittarius 
arm is only half as dense and active as the others, we also 
reduce it accordingly. 
 
At every point on each arm, we then multiply the differential 
annihilation rate by the $\gamma_{511}/e^+$ ratio to determine 
the differential 511 keV line luminosity and then divide by $4\pi 
d_{\odot}^2$, where $d$ is the distance from the sun, to calculate 
the expected differential 511 keV line flux at the earth. Finally, 
integrating along each arm out to $\pm50^o$ Galactic longitude 
within $\pm10^o$ latitude to compare with the SPI/INTEGRAL model, 
we find asymmetric 511 keV line disk fluxes of 
$(3.8\pm1.0)\times10^{-4}$ photons cm$^{-2}$ s$^{-1}$ from the 
negative longitudes, and $(2.3\pm0.6)\times10^{-4}$ photons 
cm$^{-2}$ s$^{-1}$ from the positive, and a negative-to-positive 
longitude flux ratio of 1.6$\pm$0.4. These values are all quite 
consistent (Table 7) with the corresponding best-fit 
(Weidenspointner et al.\ 2008b) SPI/INTEGRAL line fluxes of 
$(4.3\pm0.5)\times10^{-4}$ photons cm$^{-2}$ s$^{-1}$ and 
$(2.4\pm0.5)\times10^{-4}$ photons cm$^{-2}$ s$^{-1}$, and a 
negative-to-positive flux ratio of 1.8$\pm$0.4. 
 
{\em Thus we show that the observed (Weidenspointner et al.\ 
2008b) SPI/INTEGRAL asymmetry in the 511 keV line flux from the 
inner disk can be fully explained by the apparent asymmetry in the 
inner spiral arms as viewed from our Solar perspective, which 
leads to different line-of-sight intensities, without any 
difference in the positron annihilation or production rates.This asymmetry in 
the annihilation radiation from the Galactic disk is a product of the 
Norma arm dominating the negative longitudes due to its proximity to 
the inner region of the Galaxy, and the weakness of the 
Sagittarius-Carina arm due to its greater distance from the Galactic 
center from which the annihilation rate falls off exponentially from 
the Galactic center and its lower density/activity (Drimmel \& Spergel 
2001) compared to the other 3 arms. } 
 
\subsection{Disk Positron Annihilation $>$3.5 kpc} 
 
We now estimate the total outer disk annihilation rate from the 
contributions of the various phases and the propagation fraction 
within and between them. These are summarized in Table 4. 
 
As we have discussed above, we see that more than half of the 
positrons, produced in the dominant phases of the interstellar 
medium in the disk, the HI and HT, escape into adjacent phases 
before they can slow down and annihilate. Only those produced in 
the HII and WI phases have a large chance of annihilating before 
they escape, and the rest either end up in the HII, the WI, or 
the halo. {\em This result is quite robust, even allowing the 
large uncertainties in the relative likelihood of positrons 
following magnetic flux tubes that lead into these three 
annihilation traps.} 
 
Thus, we expect that the total annihilation rate in the disk 
beyond 1.5 kpc is $A_D \sim (0.59\pm0.11)\times10^{43}$ e$^+$ 
s$^{-1}$. As we discuss in detail in the summary ($\S$ 7), this 
is quite consistent with the best-fit disk annihilation rate of 
inferred from the SPI/INTEGRAL analyses (Weidenspointner et al. 
2007, 2008a). 

\section{Positron Propagation \& Annihilation in the Galactic Halo} 
 
Lastly we consider the halo, which for comparison with the 
SPI/INTEGRAL analyses we take to lie beyond 1.5 kpc from the 
Galactic center and more than 0.4 kpc above the disk. Although it 
has no direct positron production by supernovae, a large fraction 
of positrons made in all the underlying phases of the bulge and 
disk can escape into it. Thus, it too becomes a major site of 
positron annihilation, because its pervasive hot tenuous plasma, 
coming primarily out of the outer bulge and disk superbubbles 
(e.g. Norman \& Ikeuchi 1989), may be turbulent enough to produce 
a scattering mean free path much smaller than its size, so that 
most of the positrons that enter it slow down and annihilate 
before they can escape on out of the Galaxy. 
 
From the positron production in all of the regions and phases 
and their propagation fractions, discussed above, we estimate the 
positron escape into the halo and their annihilation there to be, 
 
\begin{eqnarray} 
A_{H} \sim P_{H:H}(Q_{Bi}P_{Bi:H} + Q_{Bm}P_{Bm:H}\nonumber\\ 
+ Q_{Bo}P_{Bo:H} + Q_{HT}P_{HT:H} + Q_{HI}P_{HI:H}\nonumber\\ 
+ Q_{WI}P_{WI:H} + Q_{HII}P_{HII:H}),~~~ 
\end{eqnarray} 
\vspace{0.1cm} 
 
Extensive observations of X-ray emission from the Galactic halo 
(e.g. Snowden et al.\ 1998; Pietz et al.\ 1998) have shown that 
hot tenuous plasma filling it is essentially indistinguishable 
from that in the outer bulge and superbubbles from which it is 
derived. Thus we assume a similar $T$ of 1$\times10^6$ K, a 
density of 2$\times10^{-3}$ cm$^{-3}$, and magnetic field, $B_o 
\sim$ 4 $\mu$G. This plasma has a scale height, $z_{XR}$, $\sim$ 
4.4 kpc, and a radial scale length of 15 kpc. As discussed above 
for comparison with SPI/INTEGRAL analyses, we define the halo as 
that region  beyond a radius 1.5 kpc and a $z >$ 0.4 kpc, 
including within it the outer bulge. 
 
The nature of turbulence in the halo is very uncertain, but we 
assume that a significant level of MHD turbulence can be 
maintained by erupting superbubbles. Using the parameters for 
that hot plasma, an outer turbulence scale, $l_o \sim$ 75 pc and 
($\overline{B^2_{\bot}})^{1/2}/B_o \leq$ 0.5, we also find the 
expected scattering mean free path parallel to the mean magnetic 
field, $\lambda_{\|} \sim$ 14 pc from equation (6). 
 
Assuming a minimum flux tube escape length $l_B \sim$ 10 kpc, 
roughly 2 times the scale height, then the expected positron 
escape time $t_{esc} \sim$ 200 Myr. By comparison the positron 
slowing down time, $t_{sd} \sim$ 23 Myr from equation(5), and in 
that time positrons travel $l_{sd} \sim (2\bar{\beta} c t_{sd} 
\lambda_{\|}/3)^{1/2} \sim$ 7 kpc. Since all of the positrons 
enter the halo from below, rather than being born throughout, we 
expect that they will all slow down and annihilate well before 
they can escape, since $l_{sd} < l_B$. Thus, for all of the 
positrons that enter the halo, we expect $P_{H:H} \sim$ 1. 
Because the properties of the halo are assumed to be the same and 
those of the outer bulge and superbubble HT, which are the major 
sources of positrons that diffuse into it, there is no effective 
boundary between them and no ``reflection" of those positrons 
from the halo. Moreover, as we discussed above, all positron 
annihilation in the outer bulge and superbubble HT is also 
included in the halo, because of their large scale sizes. 
 
From the estimates of the positrons entering the halo from the 
various underlying bulge and disk phases, summarized in Tables 4 
and 5, we expect a total positron annihilation rate in the halo 
beyond 1.5 kpc of $\sim (0.82\pm0.23)\times10^{43}$ e$^+$ 
s$^{-1}$. As we show in Table 6 and discuss further in the 
summary, this annihilation rate gives an average 511 keV line 
flux from the halo of $\sim (1.58\pm0.37)\times10^{-3}$ cm$^{-2}$ 
s$^{-1}$, depending on how much refractory dust survives in the 
halo. This flux together with the bulge flux of $\sim 
(0.74\pm0.14)\times10^{-3}$ cm$^{-2}$ s$^{-1}$ gives a combined 
bulge and halo flux of $\sim (2.32\pm0.40)\times10^{-3}$ cm$^{-2}$ 
s$^{-1}$ (see Table 7), which is quite consistent with that of 
$\sim (2.14\pm0.12)\times10^{-3}$ cm$^{-2}$ s$^{-1}$, determined 
from the SPI/INTEGRAL best-fit flux model (Weidenspointner et al. 
2008a). 
 
As we discuss further in the summary, not only the 511 keV line 
intensity but also the Galactic averages of both its broad/narrow 
line flux ratio and the positronium fraction are very sensitive 
to the mean dust content of the hot halo plasma. For as Jean et 
al. (2006) have shown and we list in Table 6, if the refractory 
elements are all in dust in the hot plasma, only $\sim$ 18\% of 
the resulting positron annihilation occurs via positronium 
formation, and fully $\sim$ 90\% of the 511 keV line emission is 
narrow (FWHM $\sim$ 2 keV) while the remaining 10\% is very broad 
(FWHM $\sim$ 11 keV). This gives a halo flux of $\sim 
(1.60\pm0.37)\times10^{-3}$ cm$^{-2}$ s$^{-1}$ in the narrow line 
alone which in turn gives a combined bulge and halo flux of $\sim 
(2.34\pm0.40)\times10^{-3}$ cm$^{-2}$ s$^{-1}$, which is likewise 
quite consistent with that from the SPI/INTEGRAL best fit above. 
Combined with disk flux, that would give a total Galactic bulge, 
disk and halo flux of $\sim (3.25\pm0.42)\times10^{-3}$ cm$^{-2}$ 
s$^{-1}$ which is also quite consistent with the combined 
best-fit flux of $\sim (2.87\pm0.25)\times10^{-3}$ cm$^{-2}$ 
s$^{-1}$, from the SPI/INTEGRAL analyses (Weidenspointner et al. 
2008a). This also gives Galactic average values of the 
positronium fraction of only $\sim 0.66\pm0.12$ and the 
broad/narrow 511 keV line flux ratio of only $\sim 0.17\pm0.04$, 
for which the best-fit SPI/INTEGRAL values have not yet been 
published. 
 
If on the other hand, the grains have all disintegrated, the halo 
flux would still be also as large, $\sim 
(1.41\pm0.31)\times10^{-3}$ cm$^{-2}$ s$^{-1}$, but it would all 
be in the very wide ($\sim$ 11 keV) 511 keV line, so the Galactic 
average value of the broad/narrow 511 keV line flux ratio would 
jump to $\sim 1.30\pm0.29$, and the positronium fraction would 
increase only slightly to $\sim 0.74\pm0.13$. The actual dust 
content of the halo undoubtedly lies in between and the 
SPI/INTEGRAL observations can provide a unique measure. 
 
This is also consistent with that implied by the SPI/INTEGRAL 
best-fit flux model of $A_H \sim (0.65\pm0.11)\times10^{43}$ 
e$^+$ s$^{-1}$ (Weidenspointner et al. 2008a), scaled to a 
galacto-centric distance of 8 kpc and $(e^+/\gamma_{511}) \sim 
0.65\pm0.07$ expected (Jean et al. 2006) in the halo plasma, as 
discussed above. 
\vspace*{12pt}

\section{Summary } 
 
The recent analyses by Weidenspointner et al. (2007, 2008a) of 
the SPI/INTEGRAL measurements of the Galactic 511 keV positron 
annihilation radiation luminosity suggest that the best fit to the 
observations consists of three basic components: 1) a spherical 
bulge of $\sim$ 1.5 kpc radius, 2) a thick ($\pm$ 400 
pc) Galactic disk, and 3) a spherical halo extending beyond 1.5 
kpc. Their most recent best-fit model (Weidenspointner et al. 
2008a), scaled to a galacto-centric distance of 8 kpc, gives 511 
keV line luminosities of $L_B \sim (0.57\pm0.08)\times10^{43}$ 
photons s$^{-1}$ from the bulge, $L_D \sim (0.40\pm0.06) \times 
10^{43}$ photons s$^{-1}$ from the disk, and a net $L_H \sim 
(1.0\pm0.1)\times10^{43}$ photons s$^{-1}$ from the halo beyond 
the bulge, for a total Galactic 511 keV luminosity $L \sim 
(2.0\pm0.2)\times10^{43}$ photons s$^{-1}$. 
 
This gives a positron annihilation luminosity bulge/disk ratio of 
$\sim$ 1.4$\pm$0.3. But since this is a factor of 4 higher than 
the expected positron production ratio, inferred from the stellar 
bulge/disk ratio of the assumed Galactic supernova sources, 
various exotic new sources have been invoked to explain the large 
discrepancy. 
 
{\em Here we have shown, however, that the measured 511 keV 
luminosity ratio can be fully explained by positrons from the 
decay of radionuclei made by explosive nucleosynthesis in 
supernovae, if the propagation of these relativistic positrons in 
the various phases of the interstellar medium is taken into 
account, since these positrons must first slow down to energies 
$\leq$ 10 eV before they can annihilate, and if the geometry of 
the sources is also considered. Moreover, as Jean et al. (2006) 
have shown, without propagation none of the proposed positron 
sources, new or old, can explain the two fundamental properties 
of the Galactic annihilation radiation: the fraction of the 
annihilation that occurs through positronium formation and the 
ratio of the broad/narrow components of the 511 keV line. } 
 
\subsection{Positron Production} 
 
First, we have shown, using recent estimates (Cappellaro, Evans 
\& Turatto 1999) of the SNIa and SNIp rate in our Galaxy of 
$\sim$ 0.40$\pm$0.16 SN per 100 years, that the positrons 
resulting from the $\beta^+$-decay chains of the radioactive 
nuclei, $^{56}$Ni, $^{44}$Ti, and $^{26}$Al, produced in these 
supernovae can fully account for the total Galactic positron 
production rate of $\sim (2.4\pm0.6)\times 10^{43}$ e$^+$ 
s$^{-1}$, implied by the best-fit SPI/INTEGRAL analyses 
(Weidenspointner et al. 2008a). 
 
The total Galactic production rate of positrons from long lived 
(1.04 Myr meanlife) decay of $^{26}$Al of $\sim (0.3\pm0.1) 
\times10^{43}$ e$^+$ s$^{-1}$, produced by massive stars, is 
determined directly from the SPI/INTEGRAL measurements (Diehl et 
al. 2006) of the Galactic luminosity of the 1809 keV line 
emission which accompanies the decay. The Galactic positron 
production rate the $^{44}$Ti decay chain (89 yr meanlife) of 
$\sim (0.5\pm0.2)\times10^{43}$ e$^+$ s$^{-1}$ can be determined 
from the relative Galactic abundances (Lodders 2003) of $^{44}$Ca 
and $^{56}$Fe, which are primarily produced by $^{44}$Ti and 
$^{56}$Ni decays, using the calculated SNIa rate and $^{56}$Ni 
yield, assuming that these supernovae presently produce about 
half of the $^{56}$Fe and core collapse supernovae produce the 
other half. The Galactic positron production from the much 
shorter lived (111.4 day meanlife) decay chain $^{56}$Ni, 
however, depends on an additional factor, since most of these 
decay positrons slow down and annihilate in the dense ejecta and 
only a fraction, $f_{56}$ survive into the interstellar medium. 
{\em We showed that all of the additional positron production, 
amounting to $\sim 1.6\times10^{43}$ e$^+$ s$^{-1}$, can be 
accounted for by the $^{56}$Ni decay chain with a positron 
survival fraction in SNIa ejecta of $f_{56} \sim$ 5$\pm$2\%.} 
This is the time-integrated mean of $\sim$ 5\% calculated (Chan 
\& Lingenfelter 1993 Fig. 3) from the standard deflagration 
models of SNIa (Nomoto et al. 1984, W7), assuming combed out 
magnetic fields, and it is also consistent with the mean fraction 
of $\sim 3.5\pm2$\% inferred (Milne, The \& Leising 1999) from 
SNIa light curves at late times. 
 
Although the uncertainties in both the SNIa rate and the 
normalized survival fraction are large, $\sim$ 40 to 50 \%, they 
are necessarily anticorrelated. Therefore their product yields an 
uncertainty of only $\sim$ 25\% in the production rates. 
 
Using these positron production rates and the Galactic bulge and 
disk radial distributions of the SNIa, SNIp, and massive stars, 
gave an expected positron production of 
$(0.31\pm0.07)\times10^{43}$ e$^+$ s$^{-1}$ in the inner bulge 
$<$ 0.5 kpc, $(0.37\pm0.09)\times10^{43}$ e$^+$ s$^{-1}$ in the 
middle bulge from 0.5 to 1.5 kpc, $(0.56\pm0.14)\times10^{43}$ 
e$^+$ s$^{-1}$ in the outer bulge from 1.5 to 3.5 kpc, and 
$(1.16\pm0.22)\times10^{43}$ e$^+$ s$^{-1}$ in the disk beyond 
3.5 kpc. For a positron annihilation bulge within 1.5 kpc used in 
SPI/INTEGRAL observation analyses, that gave a positron {\em 
production} bulge/disk ratio of $0.4\pm0.1$ within 1.5 kpc. But, 
as we have shown, the expected annihilation ratio is very 
different, because the interstellar medium and the resulting 
positron propagation in these regions is also very different. 
 
\subsection{Positron Propagation} 
 
We have modeled the properties of the various regions and phases 
of the interstellar medium to determine the propagation of the 
positrons as they slowed down and annihilated. In ionized plasmas 
we assumed that diffusive propagation was controlled by resonant 
scattering of the relativistic positrons by MHD waves at their 
cyclotron radius. These waves are generated by the cascade of the 
larger scale magnetic turbulence down to sufficiently small 
scales which scatter the positrons and they diffuse along 
magnetic flux tubes. Such propagation has been observed in 
interplanetary plasmas with MeV electrons and we use a 
phenomenological model based on these studies. In neutral, or at 
least nearly neutral, phases of the interstellar medium, however, 
this cascade is damped by ion-neutral collisions, and the 
positrons are expected to stream along the magnetic flux tubes 
with an isotropic pitch angle distribution. 
 
From these propagation processes, we have calculated the 
propagation, slowing down and annihilation of positrons formed in 
the various regions and phases of the interstellar medium 
described in Tables 1 and 3. From the properties of the medium in 
each phase, we determined the propagation mode, and calculated 
the diffusion mean free path in the undamped cascade mode, or the 
streaming velocity in the damped mode. From an estimate of the 
mean length of the magnetic flux tubes within a phase, compared 
to the tube length over which the positrons would propagate 
before they slowed down, we estimated the fraction of positrons 
expected to slow down and annihilate in that phase before they 
escape. We then estimated the relative fractions of the escaping 
positrons that go into each of the adjacent phases by simple 
geometric arguments. We have assumed large uncertainties of $\pm$ 
50\% in the estimated propagation fractions between phases, which 
produce comparably large uncertainties in the annihilation rates 
in each phase, although their sums within the bulge and disk are 
more tightly constrained. These propagation fractions between 
various components and the resulting distribution of their 
annihilation rates are presented in Tables 2 and 4. 
 
\subsection{Positron Annihilation Rate \& Flux} 
 
The expected positron production and annihilation rates in the 
different regions and phases of the interstellar medium are 
summarized in Tables 5 and 6. Finally, the expected positron 
annihilation rates and 511 keV line properties are compared with 
SPI/INTEGRAL analyses in Table 7. 
 
We show that $\sim 52\pm$7\% of the all the Galactic positrons 
are expected to be produced within the bulge $<$ 3.5 kpc, of 
which $\sim 28\pm$6\% are born within the positron bulge $<$ 1.5 
kpc. From the relative filling factors of the various gas and 
plasma phases in the bulge, we expect that essentially all of 
these positrons are born in the hot fully ionized plasma. From 
their propagation fractions we also expect that $\sim$ 80\% of 
them diffuse into the warm HII and cold HI envelopes of molecular 
clouds that lie within 1.5 kpc, where they slow down and 
annihilate, while the remaining $\sim$ 20\% escape into the halo 
and disk beyond. This propagation thus results in a positron 
bulge annihilation rate $<$1.5 kpc of $A_B \sim 
(1.00\pm0.20)\times10^{43}$ e$^+$ s$^{-1}$. 
 
Of the positrons that are produced in the Galactic disk beyond 
3.5 kpc, we expect from the relative filling factors that that 
about 70\% of the disk positrons are born either in the 
ubiquitous warm neutral HI medium of the disk, or in the hot 
tenuous plasmas of superbubbles, in both of which the conditions 
are such that positrons rapidly stream or diffuse along magnetic 
flux tubes. Some segments of these flux tubes thread through the 
warm ionized gas and HII shells that separate the warm neutral 
gas from the hot tenuous plasmas, while other segments have been 
blown up into the halo by the hot expanding superbubbles (e.g. 
Parker 1979), and, since there is no preferred direction, the 
positrons diffusing or streaming along them should have roughly 
equal likelihood of either going into the warm denser ionized 
shells, where they mostly slow down and annihilate, or out into 
the halo. From the estimated propagation fractions, assuming 
large uncertainties, we find that nearly half of the positrons 
enter the ionized shells and annihilate in the disk, and that 
other half escape into the overlying halo, giving a total 
annihilation rate in the disk beyond 3.5 kpc is $A_D \sim 
(0.59\pm0.11)\times10^{43}$ e$^+$ s$^{-1}$. 
 
{\em So even though roughly equal numbers of positrons are born 
in the bulge within 3.5 kpc and in the disk beyond, their 
annihilation rate is higher in the bulge than in the disk, 
because $\sim$ 80\% of those born in the interstellar bulge slow 
down and annihilate in the cloud envelopes within 1.5 kpc, while 
only $\sim$ 51\% of those born in the disk slow down and 
annihilate before they escape into the halo.} \vspace*{12pt} 
 
We now compare the spatial distribution of the Galactic positron 
annihilation radiation, expected from the $\beta^+$-decay chains 
of the radioactive nuclei, $^{56}$Ni, $^{44}$Ti, and $^{26}$Al, 
produced in supernovae, with that observed by SPI/INTEGRAL. 
 
The most direct comparison is through the 511 keV line fluxes. 
From more than four years of SPI/INTEGRAL observations 
Weidenspointner et al. (2008a, model BD) fit the 511 keV fluxes 
and luminosities to a combined bulge component within a spheroidal 
Gaussian (FWHM of 1.5 kpc) distribution and thick disk component. 
They found best-fit fluxes of $\sim (0.75\pm0.09)\times10^{-3}$ 
photons cm$^{-2}$ s$^{-1}$ from the bulge and $\sim 
(0.94\pm0.16)\times10^{-3}$ photons cm$^{-2}$ s$^{-1}$ from the 
disk. They also found a best-fit combined halo and bulge flux of 
$\sim (2.14\pm0.11)\times10^{-3}$ photons cm$^{-2}$ s$^{-1}$. 
 
For comparison we calculate in Table 5 the expected fluxes from 
the annihilation rates in the different regions and phases listed 
in Tables 2 and 4, using the 511 keV photons yields per positron 
expected in those phases (Guessoum, Jean \& Gillard 2005; Jean et 
al. 2006). Thus, we expect 511 keV line fluxes of $\sim 
(0.72\pm0.13)\times10^{-3}$ photons cm$^{-2}$ s$^{-1}$ from the 
bulge, $\sim (1.08\pm0.16)\times10^{-3}$ photons cm$^{-2}$ 
s$^{-1}$ from the disk, and $\sim (2.37\pm0.41)\times10^{-3}$ 
photons cm$^{-2}$ s$^{-1}$ from the combined halo and bulge. 
These expected fluxes, as summarized in Table 7, are all in 
excellent agreement with the best-fit fluxes observed by 
SPI/INTEGRAL (Weidenspointner et al. 2008a). 
 
We also calculated the expected 511 keV fluxes from the inner 
bulge within 0.5 kpc and the middle bulge between 0.5 and 1.5 kpc, 
$\sim (0.39\pm0.10)\times10^{-3}$ and $\sim 
(0.33\pm0.07)\times10^{-3}$ photons cm$^{-2}$ s$^{-1}$, 
respectively. As mentioned above, these boundaries were originally 
chosen for our calculations because these were the two regions 
modeled in the previous analysis by Weidenspointner et al. (2007) 
of the first two years of SPI/INTEGRAL, where they found best-fit 
fluxes of $\sim (0.38\pm0.03)\times10^{-3}$ and $\sim 
(0.41\pm0.06)\times10^{-3}$ photons cm$^{-2}$ s$^{-1}$ from the 
same regions. These expected and observed values are also 
consistent within the $1-\sigma$ uncertainties. 
 
We have compared with the earlier analysis, because the fluxes in 
the later Gaussian component analysis by (Weidenspointner et al. 
2008a) overlapped in the inner volume which was also smaller than 
0.5 kpc. The best-fit total fluxes from the analyses of 
(Weidenspointner et al. 2007) and (Weidenspointner et al. 2008a) 
within 1.5 kpc are essentially the same, $\sim 
(0.79\pm0.07)\times10^{-3}$ and $\sim (0.75\pm0.09)\times10^{-3}$ 
photons cm$^{-2}$ s$^{-1}$, respectively. 
 
We also compare the expected positron annihilation rates in the 
bulge, disk and halo, and the corresponding bulge/disk ratio, 
with those determined by Weidenspointner et al. (2008a) from their 
best-fit SPI/INTEGRAL observations. Since they used a 
galacto-centric distance of 8.5 kpc rather than the 8 kpc, which 
we assumed, we scale our annihilation rates by $(8.5/8)^2$ for 
comparison. In addition, since $\sim$ 12\% of their disk 
annihilation component lies within 1.5 kpc, we further modify our 
expected annihilation rates to compare with their model 
components. Thus, to compare with the Weidenspointner et al. 
(2008a) values, we calculate an equivalent SPI disk annihilation 
component $_{SPI}A_D \sim A_D (8.5/8)^2/(1 - 0.12) \sim 
(0.76\pm0.13)\times10^{43}$ e$^+$ s$^{-1}$. This is in very good 
agreement with the SPI/INTEGRAL value of 
$(0.81\pm0.14)\times10^{43}$ e$^+$ s$^{-1}$. Similarly, we 
calculate an equivalent SPI bulge annihilation component 
$_{SPI}A_B \sim (A_B - A_D[0.12/(1 - 0.12)])(8.5/8)^2 \sim 
(1.04\pm0.21)\times10^{43}$ e$^+$ s$^{-1}$, which is also in very 
good agreement with the SPI/INTEGRAL value of 
$(1.15\pm0.16)\times10^{43}$ e$^+$ s$^{-1}$. 
 
{\em Thus, we expect an equivalent bulge/disk ratio of the 
annihilation rate $_{SPI}A_B/_{SPI}A_D \sim (1.04\pm0.21)/ 
(0.76\pm0.13) \sim 1.37\pm0.37$, which is also in excellent 
agreement with the measured value of $(1.15\pm0.16)/(0.81\pm0.14) 
\sim 1.42\pm0.32$ from the SPI/INTEGRAL data analyses of the 
best-fit bulge and disk components (Weidenspointner et al. 
2008a).} By comparison we also note, however, that the ratio of 
the total bulge annihilation with 1.5 kpc compared to that in the 
disk beyond is $A_B/A_D \sim (1.00\pm0.19)/(0.59\pm0.11) \sim 
1.69\pm0.45$. This is nonetheless consistent with the SPI/INTEGRAL 
best-fit model when the 12\% disk component within 1.5 kpc is 
subtracted from their disk component and added to that of the 
bulge, $\sim (0.94 + 0.08)/(0.66 - 0.08) \sim 1.79\pm0.38$. 
 
We also showed that positron propagation and annihilation in the 
HII and WI gas along the spiral arms account for the asymmetry in 
the SPI/INTEGRAL 511 keV line flux (Weidenspointner et al. 2008b) 
from modeled inner disk components within Galactic longitudes 
$\pm50^o$ that show a best-fit excess of $80\pm$40\% in the 511 
keV line flux from the negative longitudes compared to that from 
the positive longitudes. As we have shown ($\S$ 4.5 and Table 7), 
this flux asymmetry can be fully explained by the apparent 
asymmetry in the inner spiral arms as viewed from our Solar 
perspective, which leads to different line-of-sight intensities, 
without any actual asymmetry in the positron annihilation or 
production rates.

From the estimated fractions of the positrons entering the halo 
from the underlying bulge and disk, summarized in Table 4, we 
also compare the total expected positron annihilation rate in the 
halo with the SPI/INTEGRAL best-fit model of a combined bulge and 
halo component by Weidenspointner et al. (2008a). To do so we 
combine the expected bulge ($<$ 1.5 kpc) and halo ($>$ 1.5 kpc) 
annihilation rates from Table 5, scaled to a galacto-centric 
distance of 8.5 kpc, to give an equivalent SPI annihilation 
component, $_{SPI}A_{H+B} \sim (A_B + A_H)(8.5/8)^2 \sim 
(1.95\pm0.28)\times10^{43}$ e$^+$ s$^{-1}$. From their combined 
halo and bulge flux, discussed above, Weidenspointner et al. 
(2008a) estimate a significantly larger combined annihilation 
rate of $\sim (3.13\pm0.2)\times10^{43}$ e$^+$ s$^{-1}$. 
 
But they use a uniform $e^+/\gamma_{511} \sim 1.82$ to convert 
from observed the 511 keV photons luminosity to an assumed 
positron annihilation rate. This conversion factor, however, was 
based on a positronium fraction of $\sim$ 97\%, determined by 
Jean et al. (2006) for warm gas in the inner Galaxy. Although 
this would be applicable to the bulge contribution, it is not 
applicable to the hot plasma of the halo contribution, for which 
Jean et al. (2006) estimate a positronium fraction of only about 
18\% to 42\%, depending on the refractory grain abundance, which 
corresponds to a $e^+/\gamma_{511} \sim 0.65\pm0.07$, as 
discussed above. Thus we use this much more appropriate 
conversion factor for flux contribution from the halo beyond the 
bulge, which from the best-fit fluxes amounts to roughly 65\% of 
the combined halo and bulge flux. We estimate an annihilation 
rate of $_{SPI}A_H \sim (0.65/1.82)(3.13 - 1.15)\times10^{43} 
\sim (0.71\pm0.26)\times10^{43}$ e$^+$ s$^{-1}$ just for the halo 
beyond the bulge, which, adding back the bulge contribution, 
gives an adjusted best-fit SPI/INTEGRAL model of $A_{H+B} \sim 
(1.86\pm0.26)\times10^{43}$ e$^+$ s$^{-1}$. Those values would be 
in good agreement with the expected equivalent SPI annihilation 
components, $_{SPI}A_H \sim (0.91\pm0.28)\times10^{43}$ and 
$_{SPI}A_{H+B} \sim (1.95\pm0.28)\times10^{43}$ e$^+$ s$^{-1}$.

\subsection{Positronium Formation} 
 
Significant propagation of the positrons also accounts for the 
other fundamental features of the Galactic annihilation radiation: 
1) the high, observed fraction of positrons annihilating via 
positronium formation of 94$\pm$7\%, a weighted average of 
94$\pm$6\% (Churazov et al. 2005), 95$\pm$3\% (Jean et al. 2006) 
and 92$\pm$9\% (Weidenspointner et al. 2006), and 2) the observed 
ratio of broad ($\sim$ 5.8 keV) to narrow ($\sim$ 1.2 keV) 511 
keV line emission from the bulge of $\sim$ 0.5$\pm$0.2 (Chuzarov 
et al. 2005; Jean et al. 2006), which gives a measure of the 
fraction of positronium that is formed by charge exchange with 
neutral HI in flight and results in broad line emission. For even 
though roughly 90\% of the positrons in the bulge ($<$ 1.5 kpc) 
are either born in, or diffuse into, the hot plasma that fills 
most of the bulge, hot plasma can not be the site of most of the 
annihilation because neither the observed positronium fraction, 
nor the broad to narrow line ratio, are at all consistent with 
annihilation in that phase (Guessoum, Ramaty \& Lingenfelter 
1991; Guessoum, Jean \& Gillard 2005). 
 
We show that the expected positron propagation, slowing down and 
annihilation in the bulge and disk quite naturally accounts for 
both of these observations. In particular, we expect (Table 5) 
that $(0.58\pm0.11)\times10^{43}$ e$^+$ s$^{-1}$ annihilate in 
the warm ionized and HII phases of the disk, 
$(0.46\pm0.08)\times10^{43}$ e$^+$ s$^{-1}$ annihilate in the 
warm HII phase of the bulge, and $(0.48\pm0.17)\times10^{43}$ 
e$^+$ s$^{-1}$ annihilate in the cold neutral phase of the bulge. 
In these phases positronium is formed $\sim$ 89\% and $\sim$ 95\% 
of the time, respectively (Guessoum, Jean \& Gillard 2005; 
Churazov et al. 2005; Jean et al. 2006), so we would expect (as 
outlined in Table 6) a combined positronium annihilation rate of 
$(1.40\pm0.74)\times10^{43}$ e$^+$ s$^{-1}$ out of a total 
positron annihilation rate of $(1.53\pm0.81)\times10^{43}$ e$^+$ 
s$^{-1}$ for a total Galactic positronium fraction $f_{Ps} sim 
92\pm$2\%. 
 
We also expect (Table 6) that of the positronium formed in the 
warm ionized and HII phases $\sim$ 56\% is in the disk and $\sim$ 
44\% is in the bulge, while only $\sim$ 2\% of that formed in the 
neutral phases is in the disk and $\sim$ 98\% is in the bulge. 
This gives an expected positronium fraction of $\sim$ 93$\pm$2\% 
in the bulge and $\sim$ 90$\pm$2\% in the disk (Table 6). {\em 
The mean positronium fraction of $f_{Ps} \sim$ 93$\pm$2\% of the 
Galactic positron bulge within 1.5 kpc is in very good agreement 
with the mean of 94$\pm$6\% measured by SPI/INTEGRAL} (Churazov 
et al. 2005; Jean et al. 2006; Weidenspointner et al. 2006) from 
the inner $\pm 20^o$ of the Galaxy. 
 
\subsection{Broad/Narrow 511 keV Line Ratio} 
 
We also consider the ratio of the broad and narrow components of 
the 511 kev line fluxes, since the bulge and disk values are 
expected to be quite different (Table 6), and compare the 
expected values with those measured by SPI/INTEGRAL (Table 7). 
 
In the warm ionized and HII phases, essentially all ($\sim$99\%) 
of the positronium is formed by radiative combination after the 
positrons are thermalized and its annihilation produces narrow 
($\sim$ 1.2 keV FWHM), 2$\gamma$ 511 keV emission 25\% of the 
time and 3$\gamma$ continuum emission the rest of the time. In 
the cold neutral phases essentially all of the positronium is 
formed by charge exchange in flight producing the broad ($\sim$ 
5.8 keV) doppler shifted 511 keV line, while in the warm neutral 
phase, which accounts for a negligible $\sim$ 2\% of the neutral 
phase positronium formation, the fraction of in-flight formation 
is $\sim$ 70\% for the typical weakly ionized ($x_e \sim 2$\%) 
gas considered here. 
 
Thus, we would expect a total Galactic positron annihilation rate 
of $(0.47\pm0.12)\times10^{43}$ e$^+$ s$^{-1}$ from positronium 
formed in-flight and an annihilation rate of 
$(0.93\pm0.11)\times10^{43}$ e$^+$ s$^{-1}$ from that formed 
thermally by radiative combination. With positronium producing two 
511 keV photons in 25\% of its annihilations, this gives a broad 
511 keV line luminosity of $(0.24\pm0.06)\times10^{43}$ $\gamma$ 
s$^{-1}$ from in-flight annihilation and a narrow line luminosity 
of $(0.46\pm0.06)\times10^{43}$ $\gamma$ s$^{-1}$ from thermal 
annihilation. However, the residual direct thermal annihilation 
of $(0.12\pm0.06)\times10^{43}$ e$^+$ s$^{-1}$ from 
non-positronium processes also produces 2 narrow 511 keV photons 
from every annihilation with a luminosity of 
$(0.26\pm0.12)\times10^{43}$ $\gamma$ s$^{-1}$, which gives a 
total narrow 511 keV line luminosity of luminosity of 
$(0.72\pm0.14)\times10^{43}$ $\gamma$ s$^{-1}$. 
 
We expect (Tables 6 $\&$ 7) that of the bulge and disk 
annihilation that 68\% occurs in the warm ionized gas versus 32\% 
in the neutral gas. Nearly all ($\sim$ 98\%) of the annihilation 
in the disk is in the warm ionized gas, while in the bulge we 
expect that 49$\pm$12\% is in the ionized gas versus 51$\pm$14\% 
in the neutral. The latter values are quite consistent with the 
SPI/INTEGRAL measurements from the bulge region, where Jean et 
al. (2006) find 51$\pm$3\% in the ionized gas versus 49$\pm$3\% 
in the neutral. 
 
We further expect (Table 6) that of the narrow line positronium 
formed in the warm ionized and HII phases $\sim$ 56\% is in the 
disk and $\sim$ 44\% is in the bulge, while again only $\sim$ 2\% 
of that formed in the neutral phases is in the disk and $\sim$ 
98\% is in the bulge. In the bulge this yields a broad 511 keV 
line luminosity of $(0.23\pm0.08)\times10^{43}$ $\gamma$ s$^{-1}$ 
from in-flight annihilation and a narrow line luminosity of 
$(0.34\pm0.06)\times10^{43}$ $\gamma$ s$^{-1}$ from thermal 
annihilation, for a broad/narrow line luminosity ratio of 
0.68$\pm$0.26. In the disk the broad line luminosity is only 
$\sim0.01\times10^{43}$ $\gamma$ s$^{-1}$ and the narrow line 
luminosity is $(0.38\pm0.06)\times10^{43}$ $\gamma$ s$^{-1}$, 
yielding a broad/narrow line luminosity ratio of only $\sim$ 
0.03, and the Galactic mean broad/narrow line luminosity ratio 
from the combined bulge and disk is 0.33$\pm$12. 
 
However, the apparent ratio that would be seen from the Earth if 
the full disk were viewed is quite different, because of different 
relative weighting, since the observed fluxes from the bulge are 
equal to the bulge luminosities divided by 4$\pi R_B^2 \sim 
7.65\times10^{45}$ cm$^2$, where $R_B$ is the distance to the 
Galactic center $\sim$ 8 kpc, and the observed fluxes from the 
disk are equal to the disk luminosities divided by only about 
0.56$\times 4\pi R_B^2$ (Weidenspointner et al. 2008a). Thus the 
broad line flux is $(0.30\pm0.08)\times10^{-3}$ $\gamma$ cm$^{-2}$ 
s$^{-1}$ from the bulge and $(0.02\pm0.01)\times10^{-3}$ $\gamma$ 
cm$^{-2}$ s$^{-1}$ from the disk, while the narrow line flux is 
$(0.44\pm0.08)\times 10^{-3}$ $\gamma$ cm$^{-2}$ s$^{-1}$ from the 
bulge and $(0.89\pm0.14)\times10^{-3}$ $\gamma$ cm$^{-2}$ s$^{-1}$ 
from the disk. This gives a total observed broad 511 keV line 
flux of $(0.32\pm0.11)\times10^{-3}$ $\gamma$ cm$^{-2}$ s$^{-1}$ 
and a narrow line flux of $(1.33\pm0.16)\times10^{-3}$ $\gamma$ 
cm$^{-2}$ s$^{-1}$, which gives an apparent broad/narrow line 
ratio of 0.24$\pm$0.08. This ratio would be observed, however, 
only if the full disk were observed, and for smaller solid angles 
the apparent ratio would be expected to increase approaching the 
bulge value of $\sim$ 0.85, when only the bulge is viewed. {\em 
Thus, we expect the apparent ratio to be quite sensitive to the 
effective sky coverage, and we suggest that this angular 
dependence can test that prediction and further probe the nature 
and structure of the interstellar medium.} 
 
Indeed, analyses of SPI/INTEGRAL observations of the Galactic 
bulge by Jean et al. (2006) give a broad line flux of 
$(0.35\pm0.11)\times10^{-3}$ $\gamma$ cm$^{-2}$ s$^{-1}$ and a 
narrow line flux of $(0.72\pm0.12)\times10^{-3}$ $\gamma$ 
cm$^{-2}$ s$^{-1}$, which gives a broad/narrow line flux ratio of 
0.49$\pm$0.17. A separate analysis by Churazov et al. (2005) 
gives a broad line flux of $(0.24\pm0.10)\times10^{-3}$ $\gamma$ 
cm$^{-2}$ s$^{-1}$ and a narrow line flux of $(0.51\pm0.07)\times 
10^{-3}$ $\gamma$ cm$^{-2}$ s$^{-1}$, which gives a broad/narrow 
line flux ratio of 0.47$\pm$0.20. Considering the large 
uncertainties in both the measurements and the model value, the 
expected bulge ratio of $\sim$ 0.68$\pm$0.26 is well within 
$1-\sigma$. 
 
{\em Moreover, within the positron bulge ($<$ 1.5 kpc), we expect 
that the broad and narrow components of the 511 keV line each 
essentially come from separate and potentially resolvable regions, 
and provide a further test.} The narrow line emission is expected 
to come effectively all from central molecular zone and inner 
tilted disk within 0.5 kpc, arising from thermal positronium and 
direct annihilation in the warm HII shells surrounding molecular 
clouds, ionized by FUV from central O stars. Essentially all of 
the broad line emission on the other hand is expected to come 
from the surrounding outer tilted disk between 0.5 and 1.5 kpc, 
arising from annihilation of positronium formed in flight in the 
cold neutral HI envelope of molecular clouds, where there is 
insufficient FUV to generate surrounding HII shells. 
 
In particular, from the positron annihilation rates in Tables 2 
\& 6, we expect from the HII shells within 0.5 kpc of the 
Galactic Center a narrow line flux of $(0.39\pm0.06)\times10^{-3}$ 
$\gamma$ cm$^{-2}$ s$^{-1}$ and no significant broad line flux, 
giving {\em an expected broad/narrow line ratio of just $\sim 0$ 
within 0.5 kpc.} Whereas from the HI gas in the surrounding 
tilted disk, we expect a broad line flux of 
$(0.30\pm0.10)\times10^{-3}$ $\gamma$ cm$^{-2}$ s$^{-1}$, and a 
narrow line flux of $(0.05\pm0.03)\times10^{-3}$ $\gamma$ 
cm$^{-2}$ s$^{-1}$ from direct thermal annihilation, giving {\em 
an expected broad/narrow line ratio of $\sim 6\pm3$ between 0.5 
kpc and 1.5 kpc, which is nearly 10 times the combined full bulge 
average. Spectral analysis of the emission from these two regions 
should be able to test this prediction.} 
 
\subsection{Halo Contributions} 
 
The 511 keV line flux from the halo makes up nearly half of the 
total Galactic annihilation line flux measured by SPI/INTEGRAL 
(Weidenspointner et al. 2008a) and from our estimates of positron 
propagation we also expect that it accounts for a very similar 
($\sim$ 49\%) of the total line flux. Thus, as we discussed above, 
since this line intensity depends rather strongly on the mean dust 
content of the hot halo plasma, we expect that the Galactic 
averages of both the positronium fraction and the broad/narrow 
511 keV line flux ratio are also very sensitive to the amount of 
dust in the halo. For as Jean et al. (2006) have shown and we 
list in Table 6, if the refractory elements in the hot plasma are 
all in dust grains, only $\sim$ 18\% of the positron annihilation 
occurs via positronium formation, and fully $\sim$ 90\% of the 
511 keV line emission is narrow with a FWHM $\sim$ 2 keV, while 
the remaining 10\% is very broad (FWHM $\sim$ 11 keV). This gives 
a halo flux of $\sim (1.60\pm0.37)\times10^{-3}$ cm$^{-2}$ 
s$^{-1}$ in the narrow line alone. That in turn gives a combined 
bulge and halo flux of $\sim (2.34\pm0.40)\times10^{-3}$ 
cm$^{-2}$ s$^{-1}$, which is quite consistent with that from the 
SPI/INTEGRAL best fit flux of $\sim (2.14\pm0.11)\times10^{-3}$ 
cm$^{-2}$ s$^{-1}$ (Weidenspointner et al. 2008a), see Table 7. 
Combined with disk flux, that would give a total Galactic bulge, 
disk and halo flux of $\sim (3.25\pm0.42)\times10^{-3}$ cm$^{-2}$ 
s$^{-1}$ which is also quite consistent with the combined 
best-fit flux of $\sim (2.87\pm0.25)\times10^{-3}$ cm$^{-2}$ 
s$^{-1}$, from the SPI/INTEGRAL analyses. This also gives 
Galactic average values of the positronium fraction of only $\sim 
0.66\pm0.12$ and the broad/narrow 511 keV line flux ratio of only 
$\sim 0.17\pm0.04$, for which the best-fit SPI/INTEGRAL values 
have not yet been published. 
 
If on the other hand, the grains have all disintegrated, the halo 
flux would still be almost as large, $\sim 
(1.41\pm0.31)\times10^{-3}$ cm$^{-2}$ s$^{-1}$, but it would all 
be in the very wide ($\sim$ 11 keV) 511 keV line, so the Galactic 
average value of the broad/narrow 511 keV line flux ratio would 
jump to $\sim 1.30\pm0.29$, and the positronium fraction would 
increase only slightly to $\sim 0.74\pm0.13$. The actual dust 
content of the halo undoubtedly lies in between and {\em since we 
expect the broad/narrow 511 keV line ratio from the halo beyond 
1.5 kpc to be extremely sensitive, ranging from $\sim$ 0.1 to 
$\infty$, SPI/INTEGRAL measurements of that ratio should provide 
a unique measure of the halo dust.} 
 
\section{Conclusion} 
 
In conclusion, we find that roughly half of the Galactic positrons from 
supernova generated radionuclei are produced in he bulge and inner disk 
within the stellar bulge of $R < 3$ kpc, and half in the stellar disk 
beyond that. Essentially all of the bulge positrons are produced in the 
very hot plasma, through which they diffuse with a calculated mean free 
path of only 1\% of the size of the region. Therefore, before they can 
either slow down and annihilate in the plasma or escape beyond, most 
of them are stopped and annihilate in the very dense ($\sim$100-1000 
H cm$^{-1}$) warm outer shells of the molecular clouds that lie within 
$R <$ 1.5 kpc. The disk positrons, on the other hand, are produced 
about equally in the warm HI gas and the hot superbubbles, which 
blow out into the halo. We calculate that roughly half of the disk 
positrons, which either stream through the low density ($\sim$0.1 
H cm$^{-1}$) neutral HI at close to $c$, or diffuse through the hot 
superbubble plasma, escape into the overlying halo before they can 
stop and annihilate in the low density (0.3-3 H cm$^{-1}$) photoionized 
outer shells of clouds and superbubbles in the disk. Thus, positron 
propagation easily explains the larger annihilation flux from the 
bulge. Moreover, it explains the high observed postronium fraction 
and narrow observed 511 keV line width, that both require annihilation 
predominantly in the warm ionized gas and not in the hot plasma in 
which nearly all of the positrons are produced, either in this or in 
other suggested sources. 
 
We expect these conclusions to be quite robust, since they are based on 
1) extensive observations of galactic supernova rates and distributions, 
2) observationally confirmed models of supernova yields of radionuclei 
and their decay-positron survival fractions, 3) wide ranging 
observations of the properties and distributions of the various phases 
of the interstellar medium and magnetic fields, and 4) the widely 
used photodissociation region (PDR) models of molecular cloud shells. 
From these we have calculated the expected propagation of positrons in 
the various phases, assuming streaming in the neutral phases and diffusion 
in the ionized gas and plasma phases, using diffusion mean free paths 
in the latter phases calculated from a two-component, phenomenological 
model of anisotropic turbulence, which is consistent with observations 
of particle propagation in our best natural laboratory, the 
interplanetary medium. 
 
Thus, as we show in Table 7, when positron propagation is considered, 
the positrons from the $\beta^+$-decay chains of the radioactive 
nuclei, $^{56}$Ni, $^{44}$Ti, and $^{26}$Al, produced in Galactic 
supernovae, can fully account for {\em all} of the features of the 
diffuse Galactic 511 keV and 3-$\gamma$ continuum annihilation 
radiation observed by SPI/INTEGRAL. {\em We have also predicted 
additional measurable features that can not only further test such 
an origin of the positrons but provide new information on the nature 
of the interstellar medium.} 
 
 
This work was supported by NASA's {\em International Gamma-Ray 
Astrophysics Laboratory} Science Program, grant NNG05GE70G.

\clearpage 
 
\begin{figure} 
\centering 
\includegraphics[width=5in]{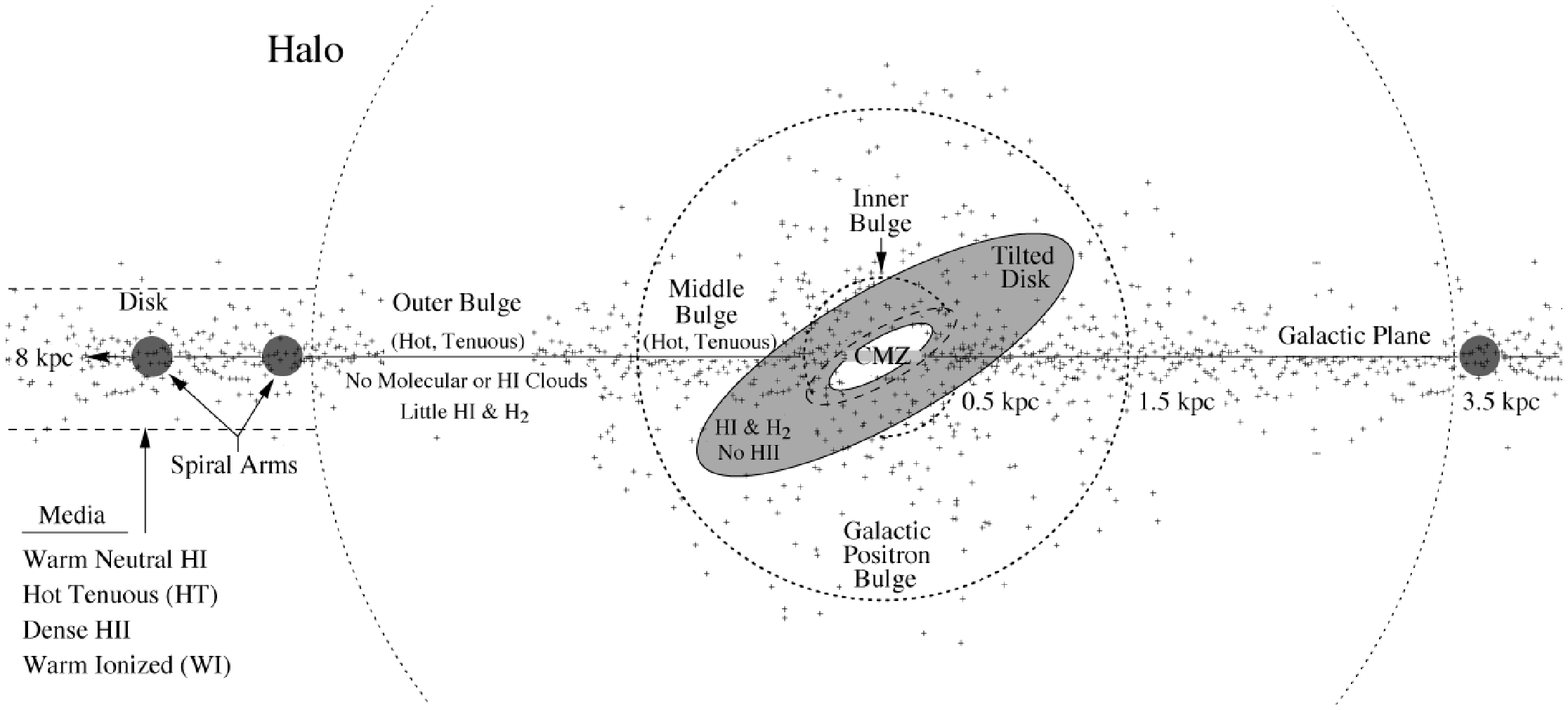} 
\caption{A schematic view of the Galactic Bulge, Disk and Halo assumed in 
these calculations, showing the stellar bulge and disk 
distributions of stars (+) and the assumed boundaries of the 
interstellar gas and plasma subdivisions, chosen for comparison 
with the SPI/INTEGRAL observational analyses.\label{f1}} 
\end{figure} 
 
\clearpage 
 
\begin{figure} 
\centering 
\includegraphics[width=5in]{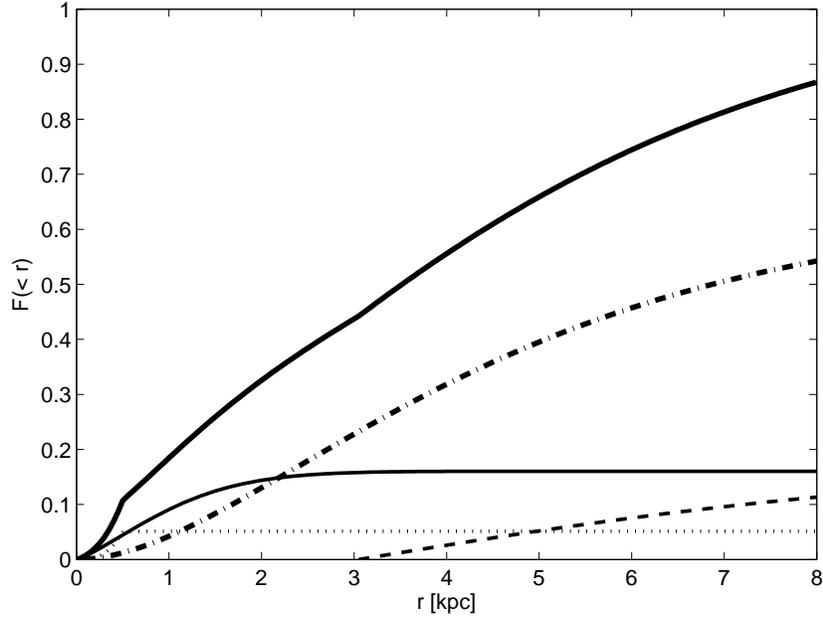} 
\caption{ 
The cummulative fraction of total (thick solid curve) Galactic positron 
production $F(<r)$ in the bulge and disk as a function of Galactic 
radius $r$ from equations (3) and (4). The assumed distribution of 
positrons from decay of $^{56}$Ni and $^{44}$Ti, produced by SNIa and 
SNIp, are shown for the stellar disk (dot-dashed curve), bulge (thin 
solid curve) and a star burst near the Galactic Center (dotted 
curve), while those from decay of $^{26}$Al, produced by massive Wolf 
Rayet stars and core collapse supernovae are shown in the short 
dashed curve. \label{f2} 
} 
\end{figure} 
 
\clearpage 
 
\begin{figure} 
\centering 
\includegraphics[width=5in]{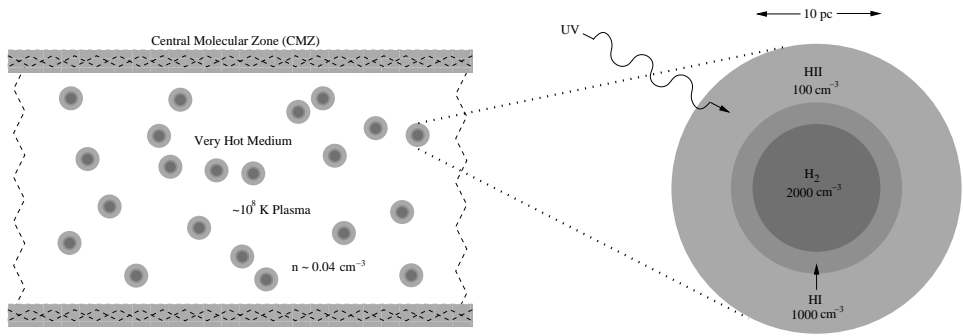} 
\caption{ 
A schematic view of molecular clouds in the Central Molecular 
Zone of the Inner Bulge within 0.5 kpc of the Galactic Center, 
together with a cross section of the HII and HI envelopes 
surrounding the H$^2$ core of a model cloud. \label{f3} 
} 
\end{figure} 
 
\clearpage 
 
\begin{figure} 
\centering 
\includegraphics[width=5in]{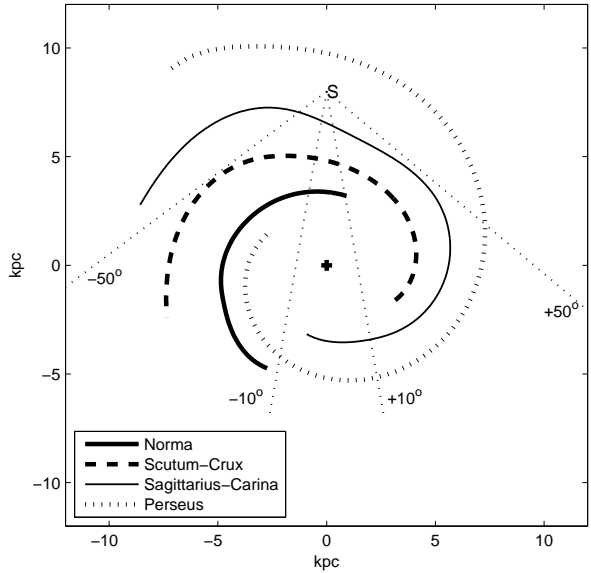} 
\caption{ The Galactic spiral arms, determined from the study by Vall\'ee 
(2005), showing the asymmetry, as viewed from our solar perspective, 
of the inner arms within 50$^\circ$ to either side of the 
positron bulge ($\pm10^\circ$), measured by SPI/INTEGRAL.The asymmetry in 
the annihilation radiation from the Galactic disk is a product of the 
Norma arm dominating the negative longitudes due to its proximity to 
the inner region of the Galaxy, and the weakness of the 
Sagittarius-Carina arm due to its greater distance from the Galactic 
center from which the annihilation rate falls off exponentially from 
the Galactic center and its lower density/activity (Drimmel \& Spergel 
2001) compared to the other 3 arms. We find 511 keV line disk fluxes 
of (3.8$\pm$1.0)$\times$10$^{-4}$ photons cm$^{-2}$s$^{-1}$ from the 
negative longitudes, and (2.3$\pm$0.6)$\times$10$^{-4}$ photons 
cm$^{-2}$s$^{-1}$ from the positive longitudes, and a negative-to-positive 
longitude flux ratio of 1.6$\pm$0.4. These values are quite consistent 
with the corresponding best fit (Weidenspointner et al. 2008b) SPI/INTEGRAL 
line fluxes.\label{f4}} 
\end{figure} 
 
\clearpage 
 
\begin{table} 
\tiny 
\caption{Assumed Properties of the Galactic Interstellar Bulge ($<$3.5 kpc)\label{t1}} 
\begin{tabular}{l|cccc|ccc|c} 
                          &          \multicolumn{4}{|c|}{\bf Inner Bulge}             &    \multicolumn{3}{|c|}{\bf Middle Bulge} &{\bf Outer Bulge}\\ 
Galactic $R$              &            \multicolumn{4}{|c|} {$<$0.5~kpc}               &     \multicolumn{3}{|c|} {0.5-1.5~kpc}     &  1.5-3.5~kpc  \\ 
 & & & & & & & & \\ 
\hline 
Phase                     &      VH       &     HII    &     HI        &     H$_2$      &       HM      &      HI     &    H$_2$    &       Bo      \\ 
Filling Factor $f$        &       1       &    0.0002  &     0.0001    &     0.0004     &        1      &   0.00004   &   0.00005   &        1      \\ 
Temperature $T$ K       &$(90-6)\times10^6$&   5000    &      150      &       30       & $5\times10^6$ &     20      &     20      & $1\times10^6$ \\ 
Density $n$ H cm$^{-3}$   &  0.04-0.007   &     100    &     1000      &   1900 H$_2$   &      0.005    &    1000     &  1000 H$_2$ &     0.002     \\ 
Ionization Fraction $x$   &       1       &      1     &$3\times10^{-4}$ &  $10^{-5}$   &        1     &$3\times10^{-4}$ & $10^{-5}$ &      1       \\ 
Slowing Down $t_{sd}$ yr &$(1-6)\times10^6$&    450    &      90       &       20       & $9\times10^6$ &     900     &     150     & $2\times10^7$ \\ 
Magnetic Field $B$ $\mu$G & $\sim$100-17  & $\sim$100  &   $\sim$100   &   $\sim$100    &    $\sim$12   &   $\sim$12  &   $\sim$12  &       4       \\ 
Fluctuation $\delta B/B$  &$\sim$0.12-0.06&     1.3    &      --       &       --       &   $\sim$0.03  &      --     &      --     &   $\sim$0.5   \\ 
Flux Tube $l_B$ pc        &    10-300     &  $\sim$7   &   $\sim$5     &      $>$7      &   $\sim$1000  &   $\sim$12   &  $\sim$5    & 1000 $\&$ 2000 \\ 
Max MHD Scale $l_o$ pc    &    10-50      &     --     &      --       &      --        &       50      &      --     &     --      &       75      \\ 
$\lambda_{\|}$ pc         &    25-30      &   0.0017   &      --       &      --        &       45      &      --     &     --      &       14      \\ 
Slowing Down $l_{sd}$ pc  &   2000-5000   &     0.3    &      12       &       3        &      7000     &      12     &      6      &      7000     \\ 
$l_B/(l_B + l_{sd})$      &   0.01-0.03   &    0.96    &     0.30      &    $>$0.70     &      0.07     &     0.50    &    0.95     & 0.13 $\&$ 0.22 \\ 
 & & & & & & & & \\ 
$Q_X~~10^{43} e^+$/s      & 0.31$\pm$0.07 &   $\sim$0  &    $\sim$0    &    $\sim$0     & 0.37$\pm$0.09 &   $\sim$0   &   $\sim$0   & 0.56$\pm$0.14 \\ 
\end{tabular} 
\end{table} 
 
\clearpage 
 
\begin{table} 
\caption{Positron Propagation \& Annihilation in the Galactic Interstellar Bulge ($<$3.5 kpc)\label{t2}} 
\tiny 
\begin{tabular}{l|cc|cc|c} 
                        &     \multicolumn{2}{|c|}{\bf Inner Bulge}   & \multicolumn{2}{|c|}{\bf Middle Bulge} &{\bf Outer Bulge}\\ 
Galactic $R$            &       \multicolumn{2}{|c|}{$<$0.5~kpc}      &   \multicolumn{2}{|c|}{0.5-1.5~kpc}     &  1.5-3.5~kpc \\ 
 & & & & & \\ 
\hline 
Phase $X$               &      VH      &      HII      &       HM      &       HI      &      Bo       \\ 
Filling Factor $f$      &       1      &    $\sim$0    &       1       &    $\sim$0    &      1        \\ 
$Q_X~~10^{43} e^+$/s    & 0.31$\pm$0.07&    $\sim$0    & 0.37$\pm$0.09 &    $\sim$0    & 0.56$\pm$0.14 \\ 
 & & & & & \\ 
\hline 
Propagation  & & & & & \\ 
Inner Bulge  & & & & & \\ 
$P_{VH:X}$            &{\bf $\sim$0.02}&  $\sim$0.96  &       --      &       --      & [$\sim$0.02] \\ 
Middle Bulge  & & & & & \\ 
$P_{HM:X}$             &      --       &   $\sim$0.22 &{\bf $\sim$0.07}&  $\sim$0.69  & [$\sim$0.02] \\ 
Outer Bulge & & & & & \\ 
$P_{Bo:X}$             &      --       &   $\sim$0.13 &   $\sim$0.04   &  $\sim$0.40 &[{\bf $\sim$ 0.14}]\\ 
  & & & & & \\ 
\hline 
Annihilation  & & & & & \\ 
Inner Bulge  & & & & & \\ 
$QP_{VH:X}~~10^{43}e^+$/s &{\bf $\sim$0.006}& 0.30$\pm$0.06 &    --   &     --     & [$\sim$0.006]\\ 
Middle Bulge  & & & & & \\ 
$QP_{HM:X}$             &      --      & 0.08$\pm$0.04 &{\bf 0.03$\pm$0.01}& 0.26$\pm$0.13 &[$\sim$0.007]\\ 
Outer Bulge & & & & & \\ 
$QP_{Be:X}$             &      --      & 0.08$\pm$0.04 &  0.02$\pm$0.01 & 0.22$\pm$0.11 &[{\bf $\sim$0.08}]\\ 
 & & & & & \\ 
\hline 
 & & & & & \\ 
$A_X~~10^{43} e^+$/s    & $\sim$0.006  & 0.46$\pm$0.08 & 0.05$\pm$0.02 & 0.48$\pm$0.17 & [$\sim$0.09]\\ 
\end{tabular} 
\end{table} 
 
\clearpage 
 
\begin{table} 
\caption{Assumed Properties of the Outer Bulge, Disk \& Halo\label{t3}} 
\tiny 
\begin{tabular}{l|c|cccc|c} 
                         & {\bf Outer Bulge} &                \multicolumn{4}{|c|}{\bf Disk}                      &  {\bf Halo}   \\ 
Galactic $R$              &   1.5-3.5 kpc    &                 \multicolumn{4}{|c|}{$>$3.5 kpc}                   &  $>$1.5 kpc   \\ 
 & & & & & & \\ 
\hline 
Phase                     &      Bo          &        HT        &        HI       &       WI      &      HII      &      H        \\ 
Filling Factor $f$        &       1          &        0.2       &       0.5       &      0.15     &     0.15      &      1        \\ 
Temperature $T$ K         & $1\times10^6$    &  $1\times10^6$   &      8000       &      8000     &     8000      & $1.6\times10^6$ \\ 
Density $n$ H cm$^{-3}$   &      0.002       &      0.002       &      0.13       &      0.30     &      2.6      &     0.002     \\ 
Ionization Fraction $x$   &       1          &        1         &      0.02       &       0.9     &      0.95     &       1       \\ 
Slowing Down $t_{sd}$ Myr &      23          &        23        &      0.7        &      0.16     &     0.018     &      23       \\ 
Magnetic Field $B$ $\mu$G &       4          &        4         &        4        &        4      &       4       &       4    \\ 
Fluctuation $\delta B/B$  &   $\sim$0.5      &    $\sim$0.5     &       --        &   $\sim$0.5   &    $\sim$1    &   $\sim$0.5   \\ 
Flux Tube $l_B$ pc        &  1000$\&$2000    &     $<$1000      &   $\sim$3000    &      160      &      85       &   $>$10000    \\ 
Max MHD Scale $l_o$ pc    &      75          &        75        &       --        &       10      &       2       &       75      \\ 
$\lambda_{\mid\mid}$ pc   &      14          &        14        &       --        &      0.2      &      0.1      &       14      \\ 
Slowing Down $l_{sd}$  pc &     7000         &       7000       &      90,000     &      120      &      15       &      7000     \\ 
$l_B/(l_B + l_{sd})$      &  0.13$\&$0.22    &     $<$0.13      &       0.03      &      0.56     &     0.85      &     $>$0.6    \\ 
 & & & & & &  \\ 
$Q_X~~10^{43} e^+$/s      & 0.56$\pm$0.14    &   0.45$\pm$0.12  &  0.45$\pm$0.15  & 0.13$\pm$0.04 & 0.13$\pm$0.04 &      0        \\ 
\end{tabular} 
\end{table}

\clearpage 
 
\begin{table} 
\caption{Positron Propagation \& Annihilation in the Outer Bulge, Disk \& Halo\label{t4}} 
\tiny 
 \begin{tabular}{l|c|cccc|c} 
                       &{\bf Outer Bulge}&      \multicolumn{4}{|c|}{\bf Disk}                            &  {\bf Halo}   \\ 
Galactic $R$            &  1.5-3.5 kpc  &       \multicolumn{4}{|c|}{$>$3.5 kpc}                          &  $>$ 1.5 kpc    \\ 
 & & & & & &  \\ 
\hline 
Phase $X$               &      Bo       &      HT       &        HI       &       WI      &      HII      &      H        \\ 
Filling Factor $f_X$    &       1       & 0.2$\pm$0.05  &   0.5$\pm$0.1   & 0.15$\pm$0.05 & 0.15$\pm$0.05 &      1        \\ 
$Q_X~~10^{43} e^+$/s    & 0.56$\pm$0.14 & 0.45$\pm$0.15 &  0.45$\pm$0.15  & 0.13$\pm$0.04 & 0.13$\pm$0.04 &      0        \\ 
 & & & & & & \\ 
\hline 
Propagation & & & & & & \\ 
$P_{Bi/Bm:X}$           &  [$\sim$0.04] &     --        &       --        &       --      &       --      &  $\sim$0.04   \\ 
$P_{Bo:X}$            &[{\bf $\sim$0.14}]&    --        &       --        &       --      & 0.14$\pm$0.07 & 0.29$\pm$0.10 \\ 
$P_{HT:X}$              &      --       &[{\bf $<$0.08}]&       --        & 0.15$\pm$0.07 & 0.15$\pm$0.07 & 0.70$\pm$0.35 \\ 
$P_{HI:X}$              &      --       &     --        & {\bf $\sim$0.02}& 0.16$\pm$0.08 & 0.16$\pm$0.08 & 0.66$\pm$0.33 \\ 
$P_{WI:X}$              &      --       &    $\sim$0    &     $\sim$0   &{\bf 0.76$\pm$0.38}&0.05$\pm$0.02& 0.19$\pm$0.09 \\ 
$P_{HII:X}$             &    $\sim$0    &    $\sim$0    &     $\sim$0    &0.03$\pm$0.01 &{\bf 0.85$\pm$0.42}&0.10$\pm$0.05\\ 
$P_{H:X}$               &    $\sim$0    &    $\sim$0    &     $\sim$0     &    $\sim$0    &    $\sim$0    & {\bf $\sim$1} \\ 
 & & & & & & \\ 
\hline 
Annihilation & & & & & & \\ 
$QP_{Bi/Bm:X}~~10^{43}e^+$/s & [$\sim$0.01] &  --       &       --        &       --      &       --      &   $\sim$0.01  \\ 
$QP_{Bo:X}$           &{\bf [$\sim$0.08]}&     --       &       --        &       --      & 0.08$\pm$0.04 & 0.16$\pm$0.08 \\ 
$QP_{HT:X}$             &      --       &[{\bf $<$0.04}]&       --        & 0.07$\pm$0.03 & 0.07$\pm$0.03 & 0.31$\pm$0.15 \\ 
$QP_{HI:X}$             &      --       &     --        &{\bf $\sim$0.01} & 0.07$\pm$0.03 & 0.07$\pm$0.03 & 0.30$\pm$0.15 \\ 
$QP_{WI:X}$             &      --       &    $\sim$0    &     $\sim$0   &{\bf 0.10$\pm$0.05}& 0.01$\pm$0.01&0.02$\pm$0.01 \\ 
$QP_{HII:X}$            &    $\sim$0    &    $\sim$0    &     $\sim$0     &    $\sim$0  &{\bf 0.11$\pm$0.05}&0.02$\pm$0.01\\ 
$QP_{H:X}$              &    $\sim$0    &    $\sim$0    &     $\sim$0     &    $\sim$0    &    $\sim$0    & {\bf $\sim$0} \\ 
 & & & & & & \\ 
\hline 
 & & & & & & \\ 
$A~~10^{43} e^+$/s      &  [$\sim$0.09] & [$\sim$0.04]  &   $\sim$0.01    & 0.24$\pm$0.07 & 0.34$\pm$0.08 & 0.82$\pm$0.23 \\ 
\end{tabular} 
\end{table} 
 
\clearpage 
 
\begin{table} 
\caption{Summary of Positron Production \& Annihilation in Positron Bulge, Galactic Disk \& Halo\label{t5}} 
\tiny 
\begin{tabular}{lcccc} 
                          & {\bf Production Rate} & {\bf Annihilation Rate} & {\bf Production} & {\bf Annihilation} \\ 
                          & $Q$ $10^{43}$ e$^+$/s &  $A$ $10^{43}$ e$^+$/s  &  $Q/Q_{Tot}$ \%  &   $A/A_{Tot}$ \%   \\ 
&&&&\\ 
Bulge $<$ 1.5 kpc         &     0.68$\pm$0.11     &     1.00$\pm$0.19       &     28$\pm$6     &     41$\pm$8      \\ 
Bulge 1.5-3.5 kpc         &     0.56$\pm$0.14     &        [0.09]*          &     23$\pm$6     &         [4]*      \\ 
Disk $>$ 3.5 kpc          &     1.16$\pm$0.22     &     0.59$\pm$0.11       &     48$\pm$15    &     25$\pm$5      \\ 
Halo $>$ 1.5 kpc          &           0           &     0.82$\pm$0.21       &         0        &     34$\pm$9      \\ 
 &&&&\\ 
Hot Plasma H$^+$          &     1.66$\pm$0.24     &     0.88$\pm$0.21       &     69$\pm$13    &     37$\pm$9       \\ 
Warm HII \& WI            &     0.27$\pm$0.06     &     1.04$\pm$0.14       &     11$\pm$3     &     43$\pm$6       \\ 
Warm Neutral HI           &     0.45$\pm$0.15     &      $\sim$0.01         &     19$\pm$7     &      $\sim$0       \\ 
Cold Neutral HI           &        $\sim$0        &     0.48$\pm$0.17       &      $\sim$0     &     20$\pm$7       \\ 
Molecular H$_2$           &        $\sim$0        &        $\sim$0          &      $\sim$0     &      $\sim$0       \\ 
 & & & &  \\ 
\underline{Hot Plasma} & & & &  \\ 
Bulge $<$ 1.5 kpc         &     0.65$\pm$0.11     &    0.06$\pm$0.02        &     27$\pm$6     &      3$\pm$1       \\ 
Bulge 1.5-3.5 kpc         &     0.56$\pm$0.14     &        [0.09]*          &     23$\pm$6     &         [4]*       \\ 
Disk $>$ 3.5 kpc          &     0.45$\pm$0.12     &        [0.04]*          &     19$\pm$5     &         [2]*       \\ 
Halo $>$ 1.5 kpc          &           0           &    0.82$\pm$0.21        &         0        &     34$\pm$9       \\ 
 & & & &  \\ 
\underline{Warm HII \& WI} & & & &  \\ 
Bulge $<$ 1.5 kpc         &      $\sim$0.01       &     0.46$\pm$0.08       &      $\sim$0     &     19$\pm$4       \\ 
Disk $>$ 3.5 kpc          &     0.26$\pm$0.06     &     0.58$\pm$0.11       &     11$\pm$3     &     24$\pm$5       \\ 
 & & & &  \\ 
\underline{Warm Neutral HI} & & & &  \\ 
Disk $>$ 3.5 kpc          &     0.45$\pm$0.15     &      $\sim$0.01         &     19$\pm$7     &      $\sim$0       \\ 
 & & & &  \\ 
\underline{Cold Neutral HI} & & & &  \\ 
Bulge $<$ 1.5 kpc         &       $\sim$0         &     0.48$\pm$0.17       &      $\sim$0     &     20$\pm$7       \\ 
\hline 
\multicolumn{2}{l}{* Included in Halo} \\ 
\end{tabular} 
\end{table} 
 
\clearpage 
 
\begin{table} 
\caption{Galactic Positronium (Ps) Fraction \& 511 keV Line Luminosities \& Fluxes\label{t6}} 
\tiny 
\begin{tabular}{lcc|cc|cc} 
       &  \multicolumn{2}{c}{\bf Bulge}   & \multicolumn{2}{c}{\bf Disk} & \multicolumn{2}{c}{\bf Halo}  \\ 
       &        HII       &       HI      &   HII \& WI  &     HI        &   Grains     &  No Grains   \\ 
       &     $<$ 0.5 kpc  &  0.5--1.5 kpc &  $>$ 3.5 kpc &  $>$ 3.5 kpc  &  $>$ 1.5 kpc & $>$ 1.5 kpc  \\ 
\hline 
Ps Fraction                 & 0.89 &  0.95  & 0.89 & 0.99 & 0.18 & 0.42 \\ 
In-Flight Ps Fraction       & 0.01 &  0.95  & 0.01 & 0.70 &   0  &   0  \\ 
Thermal Ps Fraction         & 0.88 & $\sim$0& 0.88 & 0.29 & 0.18 & 0.42 \\ 
Direct Processes Fraction   & 0.11 &  0.05  & 0.11 & 0.01 & 0.82 & 0.58 \\ 
e$^+$/$\gamma_{511}$        & 1.50 &  1.74  & 1.50 & 1.94 & 0.58 & 0.73 \\ 
Broad $\gamma_{511}$ \%     &  0   &   83   &   0  &  68  &  10  & 100  \\ 
Narrow $\gamma_{511}$ \%    & 100  &   17   & 100  &  32  &  90  &  0   \\ 
 
\multicolumn{2}{l}{{\bf Annihilation Rate}~~ $10^{43}$ e$^+$ s$^{-1}$} & & & & & \\ 
Total Positron        &  0.46$\pm$0.08  &  0.48$\pm$0.17  & ~0.58$\pm$0.11 & $\sim$0.01 & ~0.82$\pm$0.21 & 0.82$\pm$0.21 \\ 
Total Positronium     &  0.41$\pm$0.08  &  0.46$\pm$0.16  & ~0.52$\pm$0.10 & $\sim$0.01 & ~0.15$\pm$0.04 & 0.34$\pm$0.09 \\ 
In-Flight Positronium &       0         &  0.46$\pm$0.16  &       0       & $\sim$0.01 &       0       &        0      \\ 
Thermal Positronium   &  0.41$\pm$0.08  &     $\sim$0     & ~0.52$\pm$0.10 &  $\sim$0   & ~0.15$\pm$0.04 & 0.34$\pm$0.09 \\ 
Direct Processes      &  0.05$\pm$0.02  &  0.02$\pm$0.01  & ~0.06$\pm$0.01 &  $\sim$0   & ~0.67$\pm$0.17 & 0.48$\pm$0.12 \\ 
Mean P$_s$ Fraction   & \multicolumn{2}{c}{0.93$\pm$0.08} & \multicolumn{2}{c}{0.90$\pm$0.11} & \multicolumn{2}{c}{0.30$\pm$0.12}\\ 
 
\multicolumn{2}{l}{{\bf Luminosity} ~~ $10^{43}$ $\gamma$ s$^{-1}$} & & & & & \\ 
Ps 511 keV Line         & 0.20$\pm$0.04 & 0.23$\pm$0.08 & ~0.26$\pm$0.05 &  $\sim$0.01  & ~0.07$\pm$0.02 & 0.17$\pm$0.05 \\ 
Direct 511 keV Line     & 0.10$\pm$0.03 & 0.04$\pm$0.02 & ~0.12$\pm$0.04 &    $\sim$0   & ~1.35$\pm$0.34 & 0.96$\pm$0.24 \\ 
Total 511 keV Line      & 0.30$\pm$0.05 & 0.27$\pm$0.08 & ~0.38$\pm$0.06 &  $\sim$0.01  & ~1.42$\pm$0.34 & 1.13$\pm$0.25 \\ 
3$\gamma$ Continuum     & 0.88$\pm$0.18 & 1.04$\pm$0.38 & ~1.17$\pm$0.23 & 0.02$\pm$0.01 & ~0.33$\pm$0.08 & 0.77$\pm$0.19 \\ 
Broad 511 keV Line      &    $\sim$0    & 0.23$\pm$0.08 &    $\sim$0     &  $\sim$0.01  & ~0.14$\pm$0.04 & 1.13$\pm$0.25 \\ 
Narrow 511 keV Line     & 0.30$\pm$0.05 & 0.04$\pm$0.02 & ~0.38$\pm$0.06 &    $\sim$0   & ~1.28$\pm$0.30 &       0       \\ 
Broad 511 keV Line      & \multicolumn{2}{c}{0.23$\pm$0.08} & \multicolumn{2}{c}{$\sim$0.01} & \multicolumn{2}{c}{$<0.64\pm0.15>$} \\ 
Narrow 511 keV Line     & \multicolumn{2}{c}{0.34$\pm$0.05} & \multicolumn{2}{c}{0.38$\pm$0.06} & \multicolumn{2}{c}{$<0.64\pm0.15>$} \\ 
Broad/Narrow Line Ratio & \multicolumn{2}{c}{0.68$\pm$0.26} & \multicolumn{2}{c}{$\sim$0.03} & \multicolumn{2}{c}{$<1.0\pm0.2>$} \\ 
Broad/Narrow Line Ratio &                  \multicolumn{4}{c}{0.33$\pm$0.12}               &       &      \\ 
 
\multicolumn{2}{l}{{\bf Flux}~~ $10^{-3}$ $\gamma$ cm$^2$ s$^{-1}$} & & & \\ 
Total 511 keV Line      &  0.39$\pm$0.06  &  0.35$\pm$0.12  & ~0.89$\pm$0.14 &   $\sim$0.02  & ~1.77$\pm$0.43 &  1.41$\pm$0.31 \\ 
3$\gamma$ Continuum     &  1.15$\pm$0.24  &  1.36$\pm$0.50  & ~2.73$\pm$0.65 & 0.05$\pm$0.02 & ~0.41$\pm$0.11 &  0.45$\pm$0.24 \\ 
Broad 511 keV Line      &     $\sim$0     &  0.30$\pm$0.10  &     $\sim$0    &   $\sim$0.02  & ~0.17$\pm$0.05 &  1.41$\pm$0.31 \\ 
Narrow 511 keV Line     &  0.39$\pm$0.06  &  0.05$\pm$0.03  & ~0.89$\pm$0.14 &    $\sim$0    & ~1.59$\pm$0.37 &         0      \\ 
Broad/Narrow Line Ratio &     $\sim$0     &     6$\pm$3     & \multicolumn{2}{c}{$\sim$0.02} &      0.11      &     $\infty$   \\ 
Total 511 keV Line      & \multicolumn{2}{c}{0.74$\pm$0.14} & \multicolumn{2}{c}{0.91$\pm$0.14} & \multicolumn{2}{c}{$<1.59\pm0.37>$} \\ 
Broad 511 keV Line      & \multicolumn{2}{c}{0.30$\pm$0.10} & \multicolumn{2}{c}{$\sim$0.02}    & \multicolumn{2}{c}{$<0.79\pm0.19>$} \\ 
Narrow 511 keV Line     & \multicolumn{2}{c}{0.44$\pm$0.08} & \multicolumn{2}{c}{0.89$\pm$0.13} & \multicolumn{2}{c}{$<0.79\pm0.19>$} \\ 
Broad/Narrow Line Ratio & \multicolumn{2}{c}{0.68$\pm$0.26} & \multicolumn{2}{c}{$\sim$0.02}    & \multicolumn{2}{c}{$<1.0\pm0.2>$} \\ 
Broad 511 keV Line      &                  \multicolumn{4}{c}{0.32$\pm$0.10}                    & & \\ 
Narrow 511 keV Line     &                  \multicolumn{4}{c}{1.33$\pm$0.15}                    & & \\ 
Broad/Narrow Line Ratio &                  \multicolumn{4}{c}{0.24$\pm$0.08}                    & & \\ 
\end{tabular} 
\end{table} 
 
\clearpage 
 
\begin{table} 
\caption{Expected \& Observed Galactic Positron Annihilation Radiation*\label{t7}} 
\vspace*{0.25in} 
\tiny 
\begin{tabular}{lccc} 
                                &    {\bf Expected}    &     {\bf Observed}   &     {\bf Reference}             \\ 
 & & &   \\ 
\hline 
 
{\bf 511 keV Line Flux} & & &  \\ 
$10^{-3}$ $\gamma$/cm$^2$ s & & &  \\ 
Bulge Component                 &     0.74$\pm$0.14      &     0.75$\pm$0.09    & Weidenspointner et al. 2008a \\ 
Disk  Component                 &     0.91$\pm$0.14      &     0.94$\pm$0.16    & "~~~~~~~~"~~~~~~~~"~~~~~~~~"  \\ 
Halo \& Bulge Component         &     2.32$\pm$0.40      &     2.14$\pm$0.11    & "~~~~~~~~"~~~~~~~~"~~~~~~~~"  \\ 
 
Bulge Core $<$0.5 kpc           &     0.39$\pm$0.10      &     0.38$\pm$0.03    & Weidenspointner et al. 2007  \\ 
Bulge Shell 0.5-1.5 kpc         &     0.33$\pm$0.07      &     0.41$\pm$0.06    & "~~~~~~~~"~~~~~~~~"~~~~~~~~" \\ 
Bulge $<$1.5 kpc                &     0.72$\pm$0.13      &     0.79$\pm$0.07    & "~~~~~~~~"~~~~~~~~"~~~~~~~~" \\ 
 
Disk $-50^o<l<0^o$              &      $\sim$0.38        &     0.43$\pm$0.05    & Weidenspointner et al. 2008b    \\ 
Disk $+50^o>l>0^o$              &      $\sim$0.23        &     0.24$\pm$0.05    &  "~~~~~"~~~~~"                  \\ 
 
Disk Ratio ($-50^o-0^o$)/($+50^o-0^o$) &  $\sim$1.6      &     1.8$\pm$0.4      &  "~~~~~"~~~~~"                  \\ 
 & & &  \\ 
{\bf Annihilation Rate} & & &  \\ 
$10^{43}$ e$^+$/s & & &  \\ 
Bulge Component                 &    1.04$\pm$0.15     &     1.15$\pm$0.16     & Weidenspointner et al. 2008a   \\ 
Disk  Component                 &    0.76$\pm$0.11     &     0.81$\pm$0.14     & "~~~~~~~~"~~~~~~~~"~~~~~~~~"         \\ 
Halo \& Bulge Combined          &    1.95$\pm$0.28     &  3.13[1.86*]$\pm$0.20 & "~~[*adjusted e$^+/\gamma_{511}$]~~" \\ 
 
Bulge/Disk Component Ratio      &     1.4$\pm$0.4      &      1.4$\pm$0.3      & "~~~~~~~~"~~~~~~~~"~~~~~~~~"    \\ 
 & & &  \\ 
{\bf Positronium Fraction} & & &  \\ 
Inner Galaxy                    &    0.92$\pm$0.08     &     0.94$\pm$0.06    & Churazov et al. 2005            \\ 
                                &                      &     0.95$\pm$0.03    & Jean et al. 2006                \\ 
                                &                      &     0.92$\pm$0.09    & Weidenspointner et al. 2006     \\ 
 
Disk  $>$ 3.5 kpc               &    0.88$\pm$0.08     &        -- ? --       &              --                 \\ 
Halo  $>$ 1.5 kpc               &    0.65$\pm$0.07     &        -- ? --       &              --                 \\ 
 & & &\\ 
{\bf Broad $\&$ Narrow 511 keV Line} & & &  \\ 
Bulge Warm Ionized Gas \%       &       49$\pm$12      &       51$\pm$3       & Jean et al. 2006                \\ 
Bulge Cold Neutral Gas \%       &       51$\pm$14      &       49$\pm$3       &  "~~~~~"~~~~~"                  \\ 
 
Broad/Narrow Line Ratio & & &  \\ 
Bulge Region                    &    0.68$\pm$0.26     &     0.49$\pm$0.17    & Jean et al. 2006                \\ 
                                &                      &     0.47$\pm$0.20    & Churazov et al. 2005            \\ 
Bulge $<$ 0.5 kpc               &       $\sim$0        &        -- ? --       &              --                 \\ 
Bulge 0.5-1.5 kpc               &       6$\pm$3        &        -- ? --       &              --                 \\ 
Disk  $>$ 3.5 kpc               &      $\sim$0.02      &        -- ? --       &              --                 \\ 
 
\end{tabular} 
\end{table} 
\footnotesize{* For a Solar Galactocentric distance of 8.5 kpc, assumed in the 
INTEGRAL analyses, instead of 8 kpc, assumed in this study.} 
 
\end{document}